\documentclass[authoryear,preprint,review,8pt]{elsarticle}

\usepackage{natbib}
\setcitestyle{numbers,square}
\usepackage{amssymb}
\usepackage{bm}
\usepackage{amsmath}
\usepackage[lined,boxed,commentsnumbered,ruled]{algorithm2e}
\usepackage{siunitx}
\usepackage{subfigure}
\usepackage{graphicx} 
\usepackage{epstopdf}
\usepackage{booktabs}
\usepackage{xcolor}
\usepackage{changes}
\usepackage{todonotes}
\usepackage{hyperref}

\usepackage{multirow}
\usepackage{booktabs}
\usepackage{array}
\usepackage{amsmath}
\usepackage{siunitx}
\usepackage{xcolor}
\colorlet{red}{black}

\journal{Journal of Computational Physics}

\begin{document}

\begin{frontmatter}

\title{An Implicit Discrete Adjoint Gas-Kinetic Scheme for Aerodynamic Shape Optimization across all Mach Number Regimes}

 \author[label1]{Hangkong Wu}
 \author[label1]{Yuze Zhu}
 \author[label3]{Yajun Zhu}
 \author[label1,label4,label5]{Kun Xu\corref{cor1}}
\cortext[cor1]{Corresponding author}
 \ead{makxu@ust.hk}
\affiliation[label1]{organization={Department of Mathematics, Hong Kong University of Science and Technology},
             city={ Clear Water Bay, Kowloon},
             state={Hong Kong},
             country={China}}
 \affiliation[label3]{organization={Research and Development Office, Shanghai Suochen Information Technology Co., Ltd},
             city={Shanghai},
             country={China}}
 \affiliation[label4]{organization={Department of Mechanical and Aerospace Engineering, Hong Kong University of Science and Technology},
             city={ Clear Water Bay, Kowloon},
             state={Hong Kong},
             country={China}}
\affiliation[label5]{organization={Shenzhen Research Institute, Hong Kong University of Science and Technology},
             city={Shenzhen},
             country={China}}

\begin{abstract}
The gas-kinetic scheme (GKS) integrates the characteristics of flux difference scheme (FDS) and flux vector splitting (FVS) scheme, providing high accuracy in smooth regions and strong robustness near discontinuities across all Mach regimes. Leveraging these properties, an implicit discrete adjoint GKS is developed for aerodynamic shape optimization over a wide range of Mach numbers.
The adjoint solver is constructed using the source-transformation-based algorithmic differentiation tool Tapenade. To enhance computational efficiency, both the flow and adjoint GKS equations are solved using an implicit time-marching strategy, also known as the Lower–Upper Symmetric Gauss–Seidel (LU-SGS) method. The effectiveness of the implicit formulation is demonstrated through comparisons with the explicit approach.
To accurately impose solid wall boundary conditions, particularly in hypersonic regimes, kinetic boundary conditions and their adjoint counterparts are formulated for both adiabatic no-slip and isothermal walls.
Four benchmark test cases covering subsonic, transonic, supersonic, and hypersonic flows are used to verify the effectiveness of the developed adjoint-based design optimization system.

\end{abstract}


\end{frontmatter}

\section{Introduction}\label{introduction}

Aerodynamic shape optimization is a fundamental problem in aerospace design, where accurate prediction of flow physics and efficient gradient evaluation are essential for obtaining optimal configurations. Among the available approaches, including finite difference methods (FDM), adjoint-based methods~\cite{Jameson1988Aerodynamic, Giles2003, KENWAY2019100542, YUAN2024113366, YUAN2025114102} have received increasing attention for gradient evaluation due to their ability to compute sensitivities with respect to a large number of design variables at a cost that is effectively independent of the parameter dimension. The adjoint methods have been extensively applied to aerodynamic optimization problems governed by the Navier–Stokes equations~\cite{Jameson1998, aerospace10020106, Wu2024JoT}, achieving notable success. 
Despite these advances, significant challenges remain in achieving robust and efficient optimization across a wide range of Mach numbers. Conventional CFD solvers are typically based on either flux difference scheme (FDS)~\cite{Roe1981} or flux vector splitting (FVS)~\cite{StegerWarming1981} scheme. FDS-type schemes generally provide low numerical dissipation and high accuracy in smooth flow regions but may suffer from stability issues in the presence of strong discontinuities. While certain modifications, such as entropy fixes~\cite{Harten1983}, can be introduced to alleviate this issue, they may have an effect on solution accuracy. In contrast, FVS-type schemes offer improved robustness and shock-capturing capabilities, albeit at the expense of increased numerical dissipation. Consequently, developing numerical methods that achieve both high accuracy and robustness across a broad spectrum of Mach regimes is critical for design optimization. 

The gas-kinetic scheme (GKS)~\cite{XU19949,XU2001289}, derived from the Boltzmann equation, offers a unified framework for simulating continuum flows across all Mach regimes. In GKS, the time evolution of the gas distribution function naturally incorporates both particle transport and collision processes, yielding a flux function that consists of equilibrium and non-equilibrium components. These components exhibit FDS-like and FVS-like characteristics, respectively, and their relative contributions are governed by the ratio of the time step to the numerical collision time. This adaptive mechanism enables GKS to achieve low-dissipation solutions in smooth regions~\cite{YANG2022110706} while maintaining robust shock-capturing capability near discontinuities~\cite{XU2005405,ZHAO2023111921,Ji2021_CGKS3D,Zhang2023_RotatingCGKS}. Consequently, GKS offers a robust and accurate framework for simulations across a wide range of Mach regimes~\cite{PAN2016197,JXAIAA}. 
Given these advantages, it is of great interest to develop adjoint methods within the GKS framework for aerodynamic shape optimization. 

However, the development of efficient and robust adjoint GKS solvers remains limited and presents several challenges. First, GKS requires the evaluation of the gas distribution function at cell interfaces, and manually deriving the corresponding adjoint formulation is nontrivial due to its inherent complexity. Second, explicit adjoint GKS methods~\cite{Wu2026_AdjointGKS} incur substantial computational cost and are constrained by restrictive stability limits on the Courant number. This limitation becomes increasingly severe in large-scale optimization problems. Third, the accurate treatment of boundary conditions~\cite{Zhang2026_WallBC_GKS, Li2005_KineticBC_GKS}, particularly for solid walls in hypersonic flows, is challenging and necessitates a consistent formulation for both the primal and adjoint systems.

To address these challenges, an implicit discrete adjoint gas-kinetic scheme is developed in the present work for aerodynamic shape optimization across all Mach regimes. The discrete adjoint solver is constructed using the source-transformation-based algorithmic differentiation tool Tapenade~\cite{Tapenade,Wu2021,Wu2024AIAA}, ensuring strict consistency between the primal and adjoint discretizations. To improve computational efficiency, an implicit time-marching strategy is employed, and both the flow and adjoint equations are solved using the Lower–Upper Symmetric Gauss–Seidel (LU-SGS) method~\cite{Wang2018_LUSGS_HB}, allowing significantly larger Courant numbers compared with explicit approaches~\cite{Yang2023_ImplicitGKS_II, Yang2022_ImplicitGKS_I}.
In addition, kinetic boundary conditions and their adjoint counterparts are formulated for solid walls, including both adiabatic no-slip and isothermal conditions. These boundary treatments are particularly important for accurately simulating hypersonic flows. The proposed framework ensures consistent treatment of boundary fluxes in both the primal and adjoint systems, which is crucial for accurate sensitivity evaluation. 
The developed method is verified through four representative benchmark cases covering subsonic, transonic, supersonic, and hypersonic flow regimes.

\section{Gas-Kinetic Scheme}\label{methodology}

In this section, the algorithm of GKS is presented first, followed by the introduction of the kinetic boundary condition. 

\subsection{Algorithm of GKS}

The Boltzmann equation with the Bhatnagar-Gross-Krook (BGK) model~\cite{BGK1954} for the collision term is given by
\begin{equation}\label{BGK}
f_t +\boldsymbol u\cdot \boldsymbol f_x = \frac{g-f}{\tau}
\end{equation}
where $f$ denotes the velocity distribution function, $f_t$ and $\boldsymbol f_x$ represent the temporal and spatial partial derivatives of $f$ respectively, $\boldsymbol u$ is the microscopic velocity vector, $\tau$ is the collision time, and $g$ is the equilibrium distribution function, which satisfies the following Maxwellian distribution
\begin{equation}\label{Maxwel}
g= \rho \Big(\frac{\lambda}{\pi}\Big)^{\frac{K+2}{2}}e^{-\lambda [(\boldsymbol u - \boldsymbol U)^2+\xi^2]}
\end{equation}
where $\boldsymbol U$ is the macroscopic velocity vector, $\rho$ is the density, $\lambda=\rho/(2p)$ with $p$ denoting the static pressure, $K$ is the number of internal degrees of freedom, and $\xi_i(i=1,2,...,K)$ represents the internal variable for each degree of freedom, with $\boldsymbol \xi^2 = \sum_{i=1}^{K}\xi_i^2$.

The integral solution of Eq.~\ref{BGK} is given by
\begin{equation}\label{int_f}
f(\boldsymbol x,\boldsymbol u,\boldsymbol \xi,t) = \frac{1}{\tau}\int_{0}^tg(\boldsymbol x^{'},\boldsymbol u, \boldsymbol \xi,t^{'})e^{-(t-t^{'})/\tau} dt^{'} + e^{-t/\tau}f_0({\boldsymbol x_0,\boldsymbol u,\boldsymbol \xi,0})
\end{equation}
where $\boldsymbol x^{'}=\boldsymbol x - \boldsymbol u t^{'}$ is the particle trajectory, $\boldsymbol x_0$ is the initial position of a particle, $f_0$ is the initial distribution at the beginning of each time step, $\tau$ is the relaxation time. For notational simplicity, we set $\boldsymbol x=\boldsymbol 0$ in the following derivations.

According to the Chapman-Enskog expansion, $f_0$ in Eq.~\ref{int_f} can be expressed by
\begin{gather}\label{exp_f0}
\begin{aligned}
f_0(\boldsymbol x_0,\boldsymbol u,\boldsymbol \xi,0) =& g^{l}(\boldsymbol 0,\boldsymbol u,\boldsymbol \xi,0)[1-(t+\tau){\boldsymbol a}^{l}\cdot \boldsymbol u-\tau{A}^{l}]H[u_n]+\\
&g^{r}(\boldsymbol 0,\boldsymbol u,\xi,0)[1-(t+\tau){\boldsymbol a}^{r}\cdot \boldsymbol u-\tau{A}^{r}][1-H[u_n]]
\end{aligned}
\end{gather}
where $g^{l}$ and $g^{r}$ are the equilibrium distribution at the left and right hand sides of an interface, and are determined by the reconstructed macroscopic flow variables $Q^l$ and $Q^r$, $u_n$ is the normal part of $\boldsymbol u$, $H[u_n]$ is the Heaviside function, $\boldsymbol {a}^l$ and ${\boldsymbol a}^r$ are the spatial derivatives of $g^l$ and $g^r$, while $A^l$ and $A^r$ are the temporal derivatives of $g^l$ and $g^r$.

According to the Taylor expansion, $g$ in Eq.~\ref{int_f} can be expressed by
\begin{equation}\label{exp_g}
g(\boldsymbol x^{'},\boldsymbol u, \boldsymbol \xi, t^{'}) = g^c(\boldsymbol 0,\boldsymbol u,\boldsymbol \xi,0)(1-\boldsymbol a^c\cdot \boldsymbol ut^{'} + A^c t^{'} )
\end{equation}
where $g^c$ is the equilibrium distribution function computed by the kinetic-averaged macroscopic flow variables $Q^c$ at the cell interface.

In Eqs.~\ref{exp_f0} and ~\ref{exp_g}, $\boldsymbol a^{l/r/c}$ and $\boldsymbol A^{l/r/c}$ are computed using the following formulae
\begin{gather}
\begin{aligned}
\boldsymbol a^{l/r/c}=\frac{1}{g^{l/r/c}}\frac{\partial g^{l/r/c}}{\partial x}=\frac{\partial ln g^{l/r/c}}{\partial x}\\
A^{l/r/c}=\frac{1}{g^{l/r/c}}\frac{\partial g^{l/r/c}}{\partial t}=\frac{\partial ln g^{l/r/c}}{\partial t}
\end{aligned}
\end{gather}
To save space, the detailed expressions of $\boldsymbol a^{l/r/c}$ and $A^{l/r/c}$ will be omitted here; interested readers can refer to Ref. \text{\cite{XU2001289}} for the full derivations.

Substituting Eqs.~\ref{exp_f0} and ~\ref{exp_g} into Eq.~\ref{int_f}, the distribution function $f$ at $\boldsymbol x=0$ can be expressed by
\begin{gather}\label{df}
\begin{aligned}
f(\boldsymbol 0,\boldsymbol u,\boldsymbol \xi,t)=&(1-e^{-\frac{t}{\tau_n}})g^c+\\
&[(t+\tau)e^{-\frac{t}{\tau_n}}-\tau]\boldsymbol a^c\cdot \boldsymbol ug^c +\\
&(t-\tau+\tau e^{-\frac{t}{\tau_n}})A^c g^c + \\
&e^{-\frac{t}{\tau_n}}[g^l H[u_n] + g^r(1-H[u_n])]-\\
&e^{-\frac{t}{\tau_n}}(\tau + t)[\boldsymbol a^l\cdot \boldsymbol u g^l H[u_n] + \boldsymbol a^r\cdot \boldsymbol u g^r (1-H[u_n])] - \\
&\tau e^{-\frac{t}{\tau_n}}[g^lA^l H[u_n] + g^r A^r(1-H[u_n])]
\end{aligned}
\end{gather}
where $\tau_n$ is the numerical collision time, defined by
\begin{equation}
\tau_n =\frac{\mu_L}{p}+C|\frac{p_l-p_r}{p_l+p_r}|\Delta t
\end{equation}
where $\Delta t$ is the time step, $\mu_L$ is the laminar viscosity coefficient, $p_l$ and $p_r$ are the static pressure at the left and right hand side of an interface, respectively, and $C$ is a constant, typically chosen in the range $1\leq C \leq 10$. 

Based on the compatibility condition, the relationships between the macroscopic flow variables and fluxes and the distribution function are given by
\begin{align}
\boldsymbol Q   &= \int f\boldsymbol \Psi   d\Xi\label{flowvar}\\
\boldsymbol{F} &= \int f\Psi \boldsymbol{u} d \Xi\label{Flux}
\end{align}
where $\boldsymbol \Psi$ is the collision invariant and has the following form
\begin{equation}\label{invariant}
\boldsymbol \Psi = [\Psi_1, \Psi_2, \Psi_3, \Psi_4]^T=[1, u, v, \frac{1}{2}(u^2+v^2+\boldsymbol \xi^2)]^T
\end{equation}

\subsection{Kinetic Boundary Condition}

Solid wall boundary conditions primarily comprise adiabatic slip walls, adiabatic no-slip walls, and isothermal walls, among others. In this section, we focus on the treatment of adiabatic no-slip and isothermal wall boundary conditions from a kinetic-theory perspective.
\begin{figure}[h!]
    \centering
\includegraphics[width=0.4\linewidth]{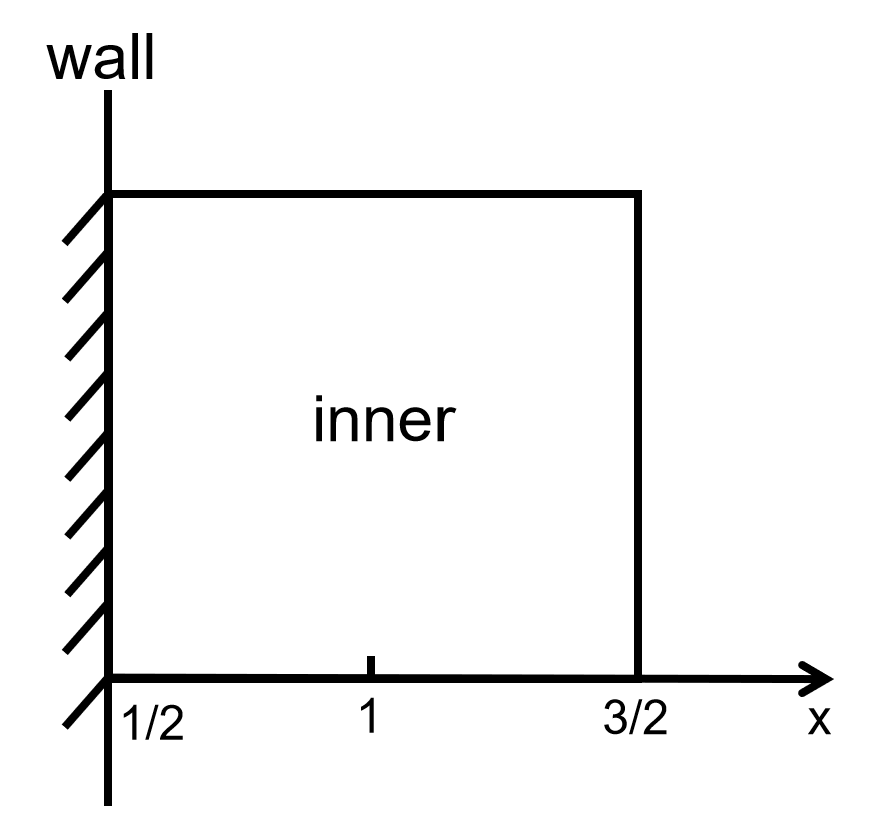}
    \caption{The schematic of the solid wall.}
   \label{wall}
\end{figure}
Assuming that the solid wall is located at $\boldsymbol{x}=1/2$, with the adjacent inner cell on the right (see Fig.~\ref{wall}).

\subsubsection{Isothermal Wall Boundary Condition}

Under the non‑penetration condition, 
the mass flux at the wall is evaluated using the following equation
\begin{equation}\label{m_bc}
F_{m_{b}} + F_{m_{i}} = 0
\end{equation}
where $F_m$ represents the mass flux, and the subscripts $i$ and $b$ represent the inner and boundary cells, respectively. 

The definition of $F_{m_{i}}$ is given by
\begin{gather}\label{m_in}
\begin{aligned}
F_{m_{i}}& = \int_{t_n}^{t_{n+1}}\int_{u<0}\int_{\xi} u f_{i} d\xi du dt 
 \\
&=\int_{t_n}^{t_{n+1}}\int_{u<0}\int_{\xi} u f^{r} d\xi du dt\\
&=\int_{t_n}^{t_{n+1}}\int_{u<0}\int_{\xi} u g^r[1-(t+\tau){\boldsymbol a}^{r}\cdot \boldsymbol u-\tau{A}^{r}]d\xi du dt
\end{aligned}
\end{gather}

The definition of $F_{m_{b}}$ is given by
\begin{gather}\label{m_w}
\begin{aligned}
F_{m_{b}}&=\int_{t_n}^{t_{n+1}}\int_{u>0}\int_{\xi}u g_b d\xi du dt\\
&=\rho_b \Big[\frac{\boldsymbol U_b}{2}\text{erfc}(-\sqrt{\lambda_b}\boldsymbol U_b) + \frac{1}{2}\frac{e^{-\lambda_b^2\boldsymbol U_b^2}}{\sqrt{\pi \lambda_b}} \Big] \Delta t
\end{aligned}
\end{gather}
where $\text{erfc}$ is the error function, defined by
\begin{equation*}
\text{erfc}(x) = 1- \text{erf}(x) = \frac{2}{\sqrt{\pi}}\int_x^{+\infty} e^{-t^2} dt
\end{equation*}
where $\rho_b$, $\lambda_b$, $\boldsymbol U_b$, and $g_b$ are the variables on the solid wall. 

For the isothermal wall boundary condition, the temperature ($T_b$) on the wall is known \textit{a priori}, which can be used to compute $\lambda_w$ 
\begin{equation}\label{iso_wall}
\lambda_b = \frac{1}{2R_gT_b}
\end{equation}
where $R_g$ where $R$ is the gas constant. 

Substituting Eqs.~\ref{m_in} and ~\ref{m_w} into Eq.~\ref{m_bc} leads to the following explicit expression of $\rho_b$
\begin{equation}\label{rho_w2}
\rho_b = -\frac{F_{m_{i}}}{\Big[\frac{\boldsymbol U_b}{2}\text{erfc}(-\sqrt{\lambda_w}\boldsymbol U_b) + \frac{1}{2}\frac{e^{-\lambda_b^2\boldsymbol U_b^2}}{\sqrt{\pi \lambda_b}} \Big]}
\end{equation}

\subsubsection{Adiabatic No-slip Wall Boundary Condition}

Under the adiabatic no-slip wall boundary condition, $\boldsymbol U_b$ is equal to zero at the wall. Therefore, equations~\ref{m_w} and ~\ref{rho_w2} will be reduced to
\begin{equation}\label{mw_reduced}
F_{m_{b}}= \frac{\rho_b}{2\sqrt{\pi\lambda_b}},
\rho_b =-2 \sqrt{\pi \lambda_b} F_{m_{i}}
\end{equation}

Furthermore, the wall temperature ($T_b$) is not known \textit{a priori}. To evaluate $\rho_b$, $T_b$ must first be determined from the energy flux conservation, which is expressed as
\begin{equation}\label{E_bc}
F_{E_b} + F_{E_{i}} = 0
\end{equation}
where $F_E$ represents the energy flux. 

The definition of $F_{E_{i}}$ is given by
\begin{gather}\label{E_in}
\begin{aligned}
F_{E_{i}} &= \int_{t_n}^{t_{n+1}}\int_{u<0}\int_{\xi} \frac{1}{2}(u^2 + v^2 + \boldsymbol \xi^2) u f_{i} d\xi du dt \\
&=\int_{t_n}^{t_{n+1}}\int_{u<0}\int_{\xi} \frac{1}{2}(u^2 + v^2 + \boldsymbol \xi^2) f^r 
u d\xi du dt \\
&=\int_{t_n}^{t_{n+1}}\int_{u<0}\int_{\xi} \frac{1}{2}(u^2 + v^2 + \boldsymbol \xi^2)u g^r[1-(t+\tau){\boldsymbol a}^{r}\cdot \boldsymbol u-\tau{A}^{r}]d\xi du dt
\end{aligned}
\end{gather}

The definition of $F_{E_b}$ is given by
\begin{equation}\label{E_w}
F_{E_{b}}=\int_{t_n}^{t_{n+1}}\int_{u>0}\int_{\xi}\frac{1}{2}(u^2 + v^2 + \boldsymbol \xi^2)ug_b d\xi du dt= \frac{3+K}{4}\frac{F_{m_b}}{\lambda_b}
\end{equation}

Substituting Eq.~\ref{m_bc} and ~\ref{E_bc} into Eq.~\ref{E_w} leads to the following explicit expression of $\lambda_b$
\begin{equation}\label{lam_w}
\lambda_b = \frac{3+K}{4}\frac{F_{m_{i}}}{F_{E_{i}}}
\end{equation}

Substituting Eq.~\ref{lam_w} into Eq.~\ref{mw_reduced} yields the wall density $\rho_b$. With all flow variables, including $\rho_b$, $\boldsymbol{U}_b$, and $\lambda_b$, determined, the reflected Maxwellian distribution $g_b$ can be constructed, leading to the following distribution at the wall:
\begin{equation}
f_b = f_{i}(1-H[u_n]) + g_{b}H[u_n]
\end{equation}

Based on $f_b$, the flux and macroscopic flow variables at the boundary can be evaluated using Eq.~\ref{Flux}.
\begin{equation}\label{QF_b}
\boldsymbol Q_b = \int f\Psi \boldsymbol d \Xi,\quad 
\boldsymbol F_b = \int f\Psi \boldsymbol u d \Xi
\end{equation}

\section{Adjoint Gas-Kinetic Scheme}

In this section, a detailed introduction to the adjoint method and the implicit time-marching scheme for solution acceleration is presented.

\subsection{Adjoint Formulation}

The aerodynamic design optimization problem can be formulated as
\begin{align}
\min_{\boldsymbol{\alpha}} \quad & I = I(\boldsymbol{Q}_i, \boldsymbol{Q}_b, \boldsymbol{\alpha}), \label{obj} \\
\text{s.t.} \quad & \boldsymbol{R}(\boldsymbol{Q}_i, \boldsymbol{Q}_b, \boldsymbol{\alpha}) = \boldsymbol{0}, \label{st1}
\end{align}
where $I$ is the objective function, $\boldsymbol{\alpha}$ denotes the design variables, and $\boldsymbol{Q}_i$ and $\boldsymbol{Q}_b$ represent the flow variables in the interior and at the boundary, respectively. The boundary variables are implicitly dependent on the interior solution and design variables through the boundary condition,
\begin{equation}
\boldsymbol{Q}_b = \boldsymbol{Q}_b(\boldsymbol{Q}_i, \boldsymbol{\alpha}).
\end{equation}

\subsubsection{Lagrangian Formulation}

To derive the adjoint equations, the Lagrangian function is introduced as
\begin{equation}
\mathcal{L} = I(\boldsymbol{Q}_i, \boldsymbol{Q}_b, \boldsymbol{\alpha}) 
- \boldsymbol{\lambda}^T \boldsymbol{R}(\boldsymbol{Q}_i, \boldsymbol{Q}_b, \boldsymbol{\alpha}),
\end{equation}
where $\boldsymbol{\lambda}$ denotes the adjoint variables.

At a local optimum, the Karush–Kuhn–Tucker (KKT) conditions require
\begin{equation}
\frac{d\mathcal{L}}{d\boldsymbol{Q}_i} = \boldsymbol{0}, 
\qquad
\frac{d\mathcal{L}}{d\boldsymbol{\alpha}} = \boldsymbol{0}.
\end{equation}

\subsubsection{Adjoint Equations}

Taking the total derivative of $\mathcal{L}$ with respect to $\boldsymbol{Q}_i$ yields
\begin{equation}
\frac{d \mathcal{L}}{d \boldsymbol{Q}_i}
=
\frac{d I}{d \boldsymbol{Q}_i}
-
\boldsymbol{\lambda}^T
\frac{d \boldsymbol{R}}{d \boldsymbol{Q}_i}.
\end{equation}

Using the chain rule, the derivatives of $I$ and $\boldsymbol{R}$ are written as
\begin{align}
\frac{d I}{d \boldsymbol{Q}_i} 
&=
\frac{\partial I}{\partial \boldsymbol{Q}_i}
+
\frac{\partial I}{\partial \boldsymbol{Q}_b}
\frac{\partial \boldsymbol{Q}_b}{\partial \boldsymbol{Q}_i}, \label{dIdQ} \\
\frac{d \boldsymbol{R}}{d \boldsymbol{Q}_i}
&=
\frac{\partial \boldsymbol{R}}{\partial \boldsymbol{Q}_i}
+
\frac{\partial \boldsymbol{R}}{\partial \boldsymbol{Q}_b}
\frac{\partial \boldsymbol{Q}_b}{\partial \boldsymbol{Q}_i}. \label{dRdQ}
\end{align}

Substituting Eqs.~\eqref{dIdQ}--\eqref{dRdQ} into the KKT condition leads to
\begin{equation}
\left(
\frac{\partial \boldsymbol{R}}{\partial \boldsymbol{Q}_i}
+
\frac{\partial \boldsymbol{R}}{\partial \boldsymbol{Q}_b}
\frac{\partial \boldsymbol{Q}_b}{\partial \boldsymbol{Q}_i}
\right)^T
\boldsymbol{\lambda}
=
\left(
\frac{\partial I}{\partial \boldsymbol{Q}_i}
+
\frac{\partial I}{\partial \boldsymbol{Q}_b}
\frac{\partial \boldsymbol{Q}_b}{\partial \boldsymbol{Q}_i}
\right)^T.
\label{adj_eqn}
\end{equation}

Equation~\eqref{adj_eqn} defines the discrete adjoint system.

\subsubsection{Sensitivity Evaluation}

The gradient with respect to the design variables is obtained from
\begin{equation}
\frac{d \mathcal{L}}{d \boldsymbol{\alpha}}
=
\frac{\partial I}{\partial \boldsymbol{\alpha}}
+
\frac{\partial I}{\partial \boldsymbol{Q}_b}
\frac{\partial \boldsymbol{Q}_b}{\partial \boldsymbol{\alpha}}
-
\boldsymbol{\lambda}^T
\left(
\frac{\partial \boldsymbol{R}}{\partial \boldsymbol{\alpha}}
+
\frac{\partial \boldsymbol{R}}{\partial \boldsymbol{Q}_b}
\frac{\partial \boldsymbol{Q}_b}{\partial \boldsymbol{\alpha}}
\right).
\label{dLda}
\end{equation}

For a given design, the gradient is generally nonzero and can be used as a convergence indicator for the optimization process. At a local optimum, the norm of the gradient approaches zero.

\subsubsection{Role of Boundary Contributions}

It is important to emphasize that both the adjoint Jacobian matrix and the sensitivity derivatives include contributions from the boundary. In particular, the terms
\begin{equation}
\frac{\partial \boldsymbol{R}}{\partial \boldsymbol{Q}_b}
\frac{\partial \boldsymbol{Q}_b}{\partial \boldsymbol{Q}_i}
\end{equation}
and
\begin{equation}
\frac{\partial \boldsymbol{R}}{\partial \boldsymbol{Q}_b}
\frac{\partial \boldsymbol{Q}_b}{\partial \boldsymbol{\alpha}}
\end{equation}
account for the implicit influence of the boundary conditions on the adjoint Jacobian matrix and the corresponding sensitivities.

For GKS, the boundary residual is evaluated from the boundary flux, which, in turn, is constructed from the gas distribution function at the wall. Specifically, the wall distribution function depends on both the macroscopic flow variables and the design variables. This dependency can be summarized as the following mapping:
\begin{equation}
(\boldsymbol{Q}_i, \boldsymbol{Q}_b, \boldsymbol{\alpha})
\;\rightarrow\;
f_b
\;\rightarrow\;
\boldsymbol{F}_b
\;\rightarrow\;
\boldsymbol{R},
\end{equation}
where $f_b$ denotes the distribution function at the boundary and $\boldsymbol{F}_b$ is the corresponding numerical flux.

Accordingly, the residual can be expressed as a nested functional dependence,
\begin{equation}
\boldsymbol{R}
=
\boldsymbol{R}(\boldsymbol{F}_b)
=
\boldsymbol{R}(\boldsymbol{F}_b(f_b))
=
\boldsymbol{R}(\boldsymbol{F}_b(f_b(\boldsymbol{Q}_i, \boldsymbol{Q}_b, \boldsymbol{\alpha}))),
\end{equation}
which introduces a multi-level coupling between the flow variables and the boundary treatment.

As a result, the boundary contribution to the adjoint Jacobian matrix and sensitivities exhibit a chain-rule structure. In particular, the terms 
\begin{equation}
\frac{\partial \boldsymbol{R}}{\partial \boldsymbol{Q}_b}
\frac{\partial \boldsymbol{Q}_b}{\partial \boldsymbol{Q}_i}
\end{equation}
and 
\begin{equation}
\frac{\partial \boldsymbol{R}}{\partial \boldsymbol{Q}_b}
\frac{\partial \boldsymbol{Q}_b}{\partial \boldsymbol\alpha}
\end{equation}
can be expanded as
\begin{equation}
\frac{\partial \boldsymbol{R}}{\partial \boldsymbol{Q}_b}
\frac{\partial \boldsymbol{Q}_b}{\partial \boldsymbol{Q}_i}
=
\frac{\partial \boldsymbol{R}}{\partial \boldsymbol{F}_b}
\frac{\partial \boldsymbol{F}_b}{\partial f_b}
\frac{\partial f_b}{\partial \boldsymbol{Q}_b}
\frac{\partial \boldsymbol{Q}_b}{\partial \boldsymbol{Q}_i}.
\end{equation}
and
\begin{equation}
\frac{\partial \boldsymbol{R}}{\partial \boldsymbol{Q}_b}
\frac{\partial \boldsymbol{Q}_b}{\partial \boldsymbol \alpha}
=
\frac{\partial \boldsymbol{R}}{\partial \boldsymbol{F}_b}
\frac{\partial \boldsymbol{F}_b}{\partial f_b}
\frac{\partial f_b}{\partial \boldsymbol{Q}_b}
\frac{\partial \boldsymbol{Q}_b}{\partial \boldsymbol\alpha}.
\end{equation}

This expression clearly shows that, compared with conventional Navier--Stokes solvers, the boundary contribution in GKS involves an additional layer of coupling through the distribution function. Consequently, the accurate evaluation of the adjoint system requires a consistent linearization of the entire boundary flux construction process, including the dependence of the distribution function on the macroscopic variables.

From a numerical perspective, this multi-level dependence makes the treatment of boundary terms particularly critical in adjoint GKS formulations. Any inconsistency in the linearization of the kinetic boundary condition may lead to significant errors in the computed sensitivities. Therefore, a fully consistent discrete adjoint formulation must account for all intermediate dependencies, ensuring that the adjoint system remains strictly consistent with the primal solver.

This requirement is especially important in aerodynamic shape optimization, where boundary contributions often dominate the sensitivity of the objective function. The accurate representation of the kinetic boundary condition in the adjoint system is thus essential for reliable and robust gradient evaluation. 

\subsection{Implicit Time Marching Method}

Within the adjoint framework, the principal task is to solve the adjoint equation. To enable an efficient iterative solution procedure, a pseudo-time derivative term is introduced on the left-hand side of Eq.~\ref{adj_eqn2}, resulting in a pseudo-time marching system
\begin{equation}\label{adj_eqn2}
\frac{\partial\boldsymbol \lambda}{\partial t} + \boldsymbol A^T \boldsymbol \lambda = \boldsymbol B^T
\end{equation}
where 
\begin{equation}
\boldsymbol A =
{\frac{\partial \boldsymbol R}{\partial \boldsymbol Q_i} } + {\frac{\partial \boldsymbol R}{\partial \boldsymbol Q_b} }  {\frac{\partial \boldsymbol Q_b}{\partial \boldsymbol Q_i} },\quad 
\boldsymbol B =
{\frac{\partial I}{\partial \boldsymbol Q_i} } + {\frac{\partial I}{\partial \boldsymbol Q_b} }  {\frac{\partial \boldsymbol Q_b}{\partial \boldsymbol Q_i} }
\end{equation}

The above equation is discretized using an implicit time-marching scheme, resulting in 
\begin{equation}
\frac{\boldsymbol \Delta \lambda}{\Delta t} + \boldsymbol A^{T}(\boldsymbol \lambda^{n} + \Delta \boldsymbol \lambda)= \boldsymbol B^{T}
\end{equation}
where $\Delta \boldsymbol \lambda=\boldsymbol \lambda^{n+1} - \Delta  \boldsymbol \lambda^{n}$.

Making some arrangements leads to
\begin{equation}
(\frac{\boldsymbol I}{\Delta t} + \boldsymbol A^{T})\Delta \boldsymbol \lambda = \boldsymbol B^T - \boldsymbol A^{T}\boldsymbol \lambda^{n}
\end{equation}

The operator $\frac{\boldsymbol I}{\Delta t} + \boldsymbol A^{T}$ is decomposed into three components, namely, the lower triangular matrix  $\boldsymbol L$, the upper matrix $\boldsymbol U$ and the diagonal matrix $\boldsymbol D$. An approximate factorization is then applied, yielding
\begin{gather}\label{LU}
\begin{aligned}
\frac{\boldsymbol I}{\Delta t} + \boldsymbol A^{T} &=
 \boldsymbol D + \boldsymbol L + \boldsymbol U \\
&= \boldsymbol D(\boldsymbol I + \boldsymbol D^{-1}\boldsymbol L + \boldsymbol D^{-1}\boldsymbol U)\\
&=\boldsymbol D(\boldsymbol I + \boldsymbol D^{-1}\boldsymbol L)(\boldsymbol I + \boldsymbol D^{-1}\boldsymbol U) - \boldsymbol L \boldsymbol D^{-1}\boldsymbol U\\
&\approx \boldsymbol D(\boldsymbol I + \boldsymbol D^{-1}\boldsymbol L)(\boldsymbol I + \boldsymbol D^{-1}\boldsymbol U)
\end{aligned}
\end{gather}

Substituting Eq.~\ref{LU} into Eq.~\ref{adj_eqn2} yields
\begin{equation}\label{LUSGS}
\boldsymbol D(\boldsymbol I + \boldsymbol D^{-1}\boldsymbol L)(\boldsymbol I + \boldsymbol D^{-1}\boldsymbol U)\Delta \boldsymbol \lambda = \boldsymbol B^T - \boldsymbol A^T \boldsymbol \lambda^{n}
\end{equation}

To solve Eq.~\ref{LUSGS}, two steps are required:
\begin{enumerate}
\item Forward sweep
\begin{equation}
\boldsymbol D(\boldsymbol I + \boldsymbol D^{-1}\boldsymbol L)\Delta \boldsymbol \lambda^{1/2} = \boldsymbol B^T - \boldsymbol A^T \boldsymbol \lambda^n
\end{equation}
and then 
\begin{equation}
\Delta \boldsymbol \lambda^{1/2}  = \boldsymbol D^{-1}(\boldsymbol B^T - \boldsymbol A^T \boldsymbol \lambda^n) -\boldsymbol D^{-1} \boldsymbol L\Delta \boldsymbol \lambda^{1/2}
\end{equation}
\item Backward sweep
\begin{equation}
(\boldsymbol I + \boldsymbol D^{-1}\boldsymbol U)\Delta \boldsymbol \lambda = \Delta \boldsymbol \lambda^{1/2}
\end{equation}
and then
\begin{equation}
\Delta \boldsymbol \lambda = \Delta \boldsymbol \lambda^{1/2} - \boldsymbol D^{-1}\boldsymbol U\Delta \boldsymbol \lambda
\end{equation}
\end{enumerate}

\section{Adjoint-Based Design Optimization System}

\begin{figure}[h!]
    \centering
\includegraphics[width=1.0\linewidth]{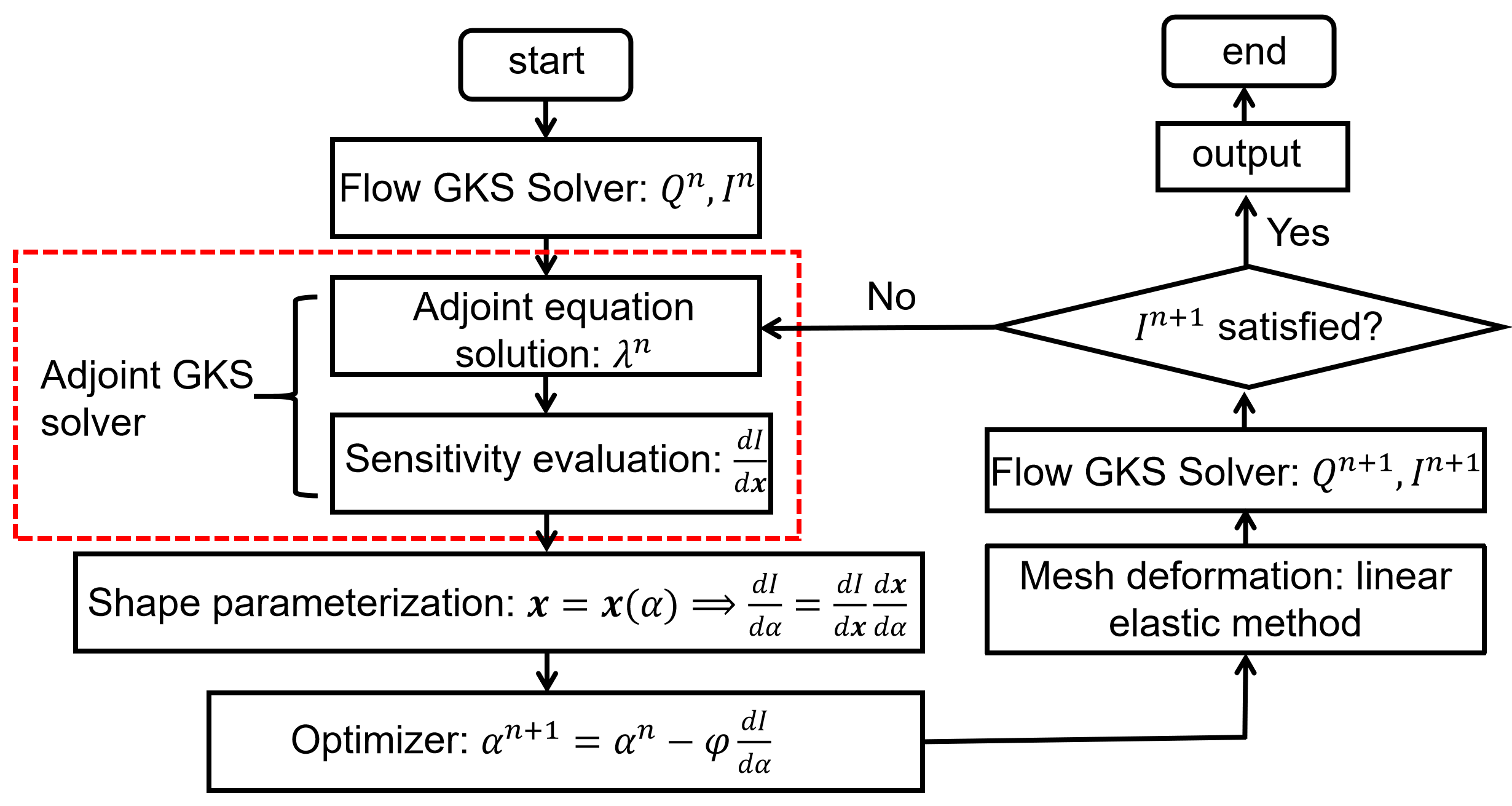}
    \caption{Adjoint-based design optimization system.}
   \label{adj_opt}
\end{figure}

Figure~\ref{adj_opt} illustrates the adjoint-based design optimization system, which comprises five components: the flow GKS solver, adjoint GKS solver, shape parameterization, optimizer, and mesh deformation. 
The flow solver computes the flow variables $\boldsymbol{Q}^n$ and the objective function $I^n$.
The adjoint solver involves two steps: solution of the adjoint equations and evaluation of the sensitivities, yielding the adjoint variables and corresponding sensitivity information. The shape parameterization establishes the mapping between grid coordinates and design variables, enabling sensitivity evaluation via the chain rule. In this work, Hicks–Henne functions~\cite{HicksHenne1978, Wu2023JoT} are employed to parameterize blade perturbations.
The optimizer updates the design variables using the steepest descent method with a constant step size. As the blade geometry evolves, the mesh is updated accordingly; to preserve mesh quality, a linear elastic deformation method is adopted.

\section{Test Cases}\label{test_case}

Four test cases are used to assess the performance of the adjoint GKS solver for shape optimization across a broad range of Mach regimes.

\subsection{Subsonic Case 1: Flow Past a Cylinder}

The first case examines flow past a cylinder at $Ma_{\infty}=0.15$ and $Re_{\infty}=40$, where an adiabatic no-slip kinetic boundary condition is imposed on the solid wall. 

\begin{figure}[h!]
\centering
\subfigure[Flow]{
	\includegraphics[width=2.2in]{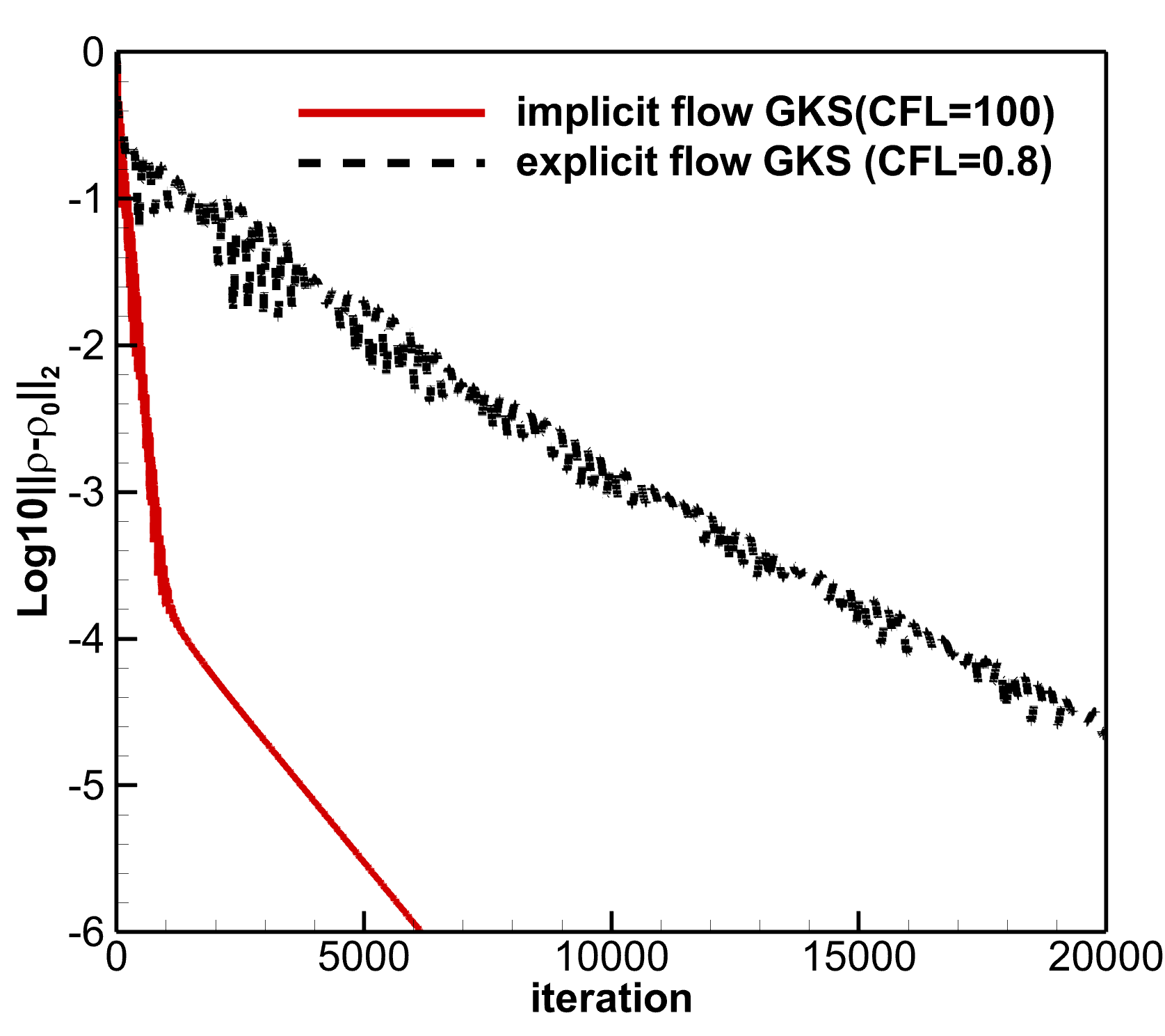}
 \label{cylinder_flowres}
 }
 \subfigure[Adjoint]{
	\includegraphics[width=2.2in]{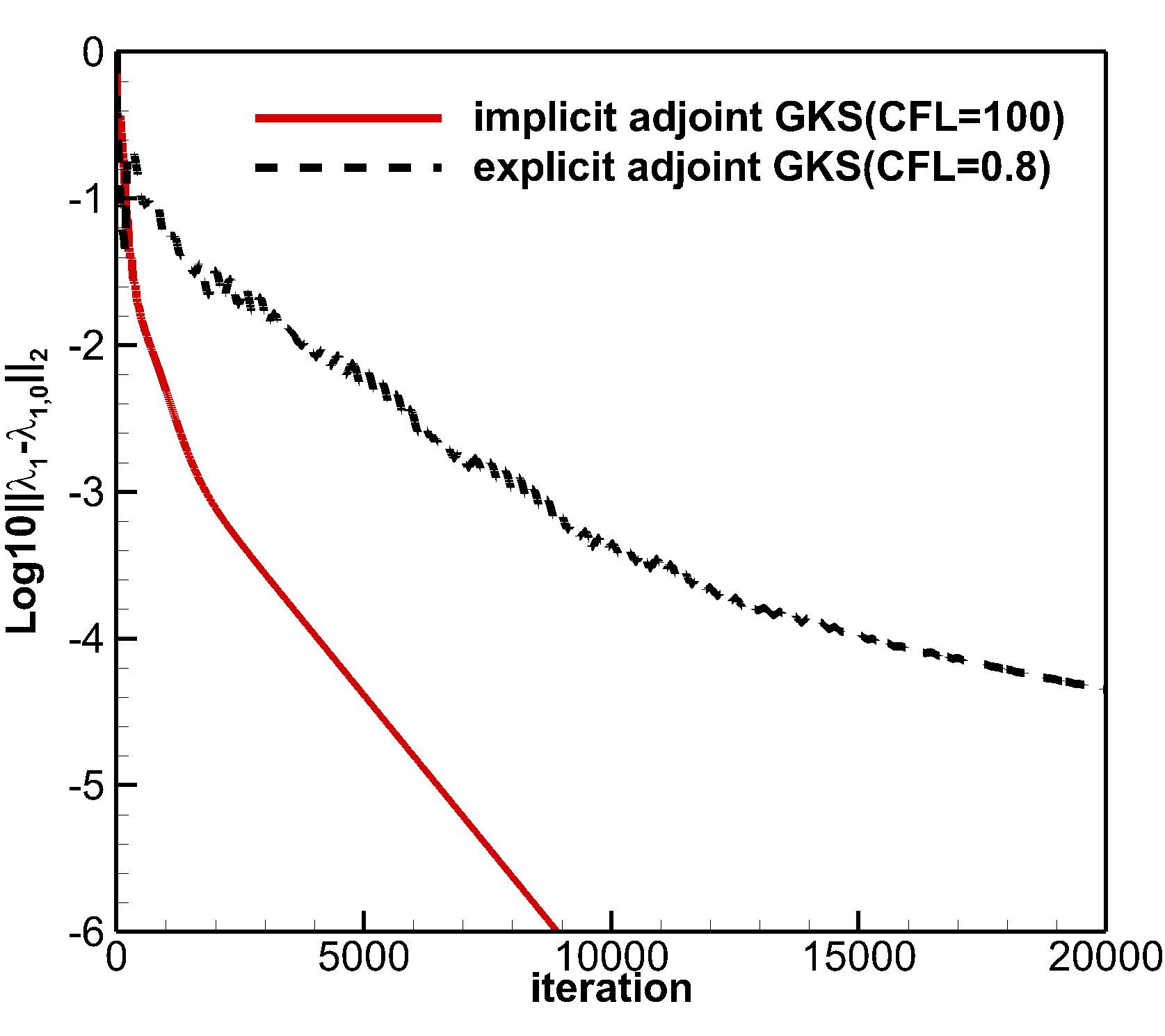}
 \label{cylinder_adjres}
 }
	\caption{Convergence histories of the residuals related to the mass equation for the cylinder case: a) flow GKS solver; b) adjoint GKS solver.}
\end{figure} 
To demonstrate the advantage of the implicit time-marching method over its explicit counterpart, the residual convergence histories of the mass equation for both the flow and adjoint GKS solvers are presented in Figs.~\ref{cylinder_flowres} and~\ref{cylinder_adjres}. The explicit method is limited to a maximum Courant number of approximately 0.8, whereas the implicit method allows a Courant number as large as 100.
For the flow GKS solver, the implicit method achieves convergence within about 6000 iterations, compared to more than 20000 iterations required by the explicit method. A similar trend is observed for the adjoint solver. Overall, the implicit approach accelerates convergence by a factor exceeding 3.5. 

\begin{figure}[h!]
\centering
\subfigure[Flow]{
	\includegraphics[width=2.2in]{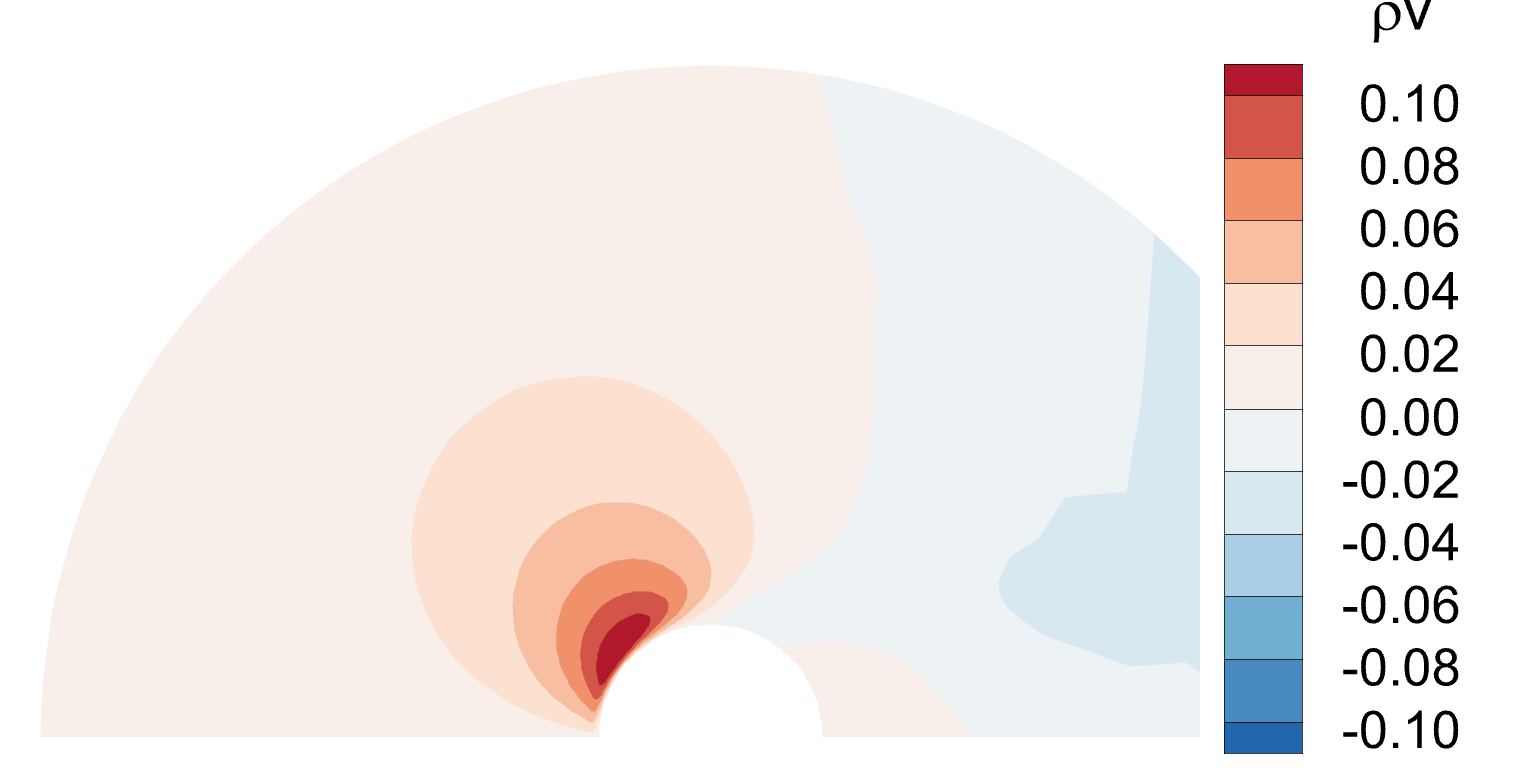}
 \label{cylinder_flow}
 }
 \subfigure[Adjoint]{
	\includegraphics[width=2.2in]{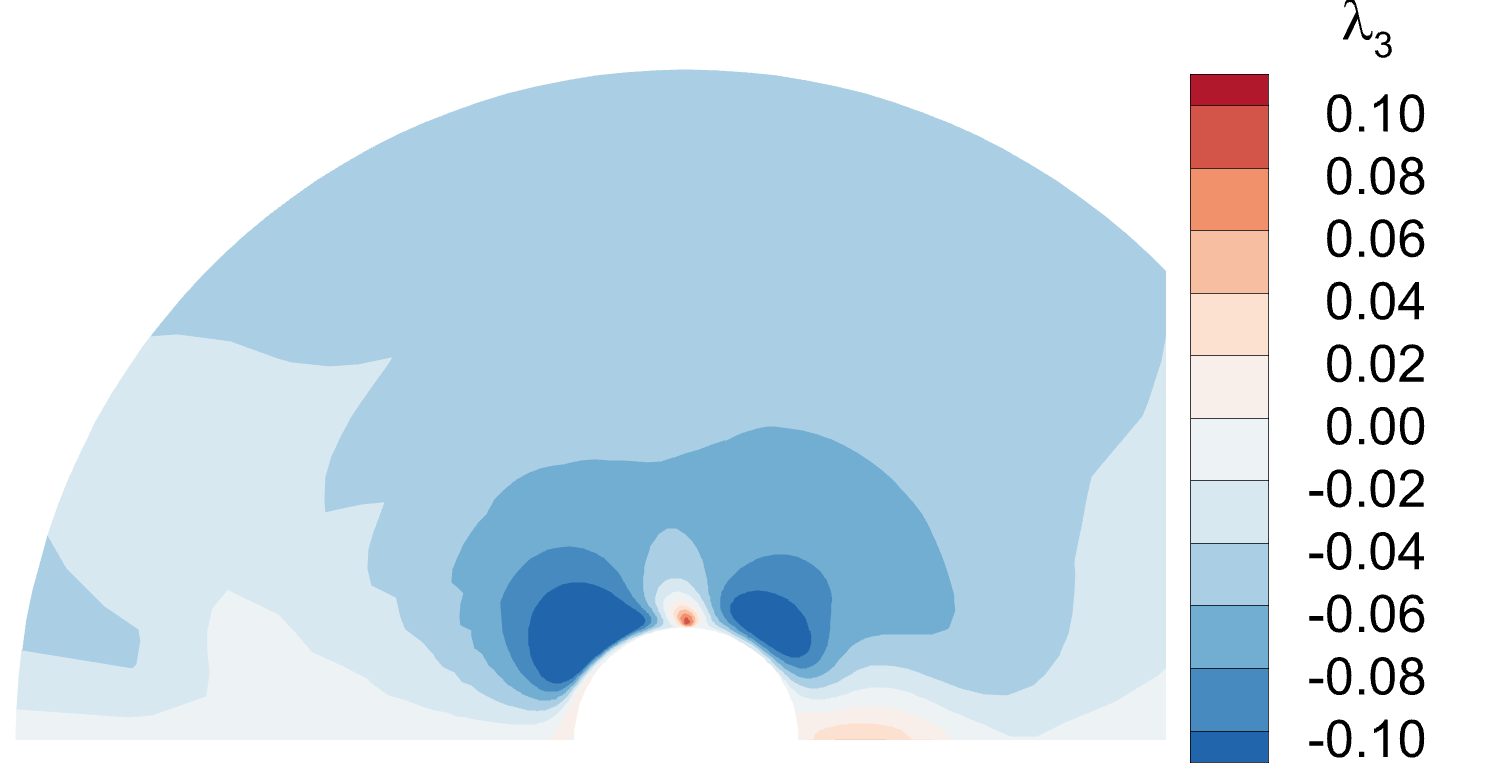}
 \label{cylinder_adj}
 }
	\caption{Flow and adjoint variable contours related to the y-moment equation for the cylinder case.}
\end{figure}

\begin{figure}[h!]
\centering
\subfigure[]{
	\includegraphics[width=2.2in]{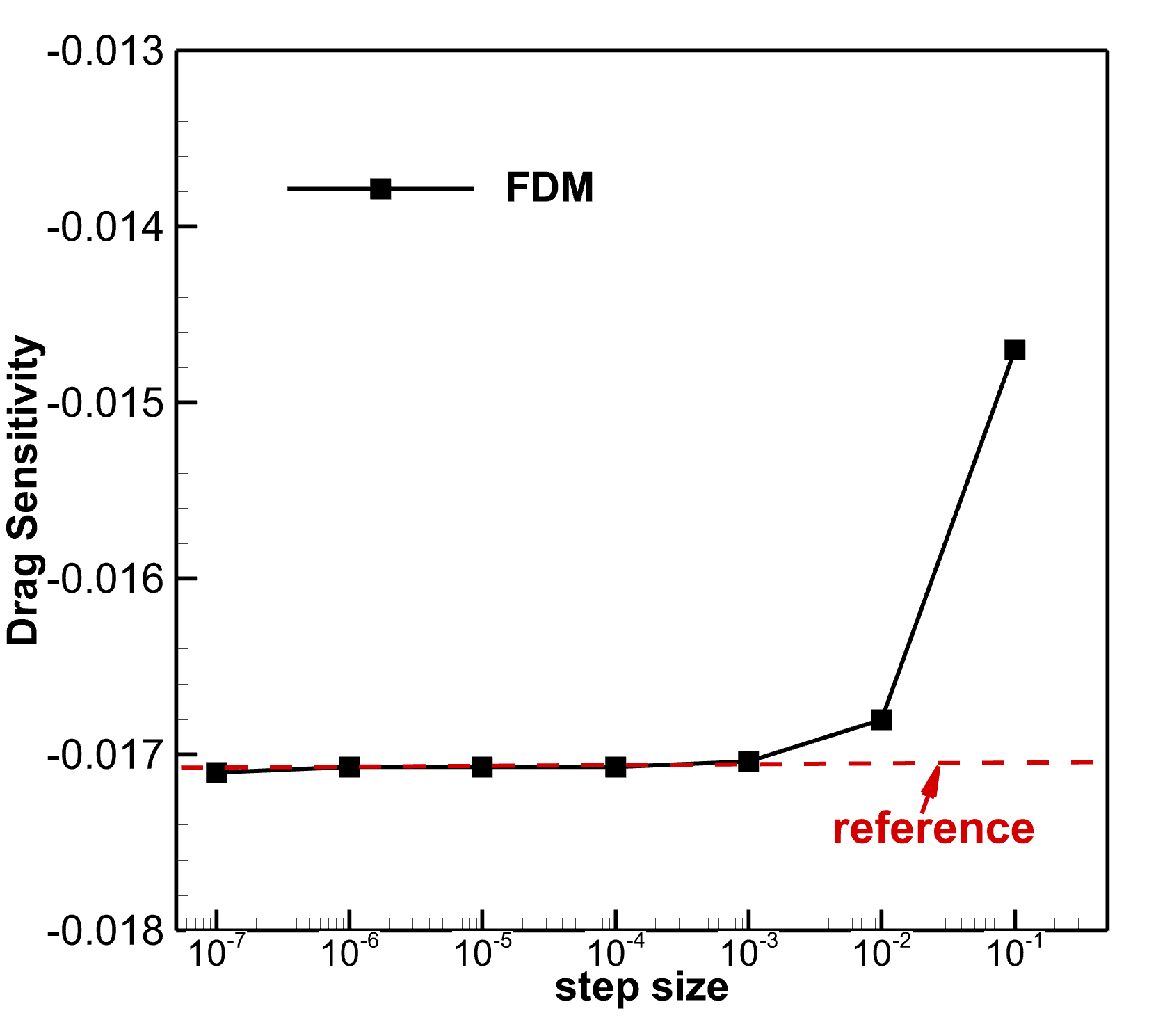}
 \label{cylinder_FDM}
 }
 \subfigure[]{
	\includegraphics[width=2.2in]{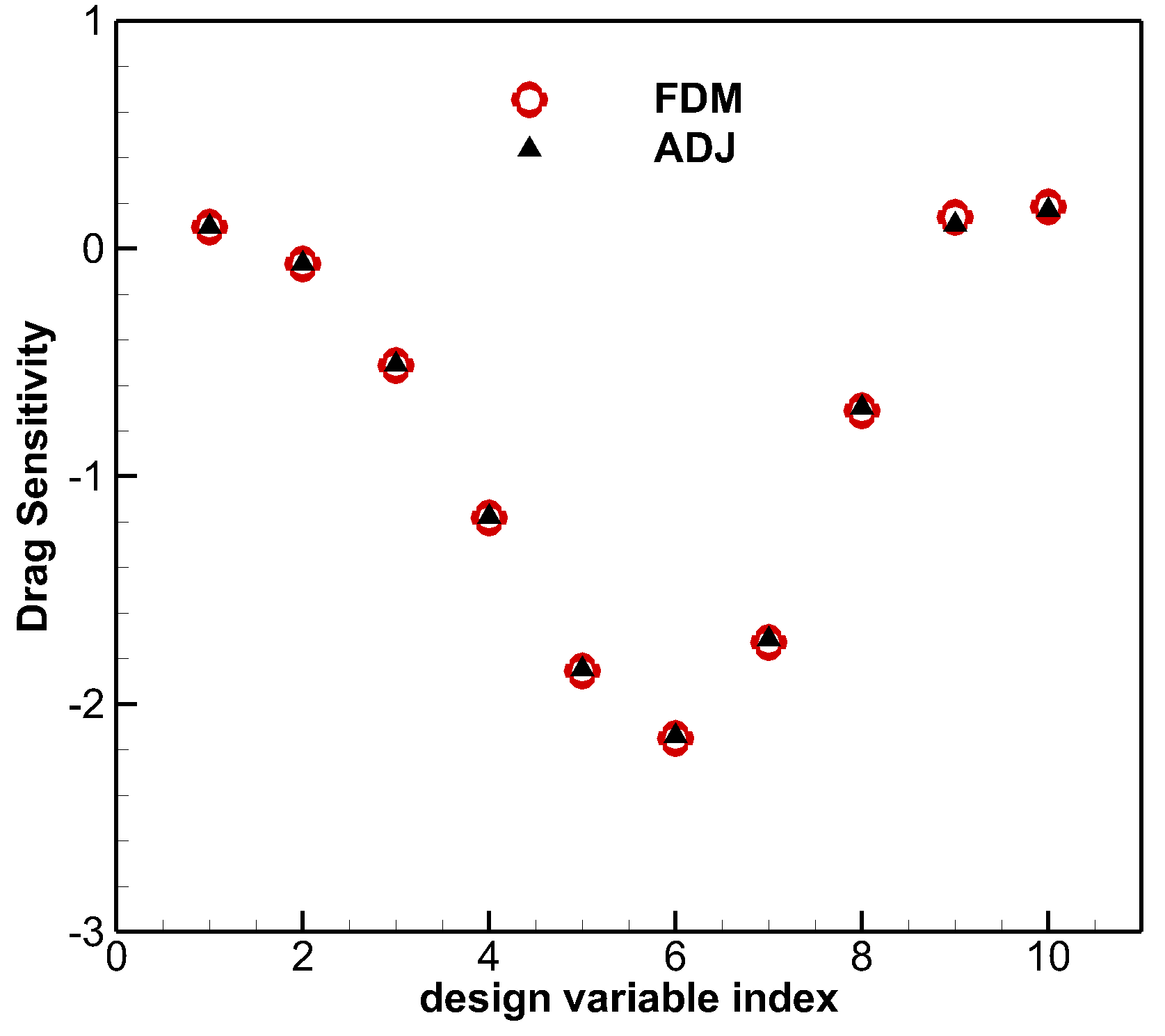}
 \label{cylinder_sen}
 }
	\caption{Verification of the adjoint sensitivities for the cylinder case: a) step size independence study; b) comparison between the adjoint method and FDM.}
\end{figure}
Prior to design optimization, the adjoint GKS solver is verified from both qualitative and quantitative perspectives. Figures~\ref{cylinder_flow} and~\ref{cylinder_adj} present the flow and adjoint fields associated with the y-momentum equation. It is evident that the adjoint field exhibits a structure opposite to that of the flow field.
The adjoint sensitivities are further verified against those obtained by the finite difference method. The objective function is the drag and the design variables are the Hicks-Henne coefficients. Since FDM sensitivities depend on the perturbation step size, a step-size independence study is conducted, as shown in Fig.~\ref{cylinder_FDM}. When the step size varies from $10^{-7}$ to $10^{-3}$, the computed sensitivities exhibit negligible variation, indicating convergence. Accordingly, a step size of $10^{-4}$ is adopted for the subsequent verification.
Figure~\ref{cylinder_sen} compares the sensitivities obtained from the adjoint and finite difference method. Good agreement between the two approaches is observed.

\begin{figure}[h!]
\centering
\subfigure[Objective Function]{
	\includegraphics[width=2.0in]{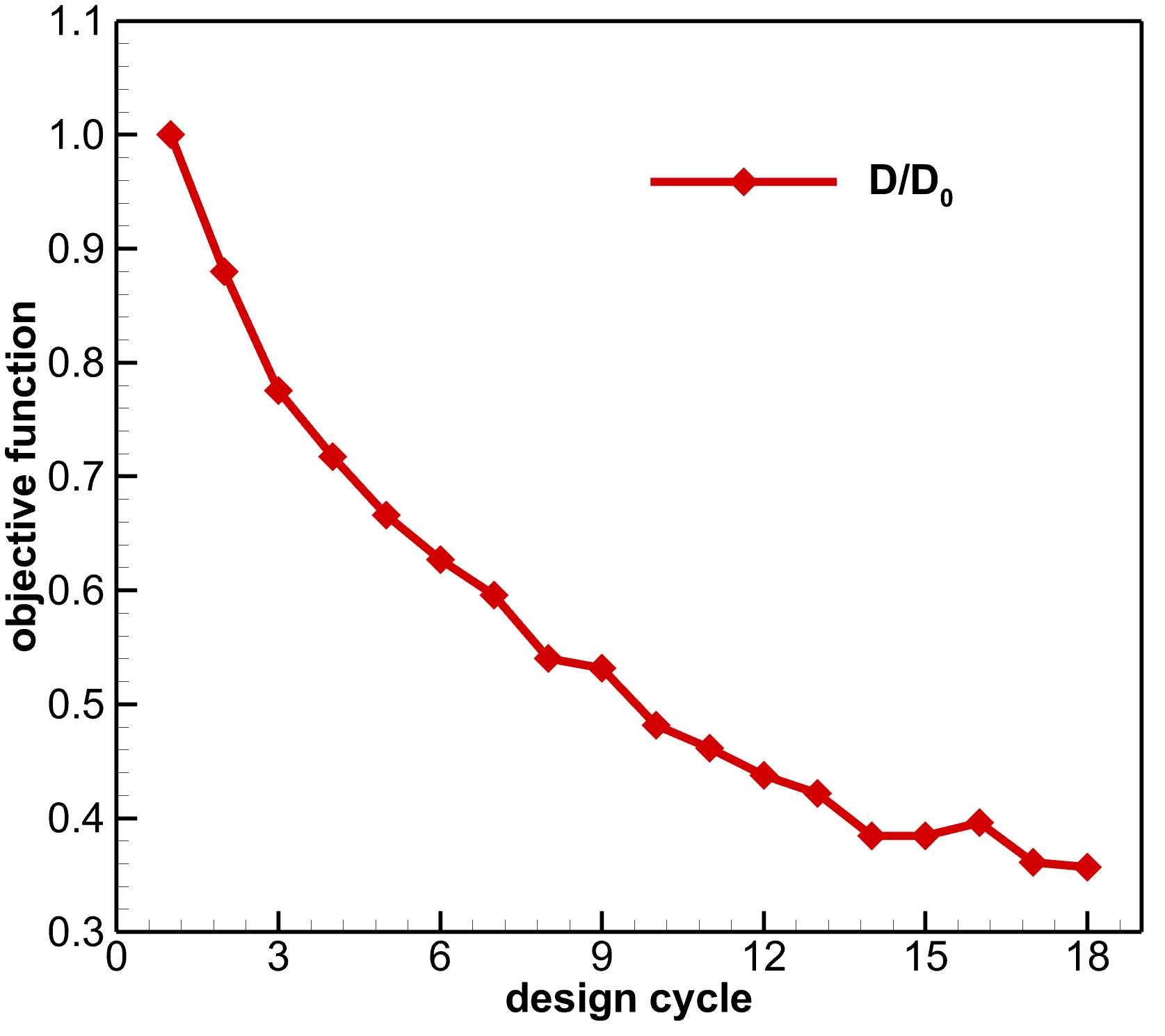}
 \label{cylinder_obj}
 }
 \subfigure[Shape]{
	\includegraphics[width=2.0in]{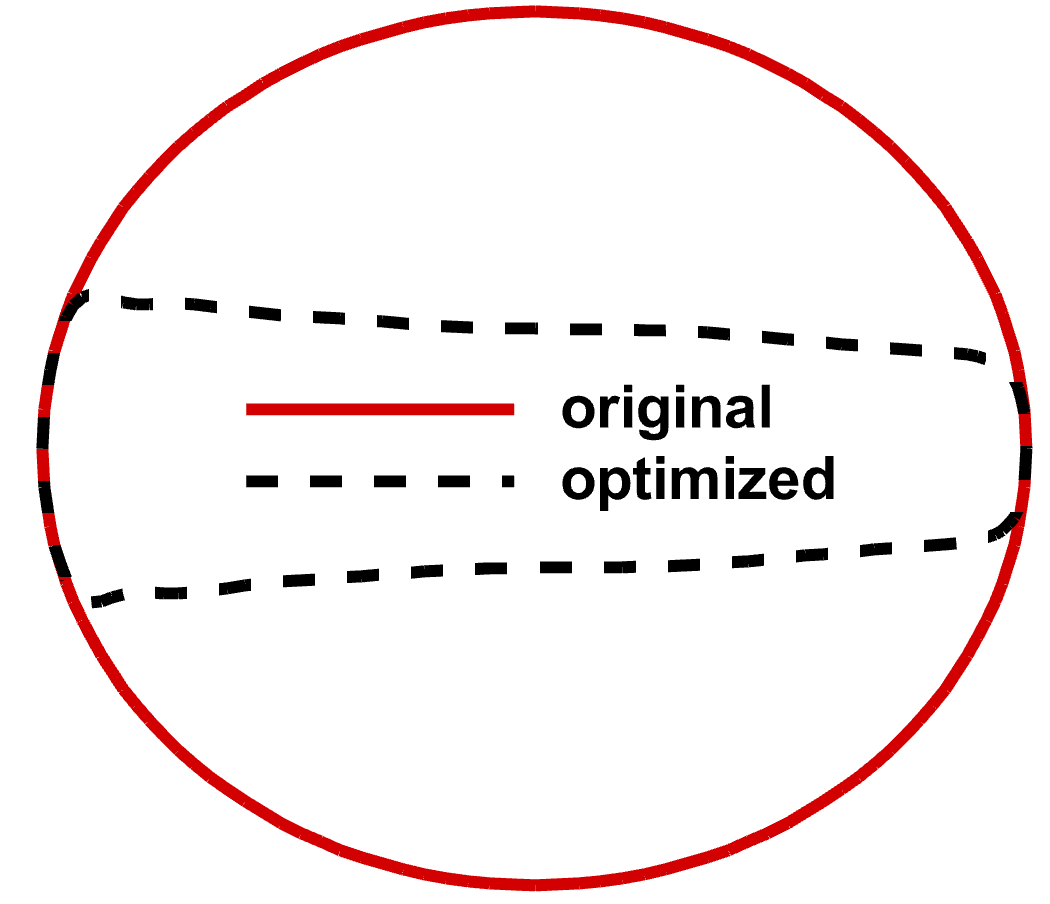}
 \label{cylinder_shape}
 }
	\caption{Aerodynamic shape optimization of the cylinder case: a) evolutionary histories; b) original and optimized shapes.}
\end{figure}
\begin{figure}[h!]
\centering
\subfigure[Pressure Contours]{
	\includegraphics[width=2.2in]{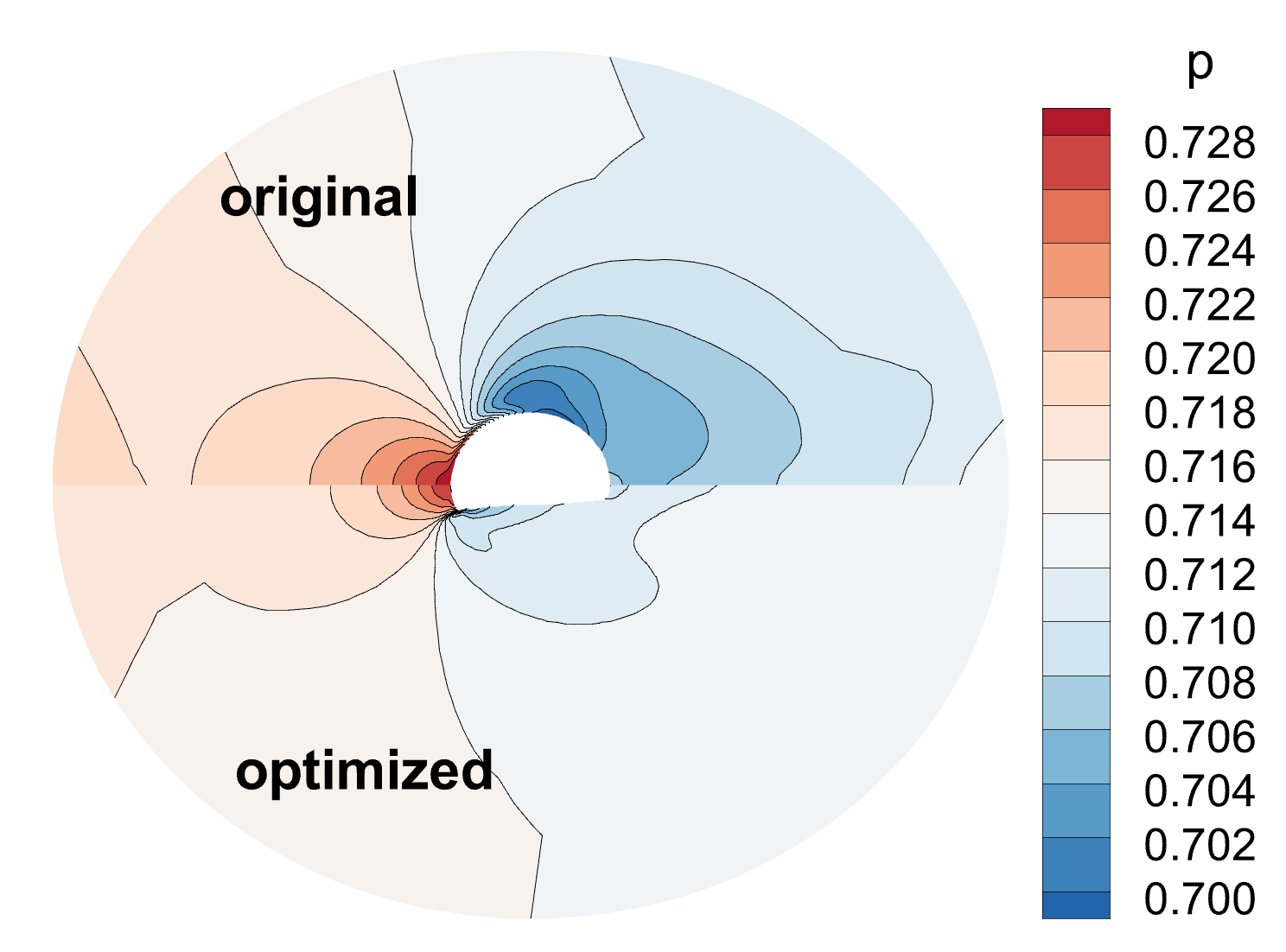}
 \label{cylinder_Pcomp}
 }
 \subfigure[Drag]{
	\includegraphics[width=1.9in]{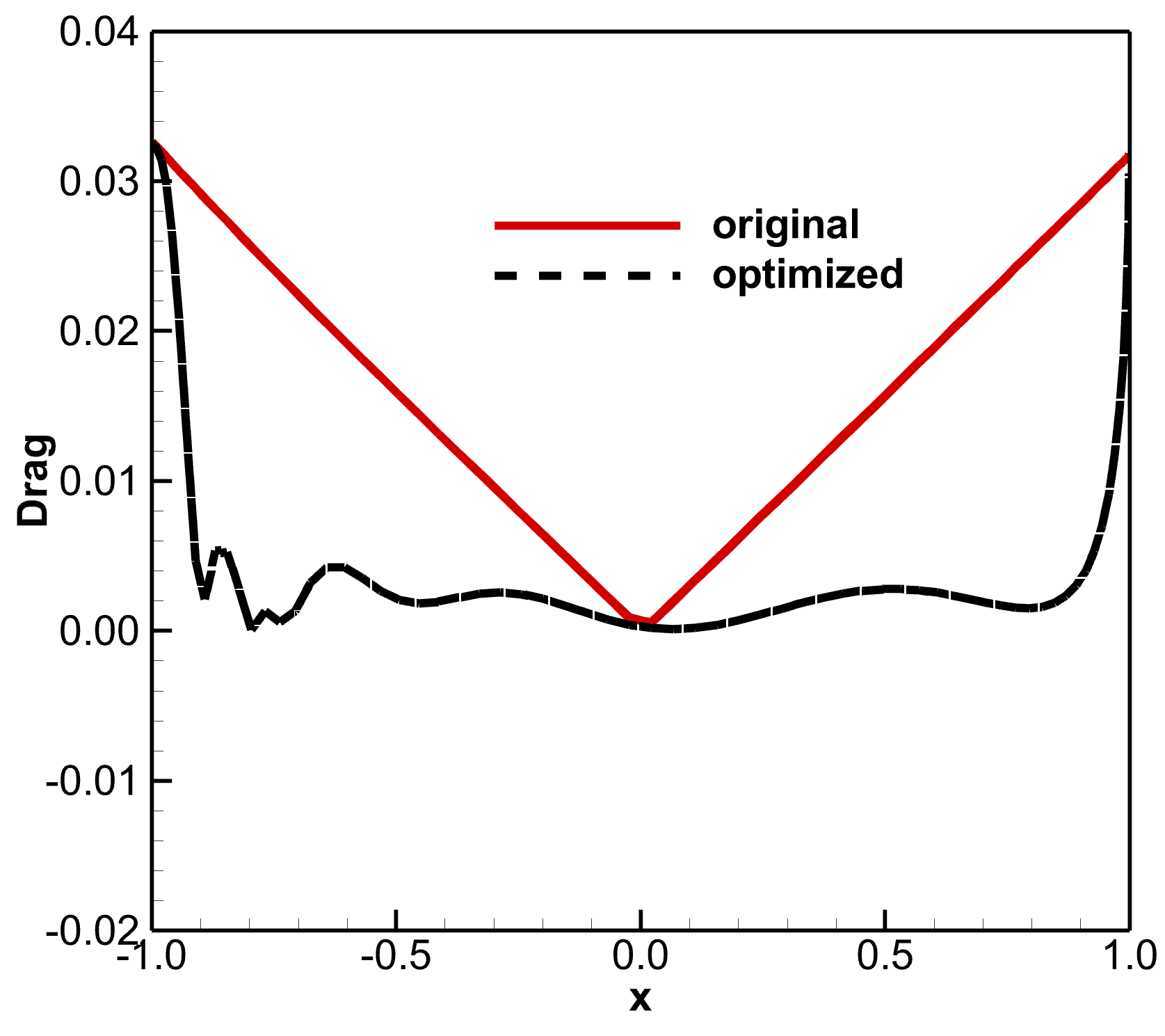}
 \label{cylinder_Dcomp}
 }
	\caption{Static pressure contours in the whole computational domain and Drag on the shape surface: a) pressure; b) drag.}
\end{figure}

In the optimization, the objective function is the surface drag. Ten design variables are uniformly distributed along the upper surface, and the lower surface deformation is prescribed in an opposite manner to enforce geometric symmetry. Figure~\ref{cylinder_obj} presents the evolution of the objective function during the optimization process, showing a reduction of approximately 68\% after 18 design cycles. Figure~\ref{cylinder_shape} compares the original and optimized shapes, showing that the optimized geometry becomes more slender. This streamlined configuration effectively reduces pressure drag. 
Figures~\ref{cylinder_Pcomp} and~\ref{cylinder_Dcomp} compare the pressure contours and surface drag distributions before and after optimization, respectively. A noticeable reduction in drag is observed over the entire surface. In particular, the drag in the middle region is significantly reduced and approaches nearly zero.

\subsection{Transonic Case 2: RAE 2822}

The second case is the transonic RAE 2822 airfoil. The grid is a single‑block two-dimensional C‑grid of $369\times65$ points, including 305 on the airfoil surface.
The freestream parameters are set as follows: the total temperature of 255.56$K$, the total pressure of 108987.82$P_a$, Mach number of 0.729, Reynolds number based on the chord of 6.5 million, and the angle of attack of $2.31^\circ$.
To compare the numerical results with the experimental data, the pressure coefficient on the blade surfaces is defined as
\begin{equation}
C_p = \frac{p_\infty-p}{
\frac{1}{2}\rho_\infty U_\infty^2}
\end{equation}

\begin{figure}[h!]
\centering
\subfigure[]{
	\includegraphics[width=2.2in]{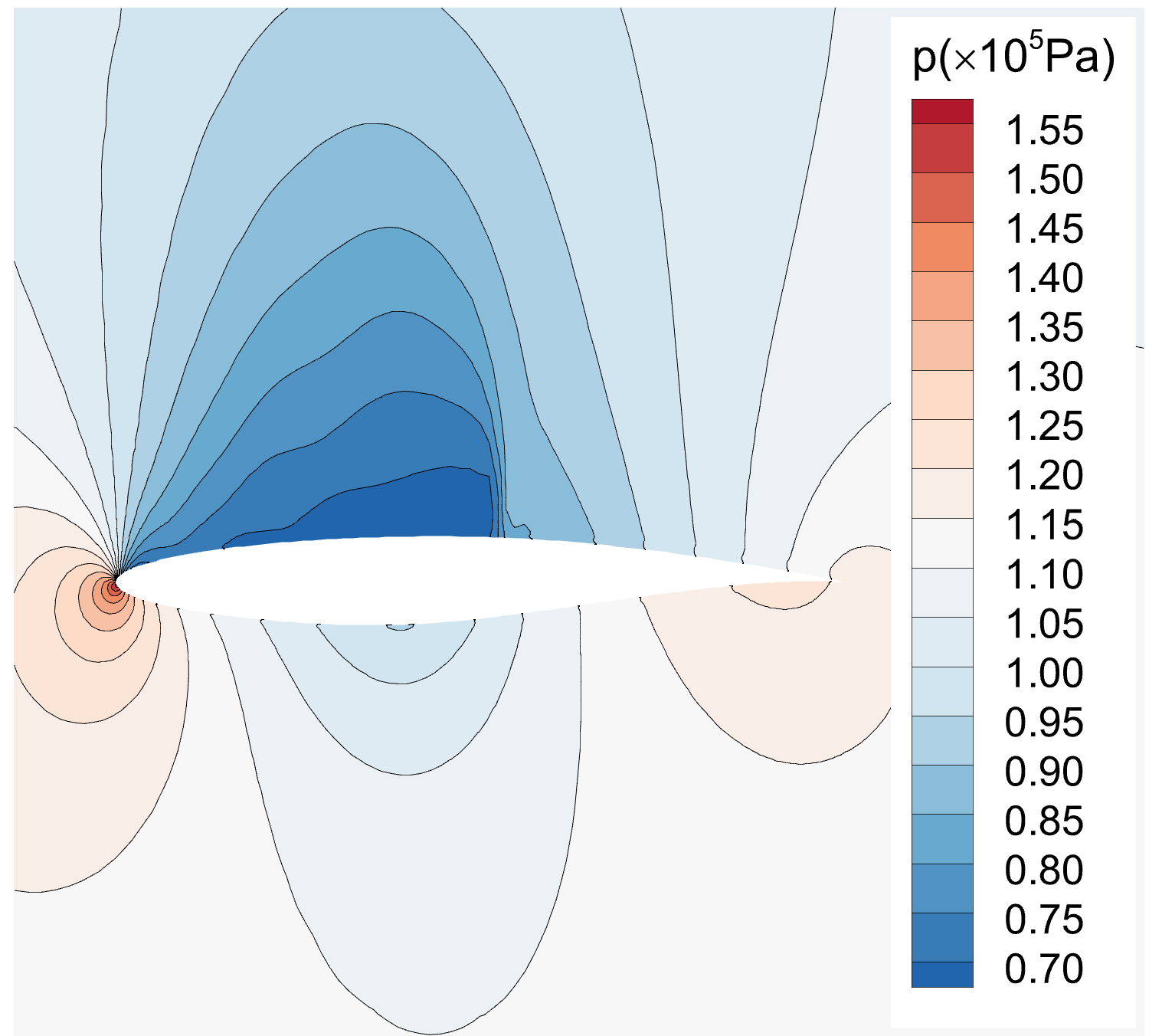}
 \label{RAE_cp0}
 }
 \subfigure[]{
	\includegraphics[width=2.2in]{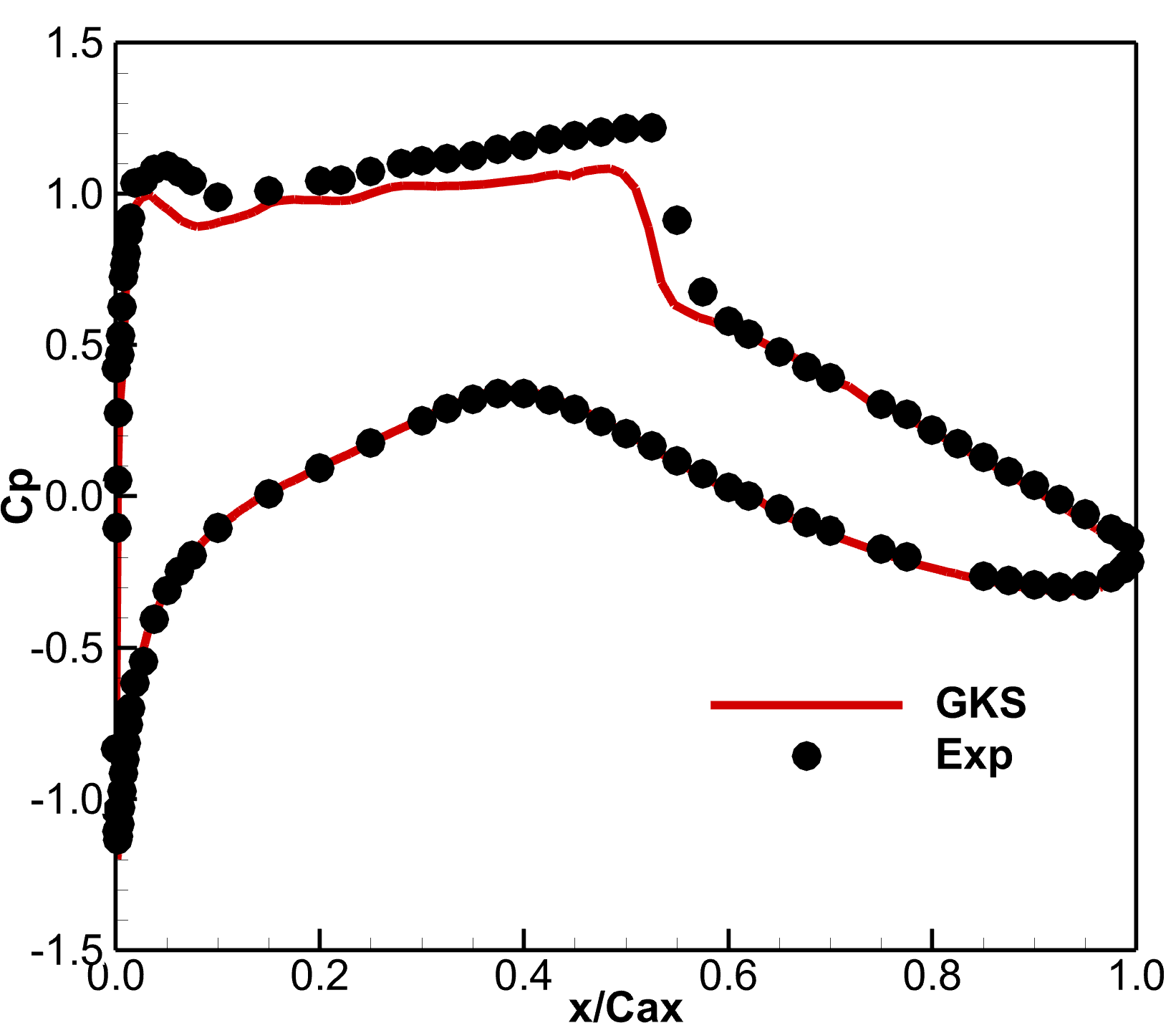}
\label{RAE_cp1}
}
	\caption{The distributions of $p$ and $C_p$ for the transonic RAE 2822 airfoil: a)in the whole computational domain; b) on airfoil surfaces.}
\end{figure}

Figures~\ref{RAE_cp0} and \ref{RAE_cp1} compare the pressure coefficients from the experimental data with those obtained from the flow GKS solver. On the lower surface, the numerical results almost coincide with the experimental measurements. On the upper surface, the strength and location of the shock wave predicted by the flow GKS solver show small deviations from the experiment, which may be attributed to the influence of the Spalart-Allmaras turbulence model~\cite{SA1992} adopted in this work.

\begin{figure}[h!]
\centering
\subfigure[Flow]{
	\includegraphics[width=2.2in]{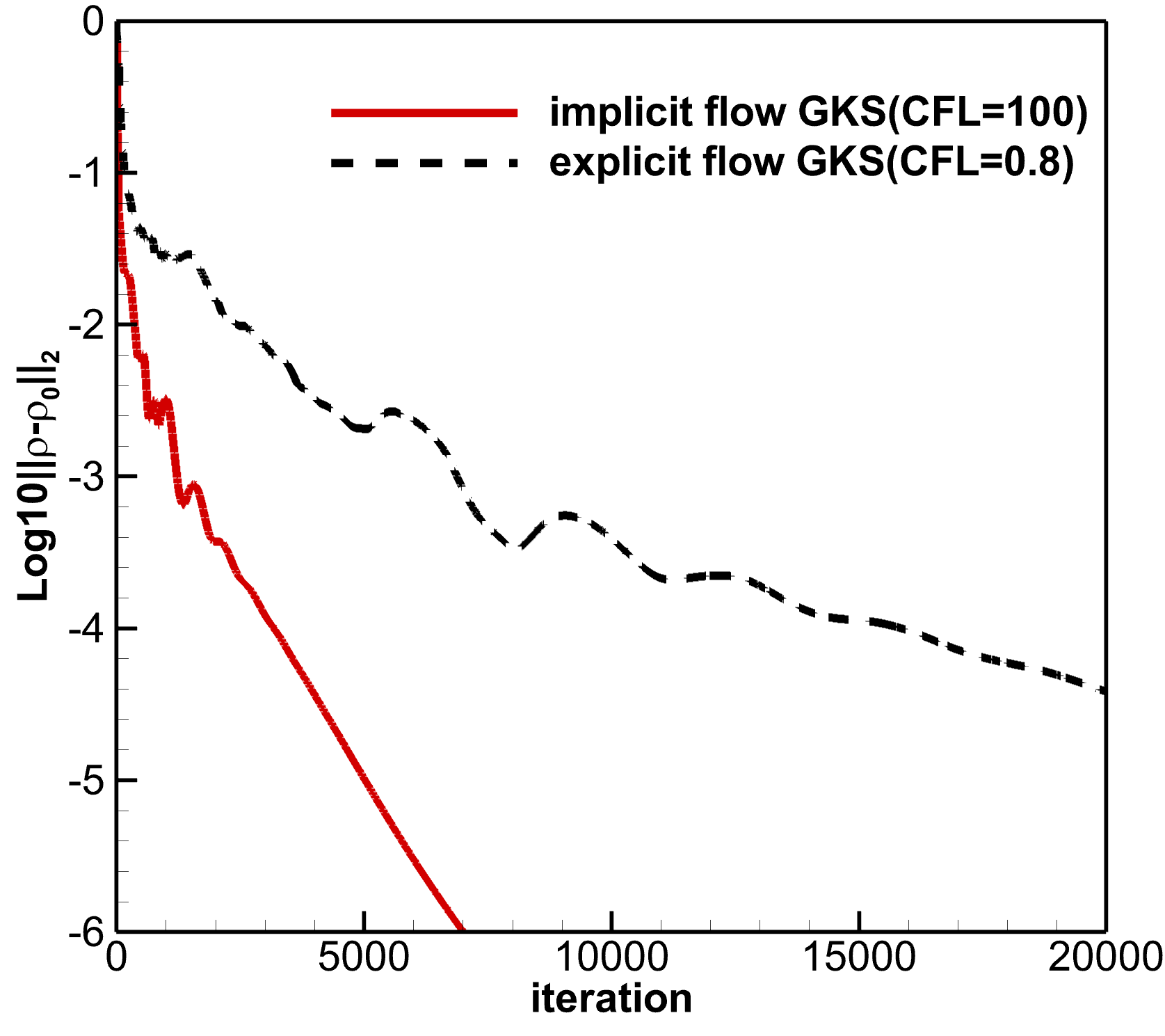}
 \label{RAE_flowres}
 }
 \subfigure[Adjoint]{
	\includegraphics[width=2.2in]{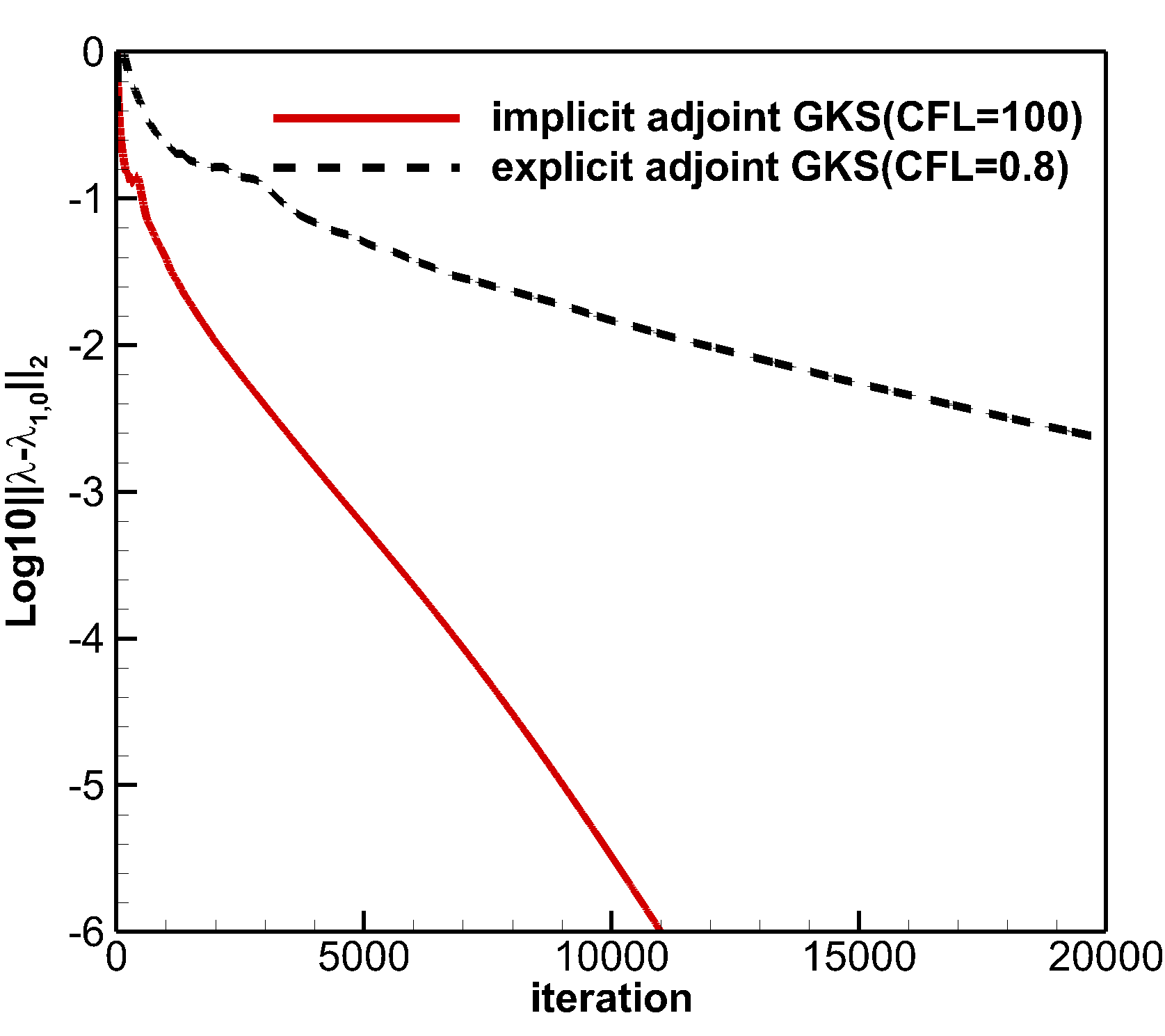}
 \label{RAE_adjres}
 }
	\caption{Convergence histories of the residuals related to the mass equation for the RAE 2822 case: a) flow GKS solver; b) adjoint GKS solver.}
\end{figure} 
Figures~\ref{RAE_flowres} and~\ref{RAE_adjres} present the residual convergence histories of the mass equation for both the flow and adjoint GKS solvers. For both systems, a monotonic decay of the residuals is observed, indicating stable and robust convergence behavior. Compared with the explicit scheme, the implicit time-marching method significantly accelerates the convergence rate, achieving an approximate fourfold reduction in the number of iterations required to reach the same residual level. This improvement highlights the enhanced efficiency of the implicit formulation for both flow and adjoint GKS solvers. 

\begin{figure}[h!]
\centering
\subfigure[Flow]{
	\includegraphics[width=3.5in]{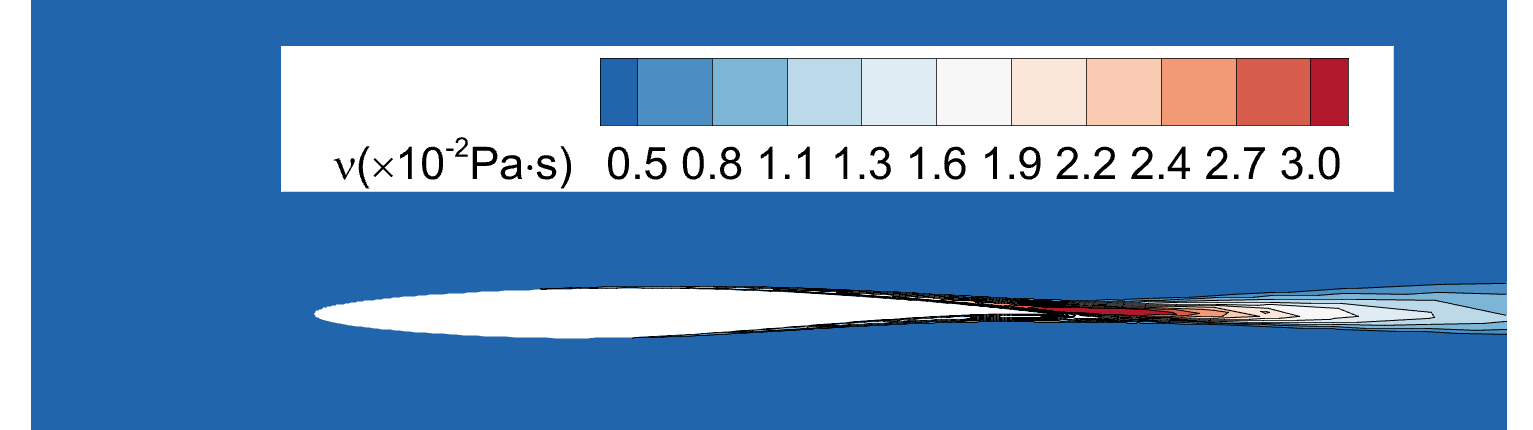}
\label{RAE_flow}
}
\subfigure[Adjoint]{
	\includegraphics[width=3.6in]{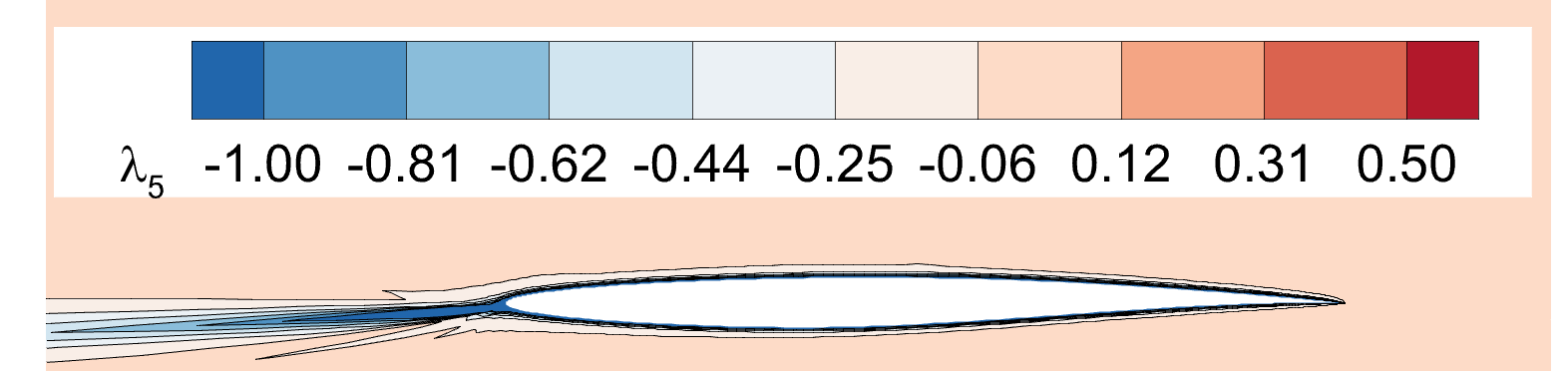}
\label{RAE_adj}
}
	\caption{Both the flow and adjoint variable contours related to the turbulence model equation for the RAE 2822 airfoil.}
\end{figure}
\begin{figure}[h!]
\centering
\subfigure[]{
	\includegraphics[width=2.2in]{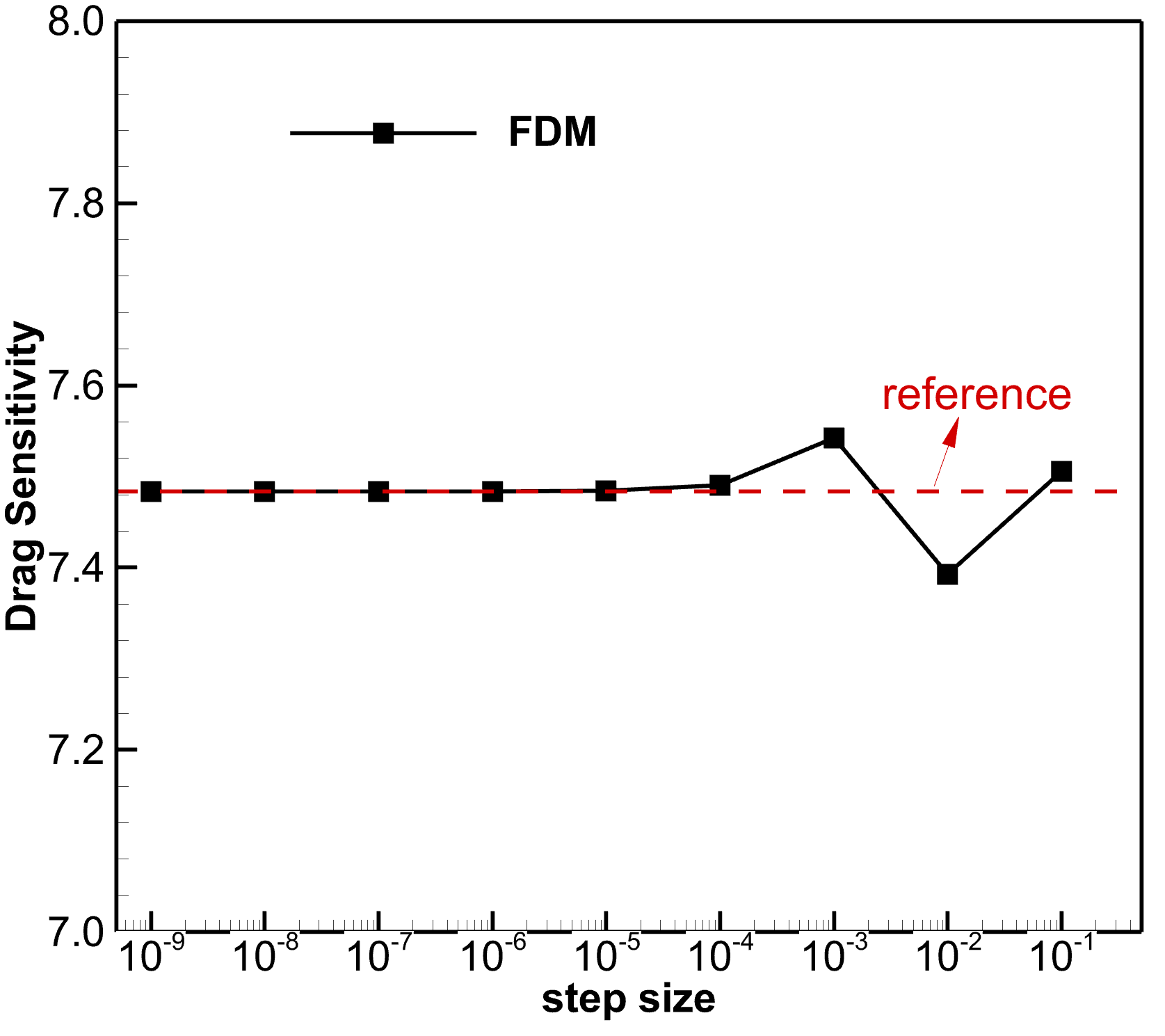}
\label{RAE_Dfdm}
}
\subfigure[]{
	\includegraphics[width=2.2in]{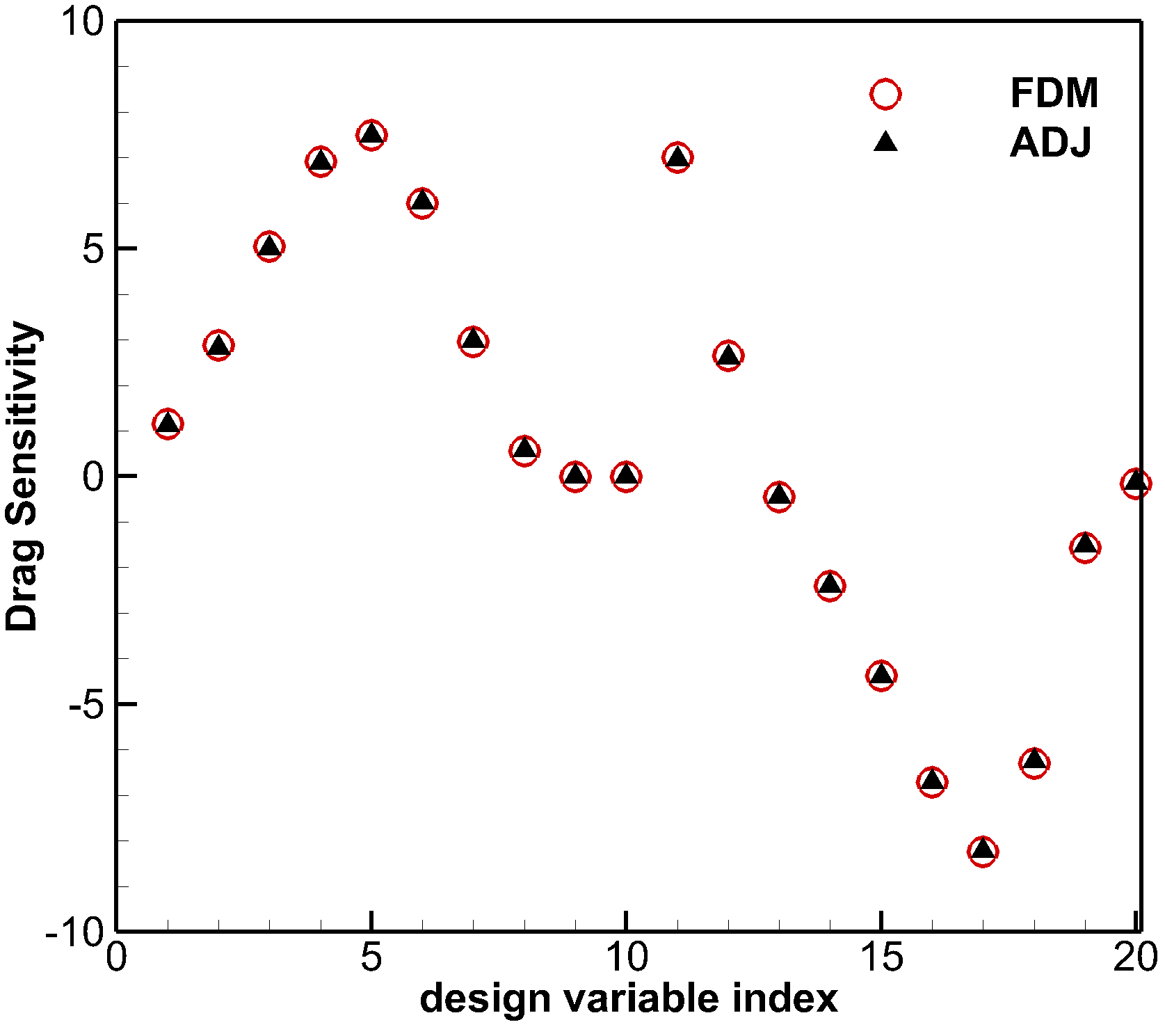}
\label{RAE_Dsen}
}
	\caption{Drag sensitivity verification for the RAE 2822 case: a) step size independence study; b) sensitivity verification.}
    \label{RAE_drag_sen}
\end{figure}
\begin{figure}[h!]
\centering
\subfigure[]{
	\includegraphics[width=2.2in]{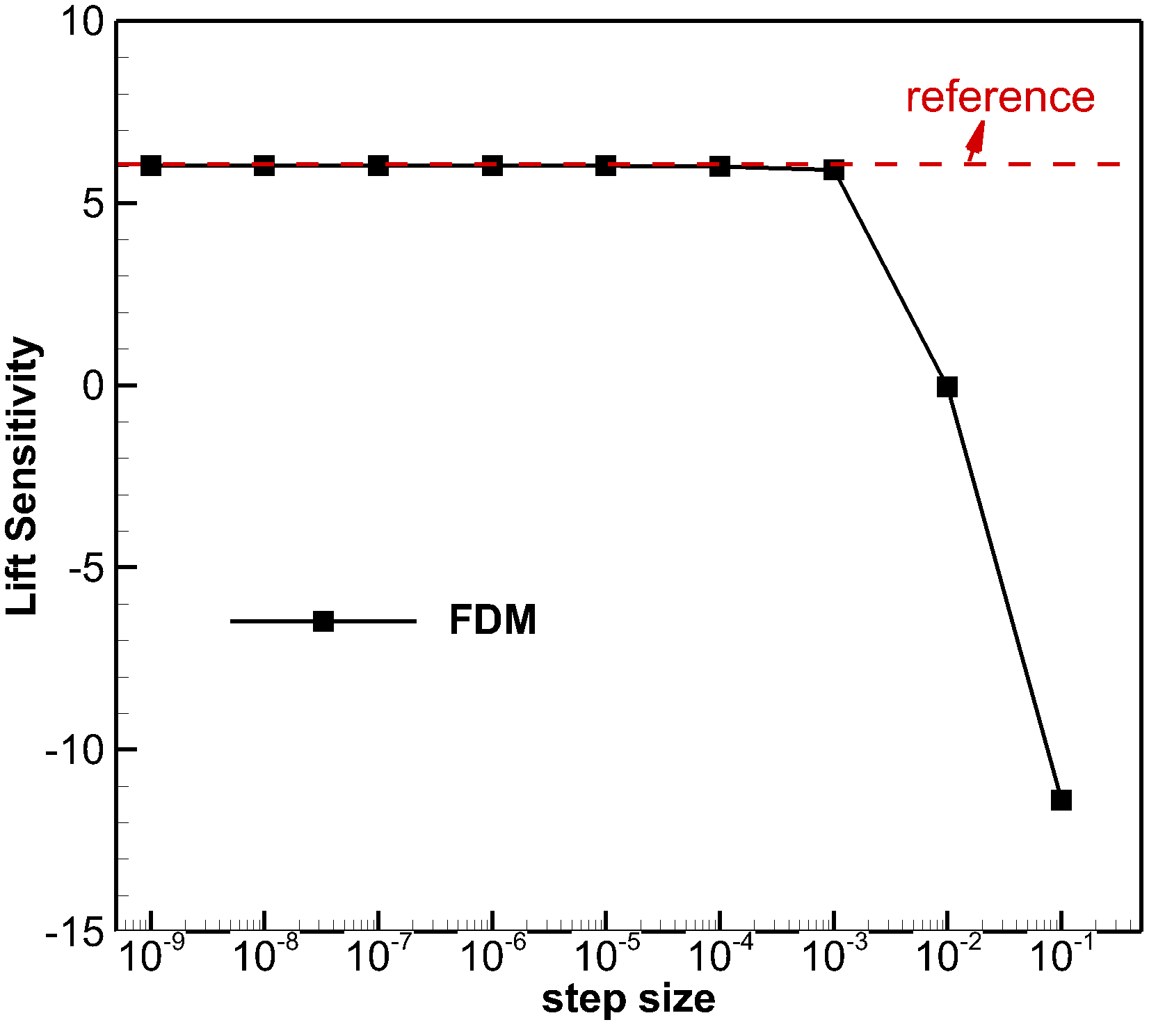}
\label{RAE_Lfdm}
}
\subfigure[]{
	\includegraphics[width=2.2in]{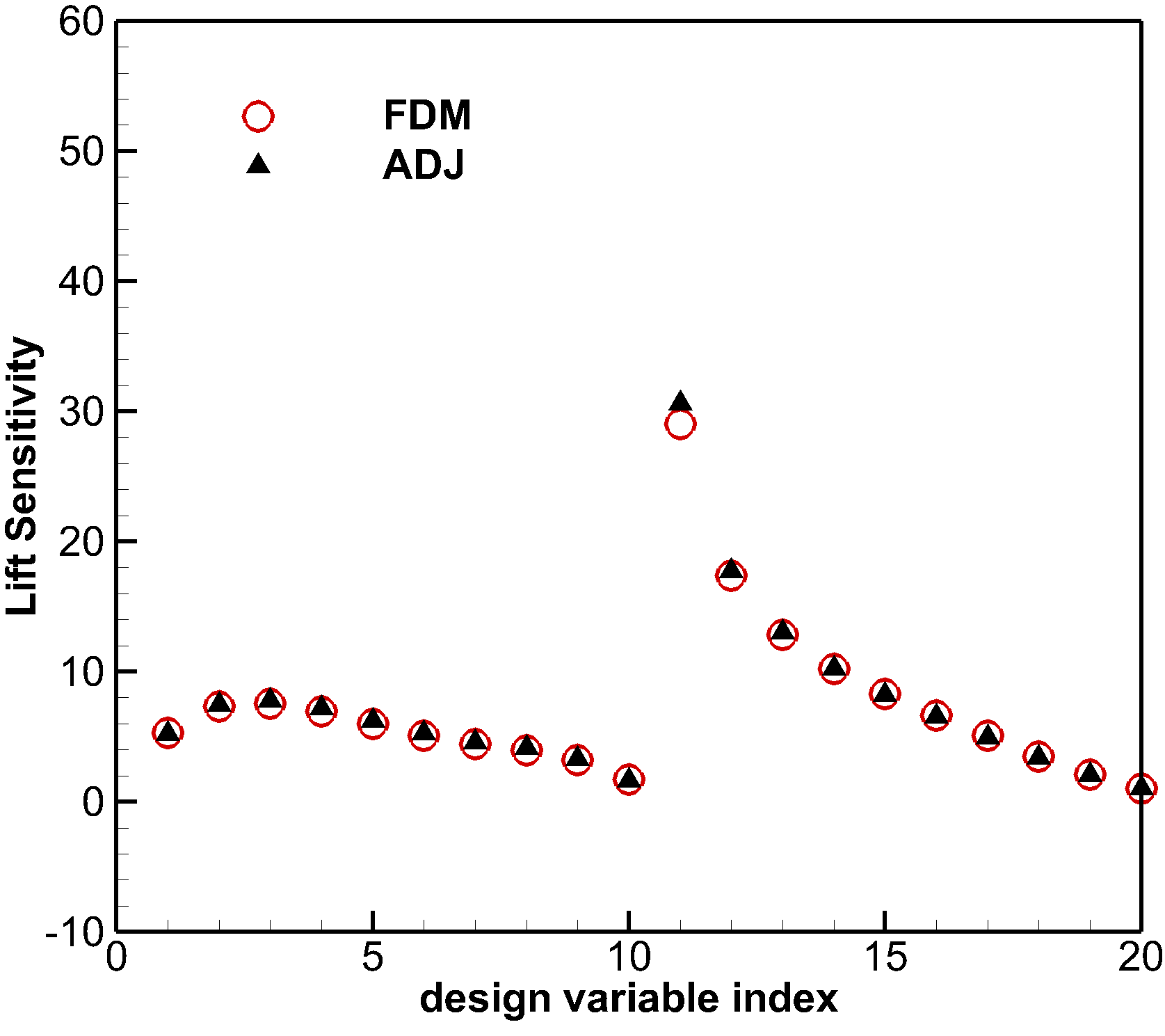}
\label{RAE_Lsen}
}
	\caption{Lift sensitivity verification for the RAE 2822 case: a) step size independence study; b) sensitivity verification.}
    \label{RAE_lift_sen}
\end{figure}

Figures~\ref{RAE_flow} and \ref{RAE_adj} show the flow and adjoint fields associated with the turbulence‑model equation. As expected, the adjoint field is opposite in structure to the flow field: regions that exhibit downstream influence in the flow field correspond to upstream influence in the adjoint field. The quantitative validation of the adjoint GKS solver for the transonic case is presented in Figs.~\ref{RAE_drag_sen} and~\ref{RAE_lift_sen}. The drag and lift sensitivities obtained from the finite difference method (FDM) show good agreement with those computed using the adjoint method.

In this optimization, the considered objective function is the lift-to-drag ratio, as shown in the following equation
\begin{equation}
I = -\frac{L/L_0}{D/D_0}
\end{equation}
where $L$ represents the lift and $D$ denotes the drag. The subscript 0 represents the initial value of  performance. 

\begin{figure}[h!]
\centering
\subfigure[Objective Function]{
	\includegraphics[width=2.2in]{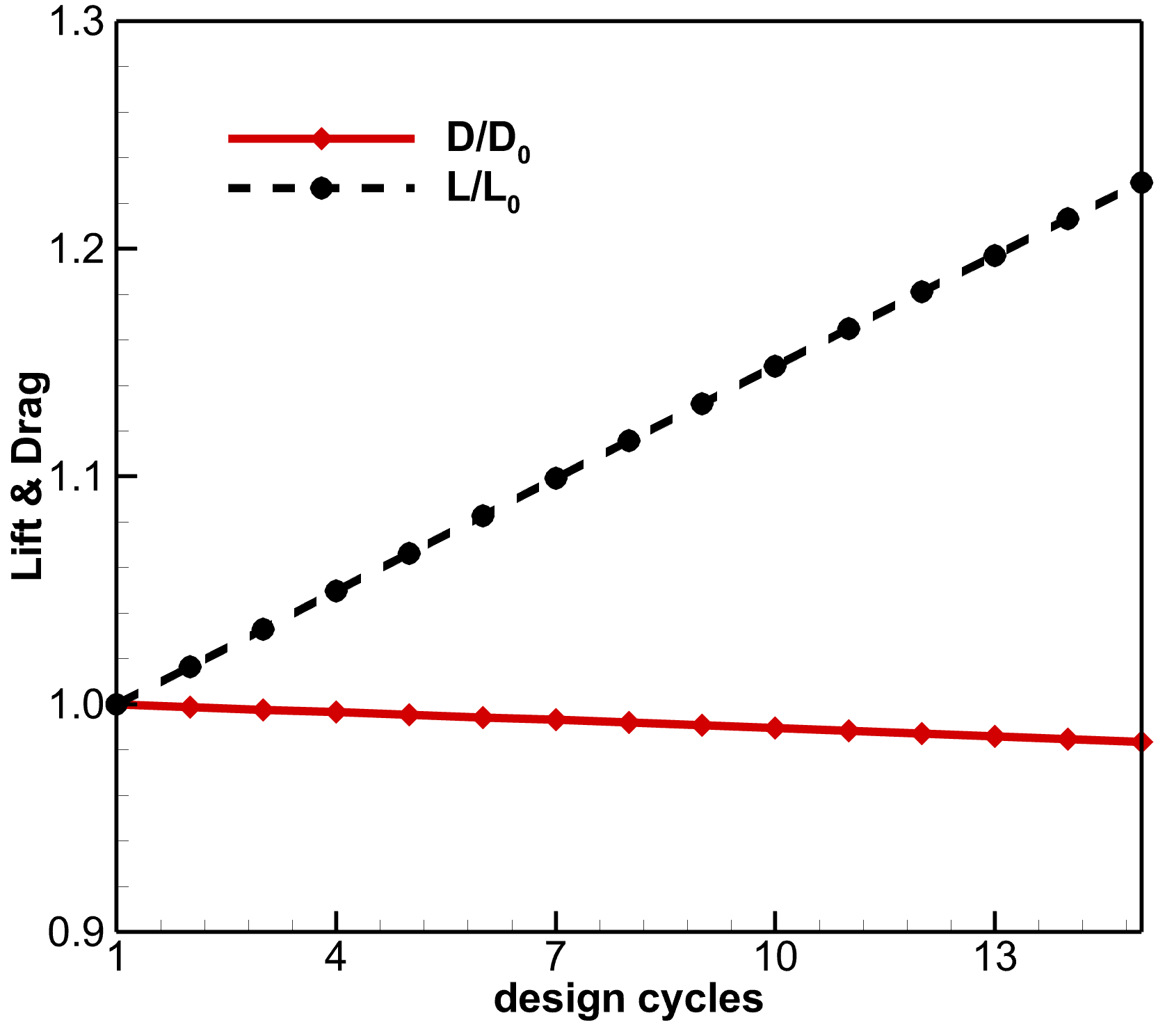}
 \label{RAE_objfun}
 }
 \subfigure[Airfoil]{
	\includegraphics[width=2.2in]{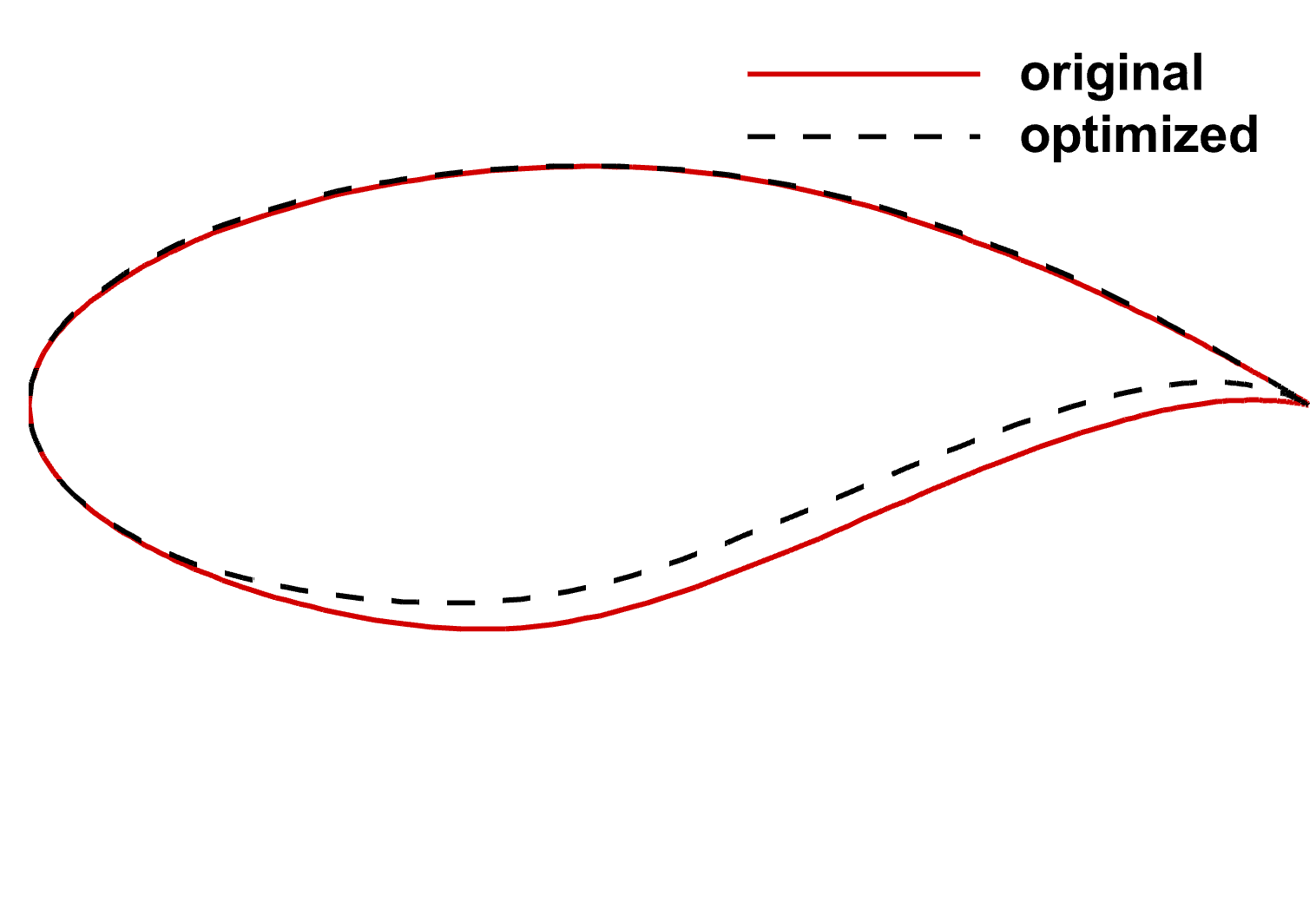}
 \label{RAE_airfoil}
 }
	\caption{Aerodynamic shape optimization of the RAE 2822: a) evolutionary histories; b) original and optimized airfoils.}
\end{figure}

\begin{figure}[h!]
\centering
\subfigure[Pressure]{
	\includegraphics[width=2.2in]{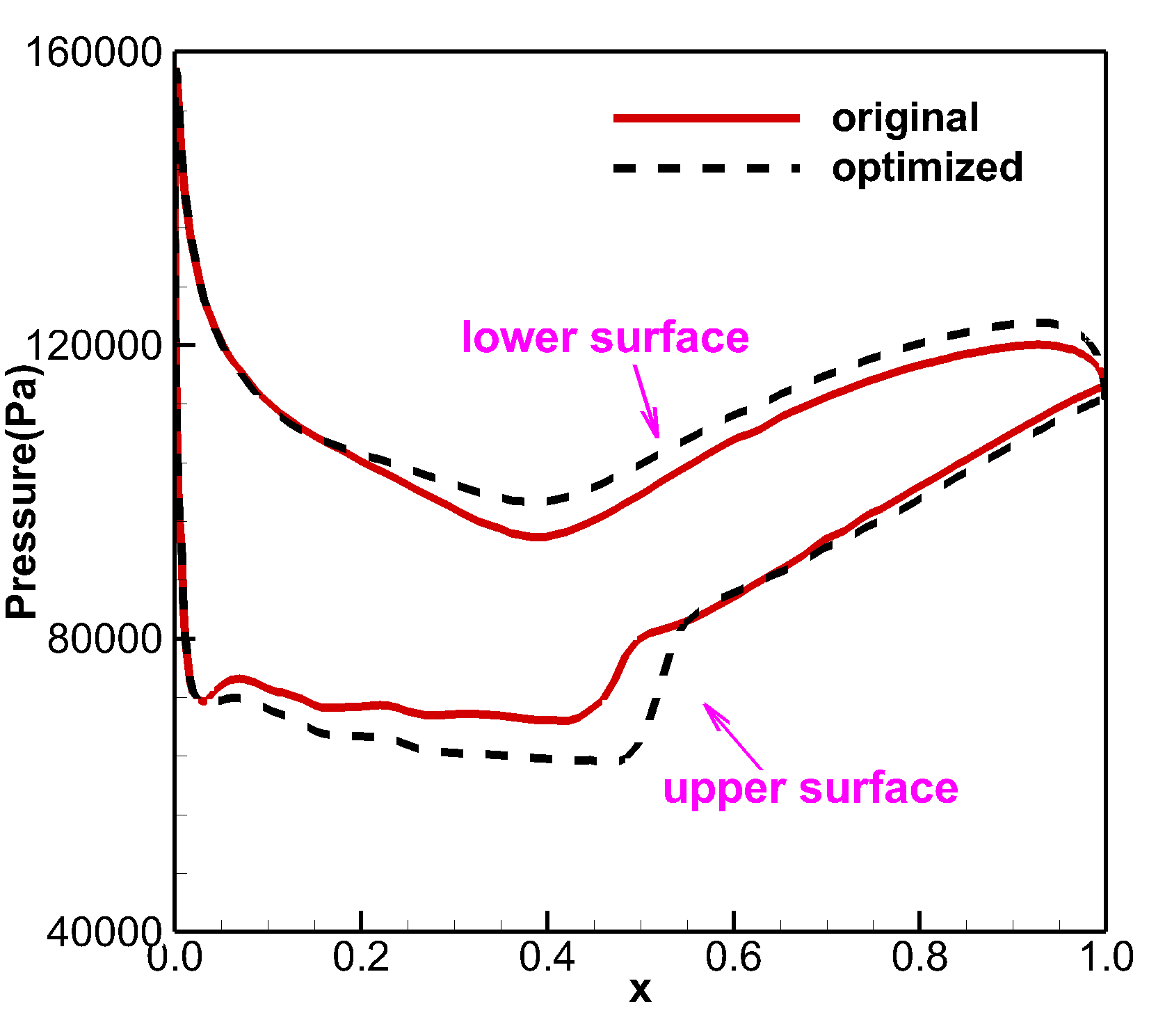}
 \label{RAE_p}
 }
 \subfigure[Drag]{
	\includegraphics[width=2.2in]{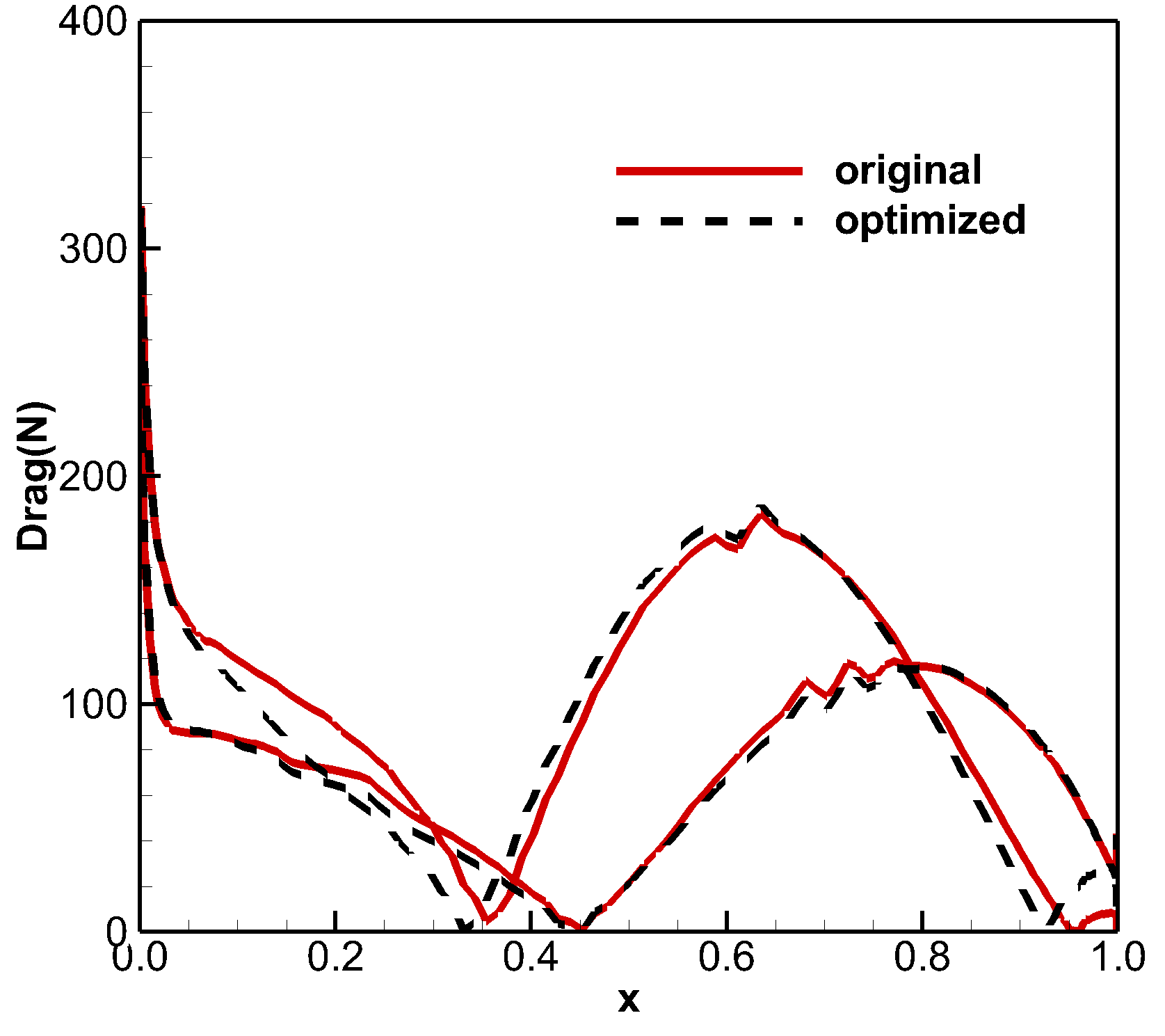}
 \label{RAE_Drag}
 }
	\caption{Comparison of static pressure and drag on airfoil surfaces for the RAE 2822: a) pressure; b) drag.}
\end{figure}

Figure~\ref{RAE_objfun} presents the evolution of the drag and lift during the optimization process. After 15 design cycles, the drag is reduced by approximately 3\%, while the lift is increased by about 22\%. Figure~\ref{RAE_airfoil} compares the original and optimized airfoil shapes. Compared to the initial configuration, the optimized airfoil exhibits increased camber, leading to enhanced acceleration of the flow over the upper surface and deceleration over the lower surface. Consequently, the velocity increases on the upper surface and decreases on the lower surface.
The pressure distribution, shown in Fig.~\ref{RAE_p}, exhibits an opposite trend to the velocity field. Specifically, the pressure increases on the lower surface and decreases on the upper surface, which contributes to the improvement in lift. Figure~\ref{RAE_Drag} compares the surface drag distributions between the original and optimized configurations. The drag reduction is primarily observed in the chordwise range of $x \in [0.1, 0.4]$. 

\subsection{Supersonic Case 3: NACA 0012}

The third case considers a NACA 0012 airfoil under supersonic conditions. The freestream Mach number is $M_\infty = 1.5$, the angle of attack is $\alpha = 0^\circ$, and the Reynolds number is $Re = 6.5 \times 10^6$. The freestream temperature and pressure are set to 300 K and 101325 Pa, respectively.

\begin{figure}[h!]
\centering
\subfigure[Flow]{
	\includegraphics[width=2.2in]{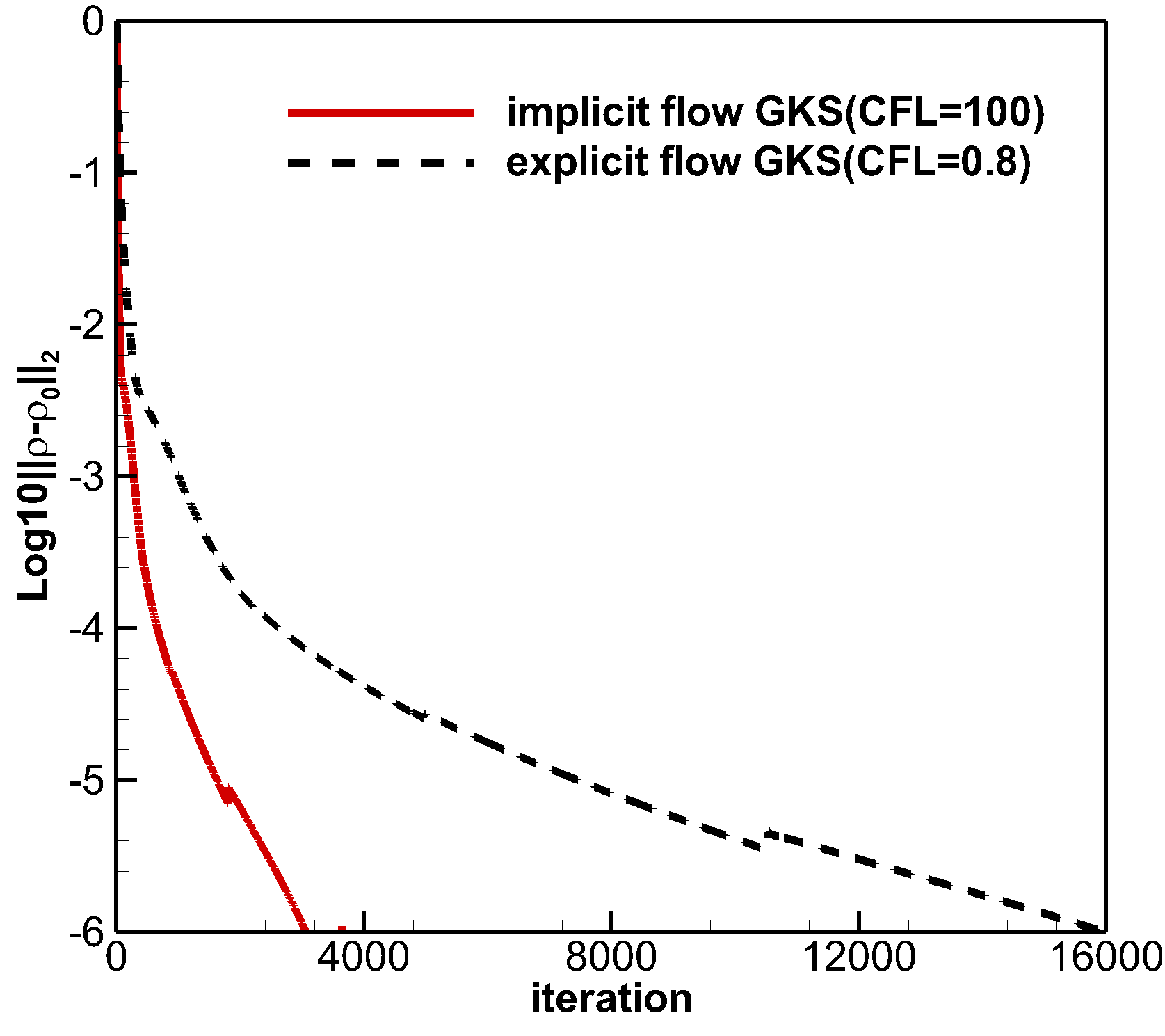}
 \label{naca_flowres}
 }
 \subfigure[Adjoint]{
	\includegraphics[width=2.2in]{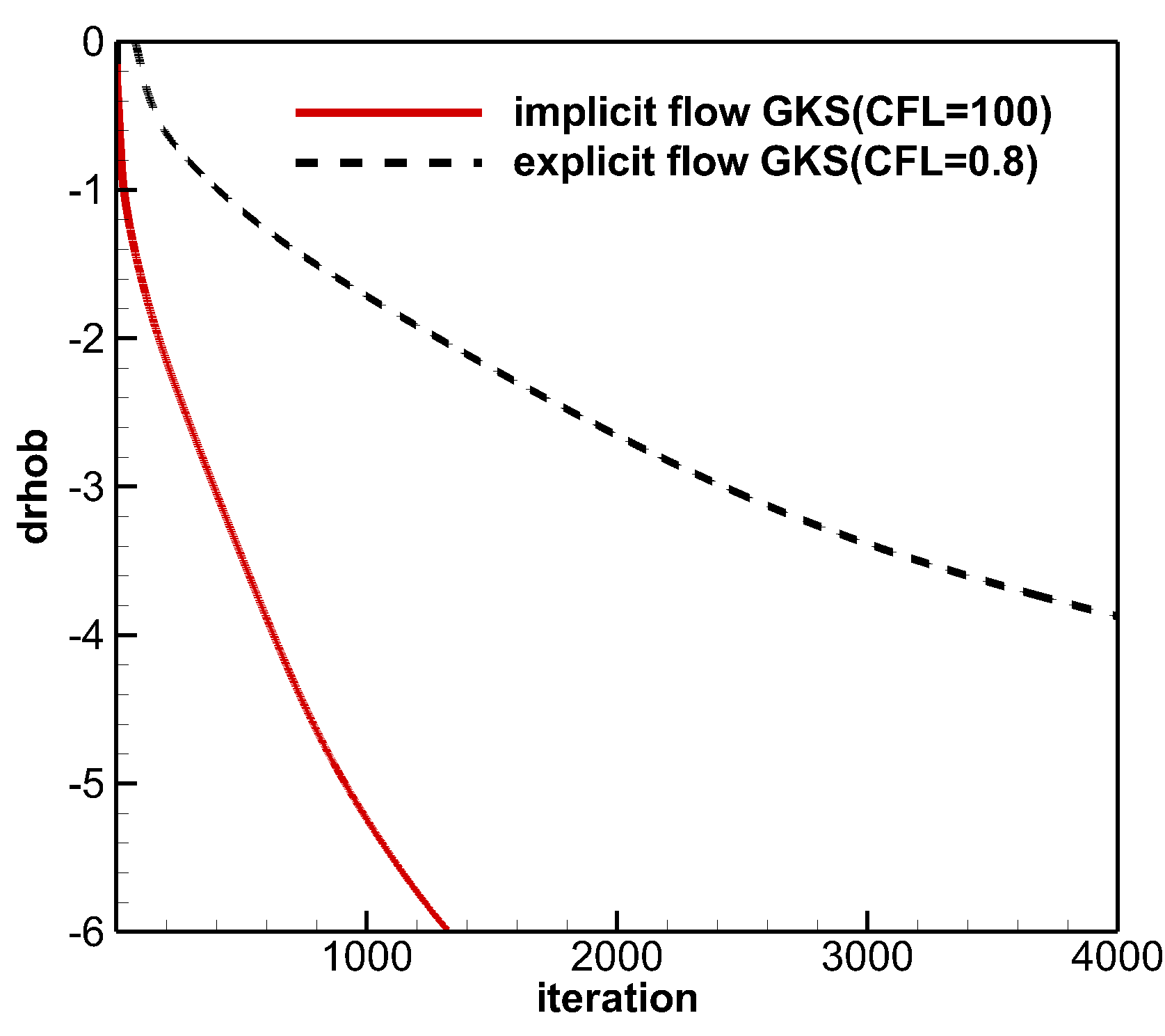}
 \label{naca_adjres}
 }
	\caption{Convergence histories of the residuals related to the mass equation for the NACA 0012 case: a) flow GKS solver; b) adjoint GKS solver.}
\end{figure} 
Figures~\ref{naca_flowres} and~\ref{naca_adjres} present the residual convergence histories of the flow and adjoint GKS solvers. When the residual is reduced to the order of $10^{-6}$, the implicit method requires only about 3500 iterations, whereas the explicit method requires more than 10000 iterations. Compared with the explicit approach, the implicit method reduces the computational cost by a factor of over 4.

\begin{figure}[h!]
\centering
\subfigure[Flow]{
	\includegraphics[width=2.2in]{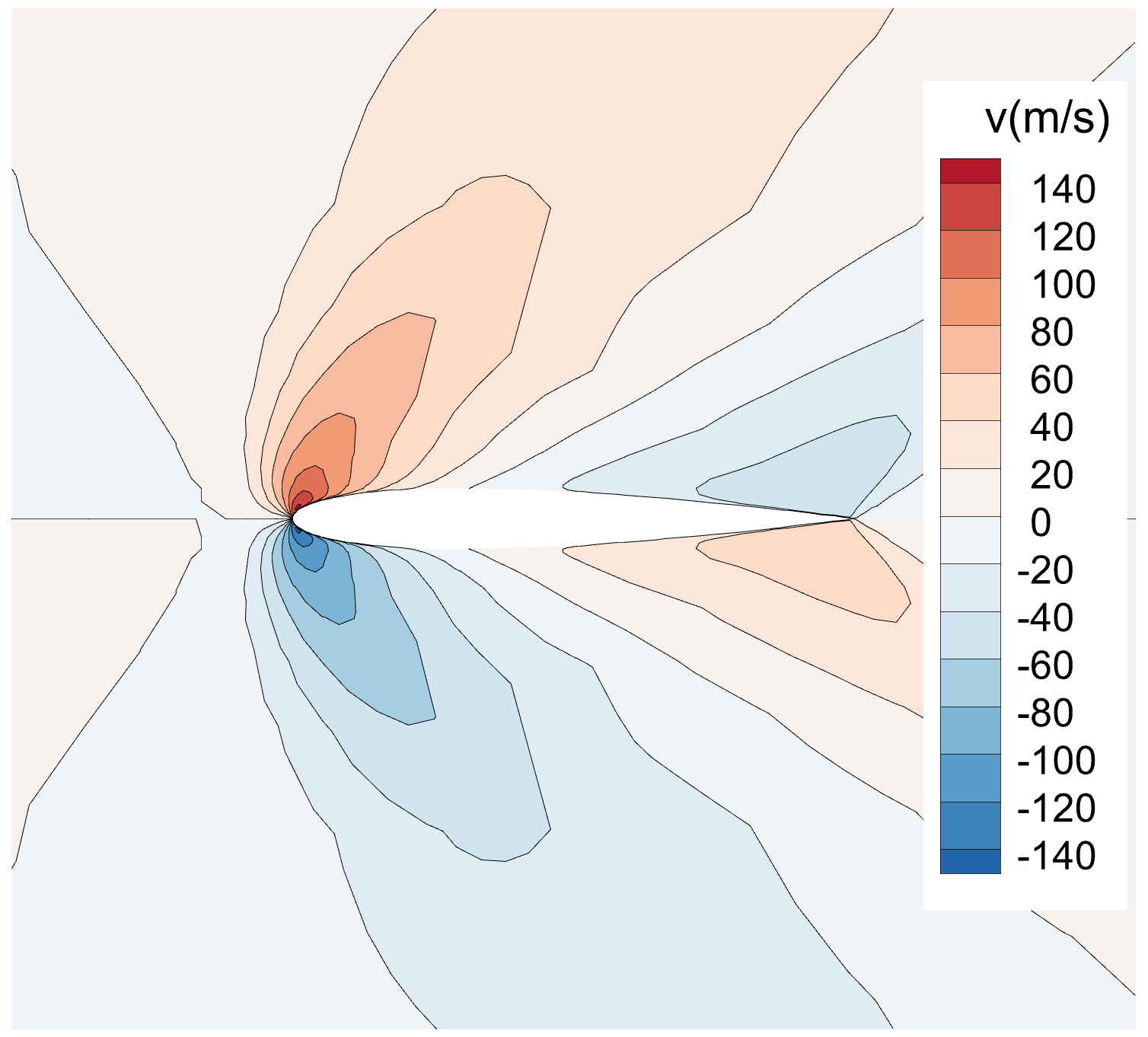}
 \label{naca_flow}
 }
 \subfigure[Adjoint]{
	\includegraphics[width=2.2in]{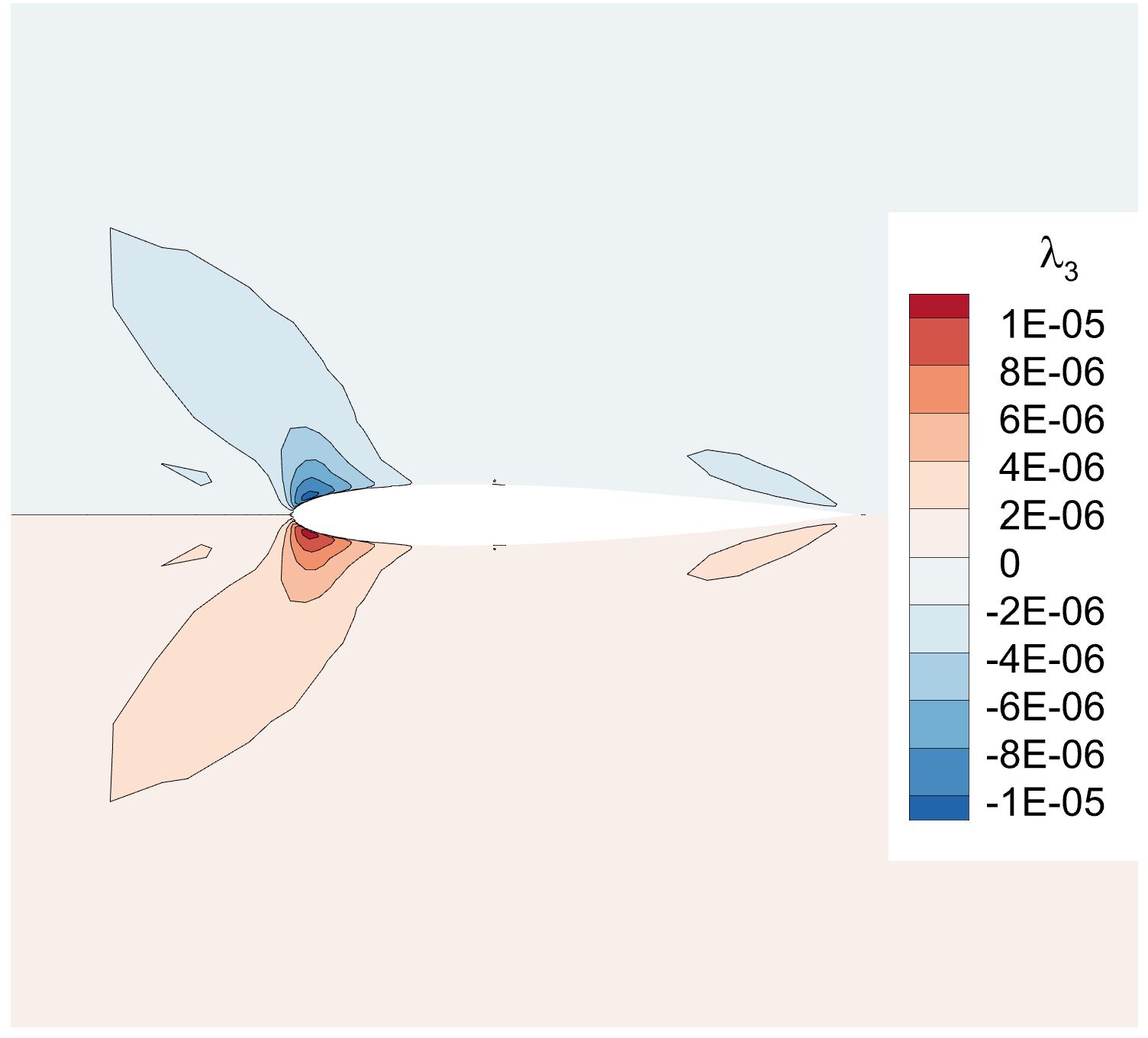}
 \label{naca_adj}
 }
	\caption{Flow and adjoint variable contours related to the y-moment equation for the NACA 0012 case.}
\end{figure}
Figures~\ref{naca_flow} and~\ref{naca_adj} present the flow and adjoint fields associated with the $y$-momentum equation. The adjoint field exhibits a structure that is opposite to that of the flow field. Near the leading edge, the flow variables are positive in the upper region and negative in the lower region, while the adjoint variables display the opposite trend. 

\begin{figure}[h!]
\centering
\subfigure[Flow]{
	\includegraphics[width=2.2in]{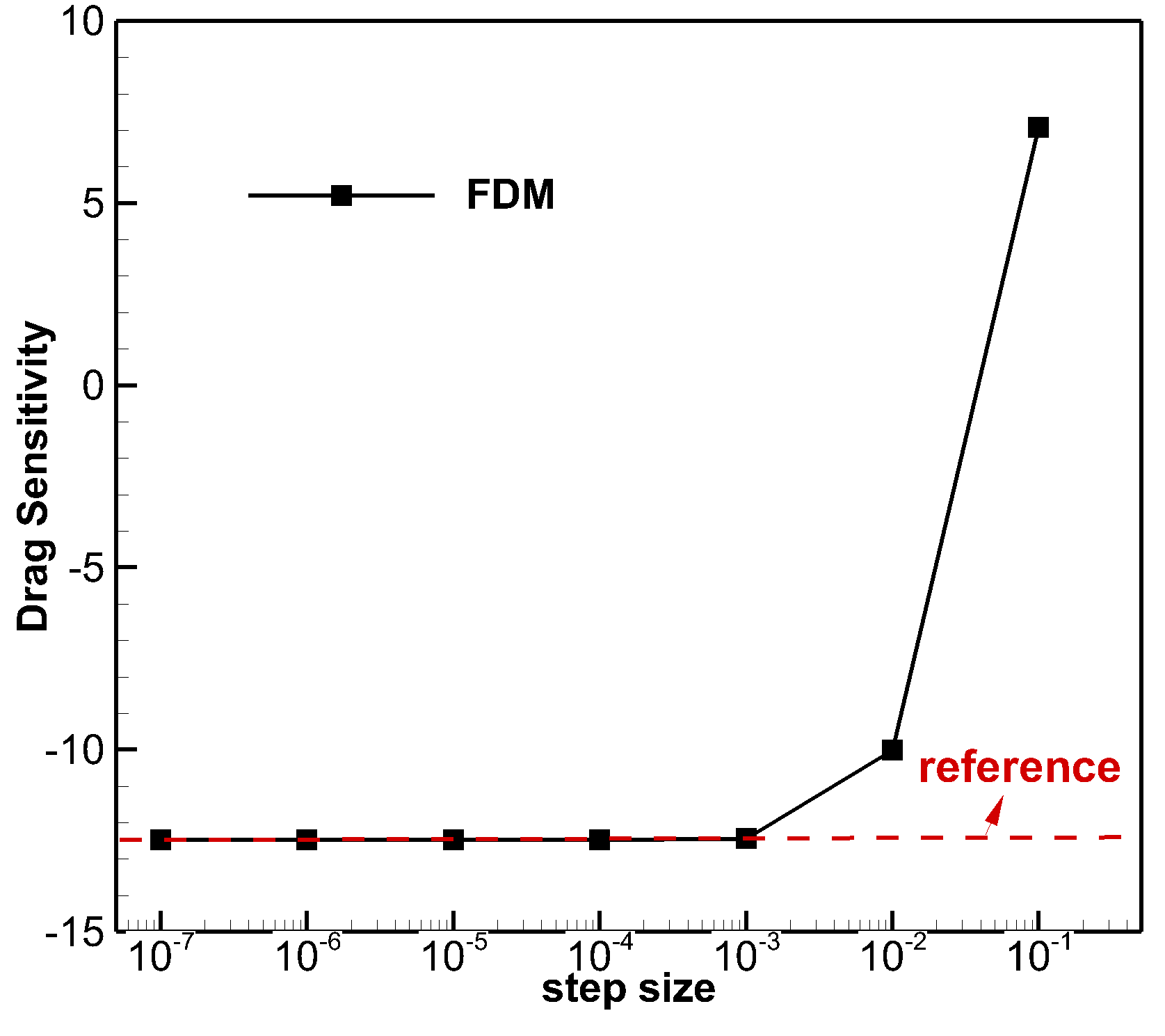}
 \label{naca_FDM}
 }
 \subfigure[Adjoint]{
	\includegraphics[width=2.2in]{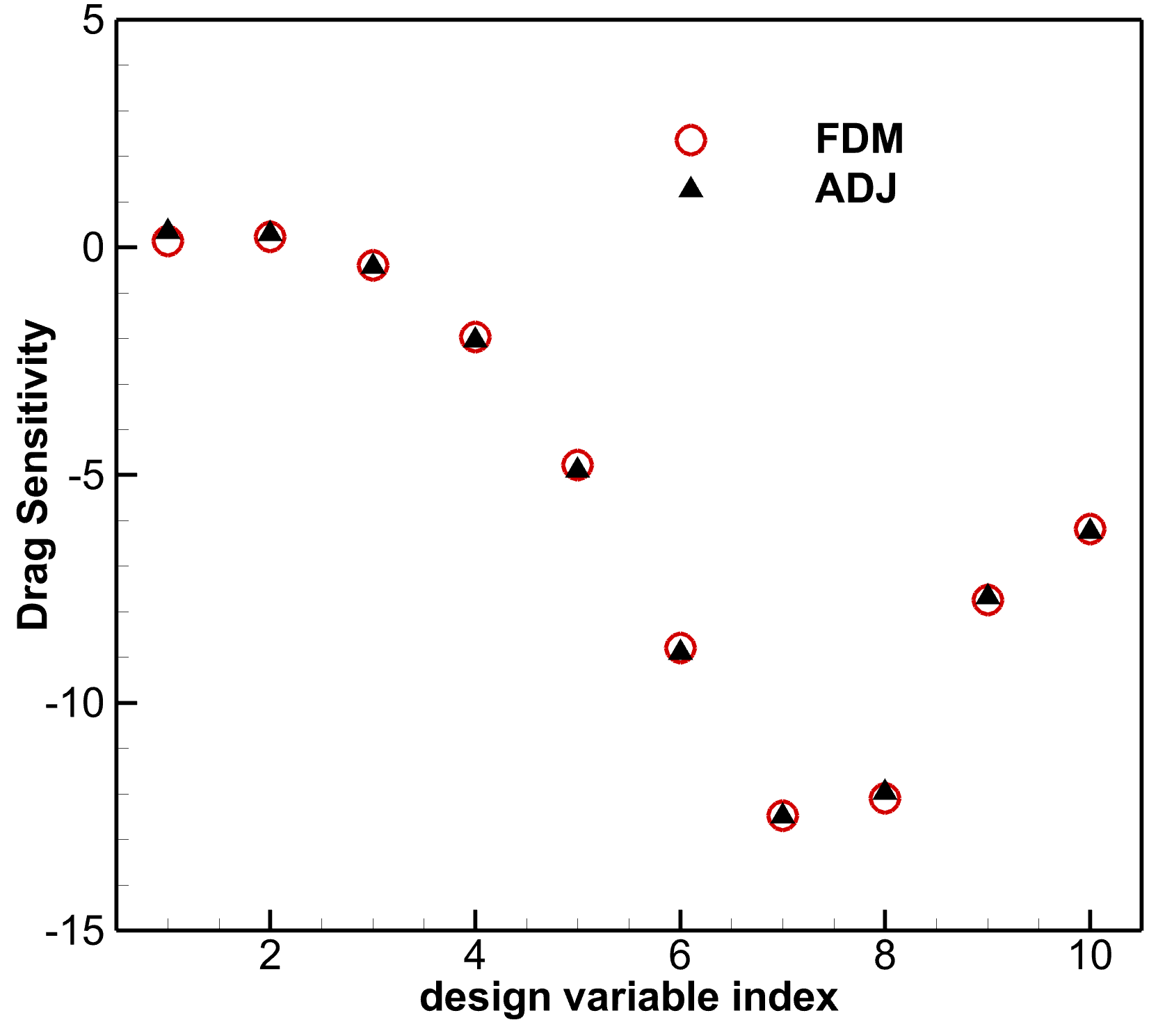}
 \label{naca_sen}
 }
	\caption{Verification of the adjoint sensitivities for the NACA 0012 case: a) step size independence study; b) comparison between the adjoint method and FDM.}
\end{figure}
\begin{figure}[h!]
\centering
\subfigure[Objective Function]{
	\includegraphics[width=2.2in]{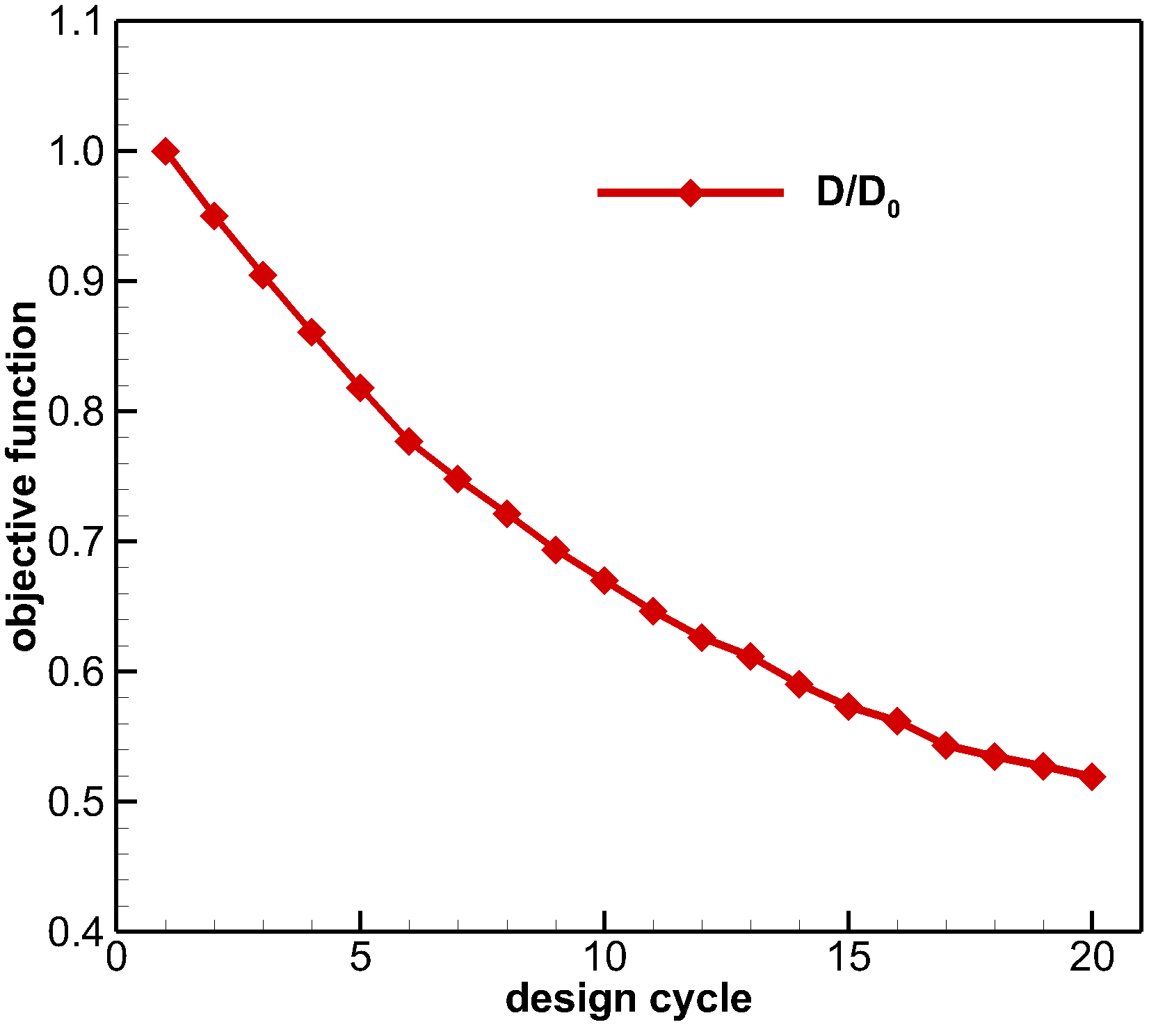}
 \label{naca_obj}
 }
 \subfigure[Shape]{
	\includegraphics[width=2.2in]{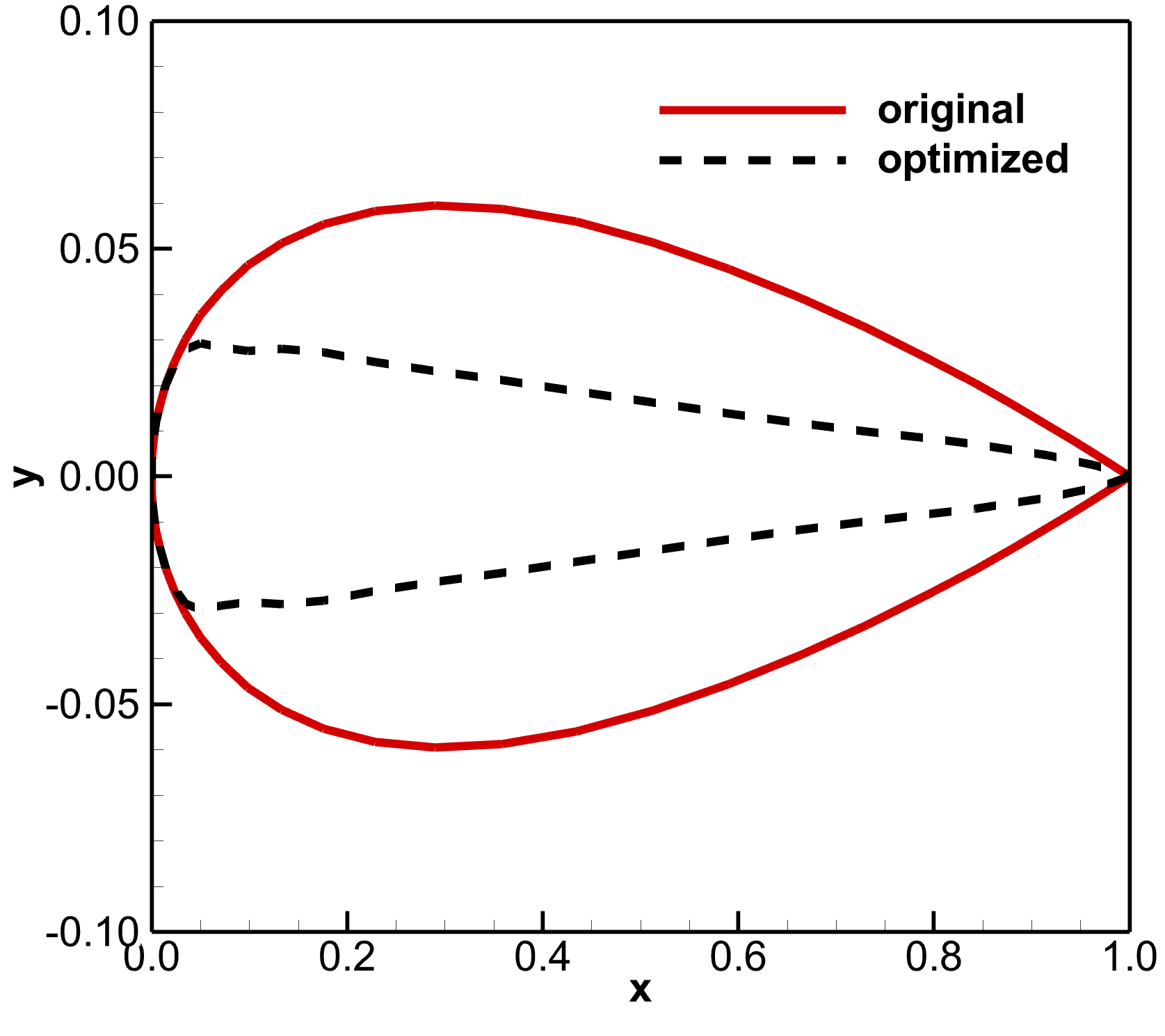}
 \label{naca_airfoil}
 }
	\caption{Aerodynamic shape optimization of the NACA 0012: a) evolutionary histories; b) original and optimized airfoils.}
\end{figure}

\begin{figure}[h!]
\centering
\subfigure[Original]{
	\includegraphics[width=2.2in]{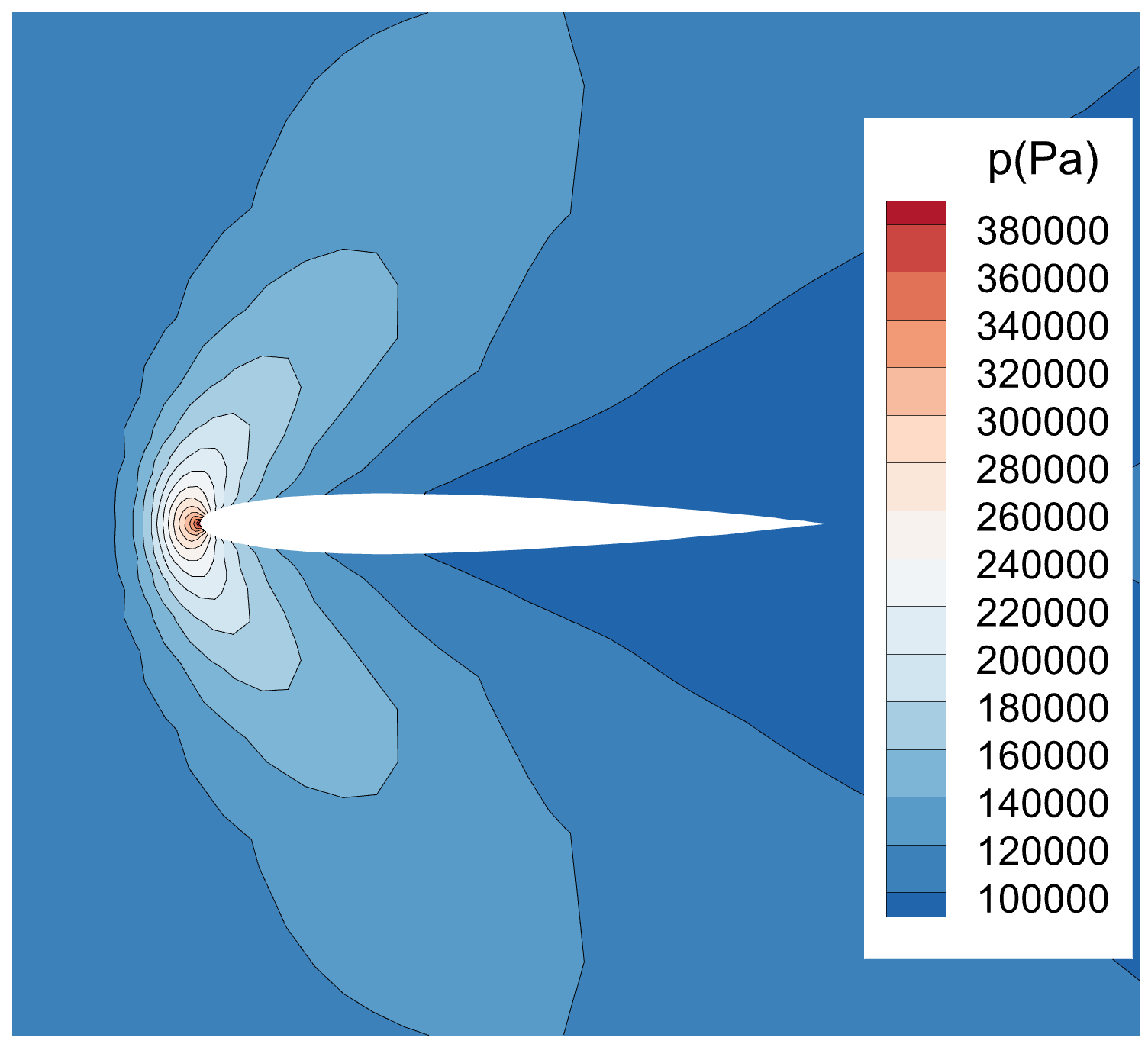}
 \label{naca_porg}
 }
 \subfigure[Optimized]{
	\includegraphics[width=2.2in]{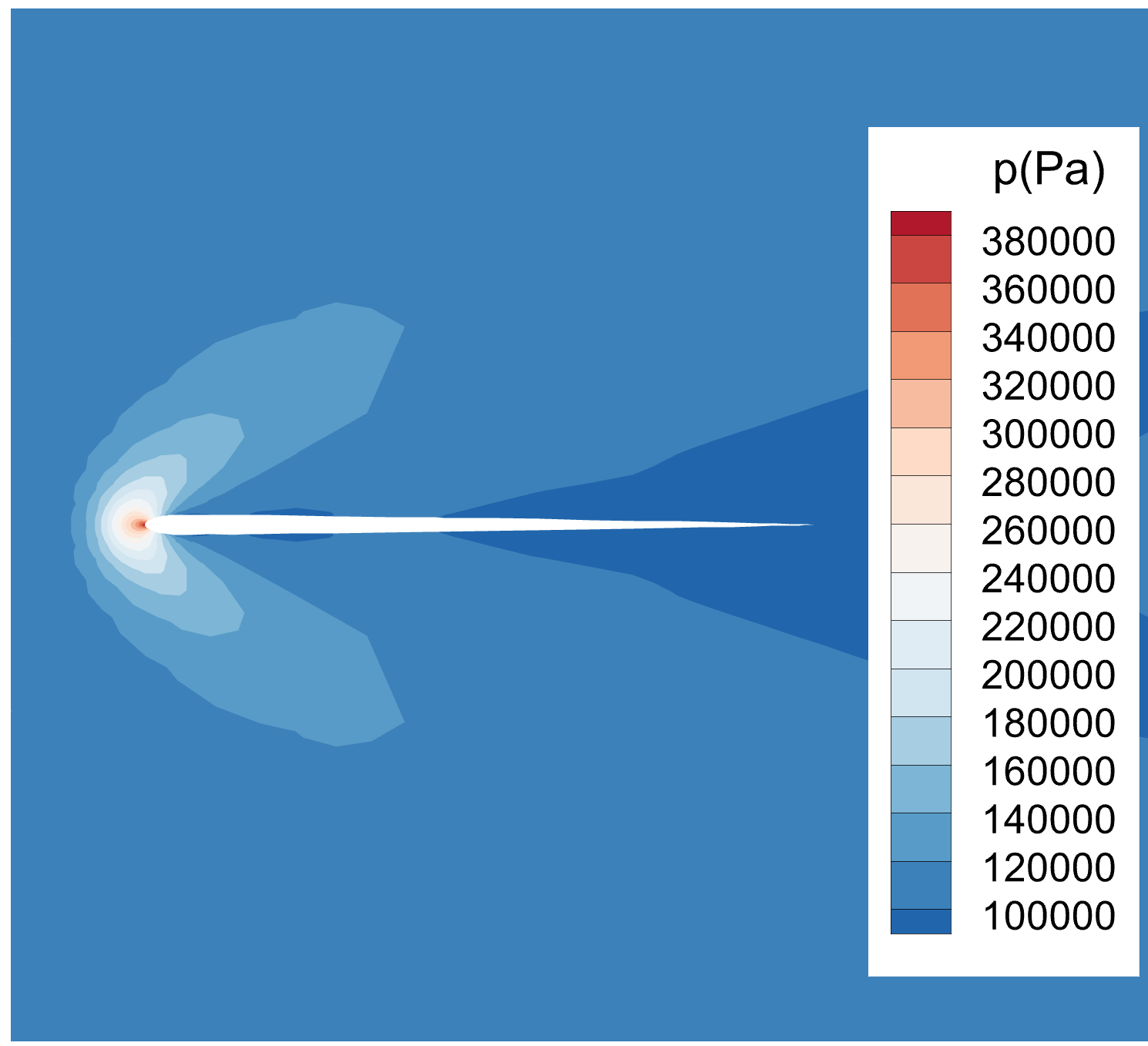}
 \label{naca_popt}
 }
	\caption{Static pressure contours in the whole computational domain: a) original; b) optimized.}
\end{figure}
\begin{figure}[h!]
\centering
\subfigure[Pressure]{
	\includegraphics[width=2.2in]{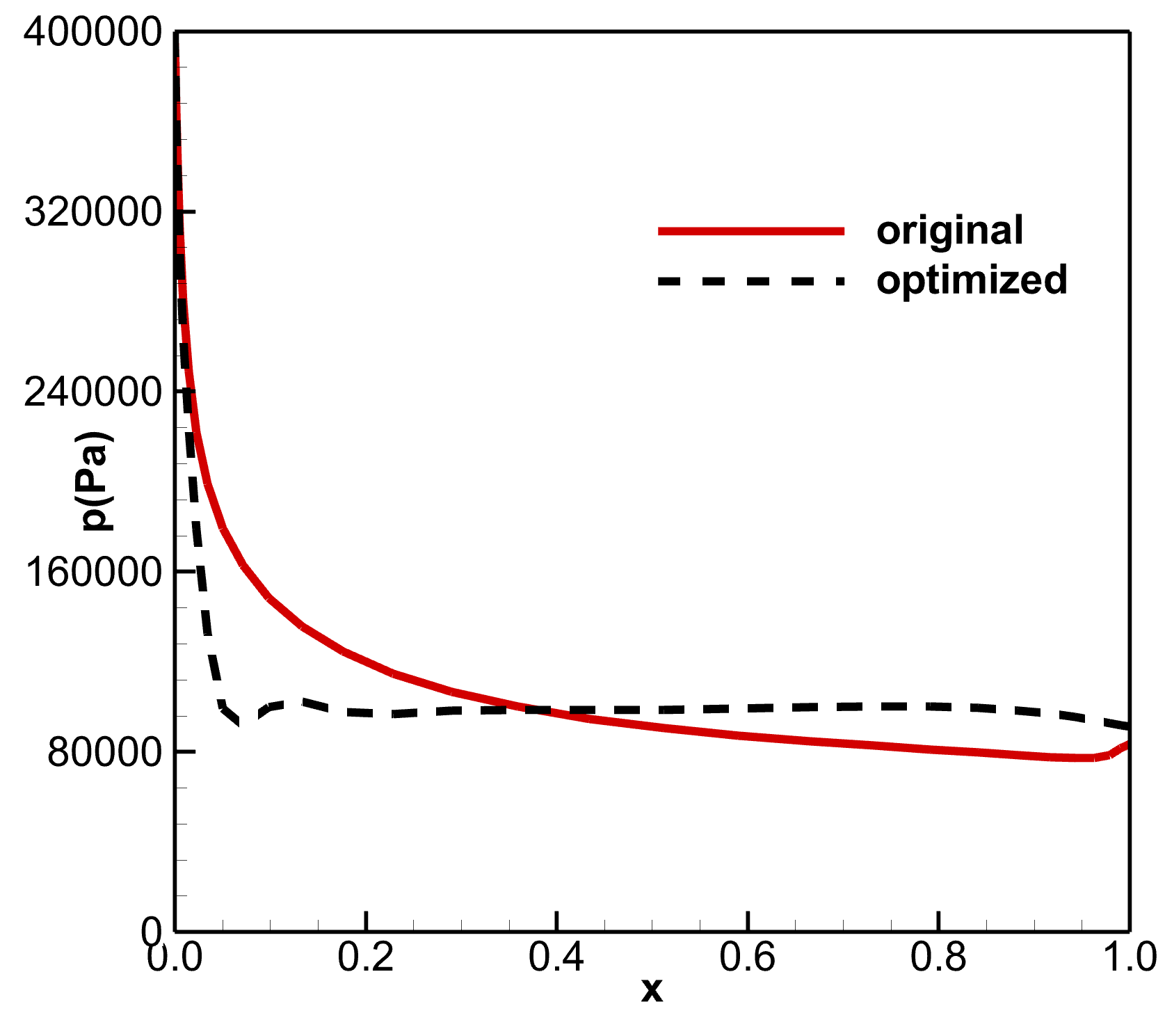}
 \label{naca_Pcomp}
 }
 \subfigure[Drag]{
	\includegraphics[width=2.2in]{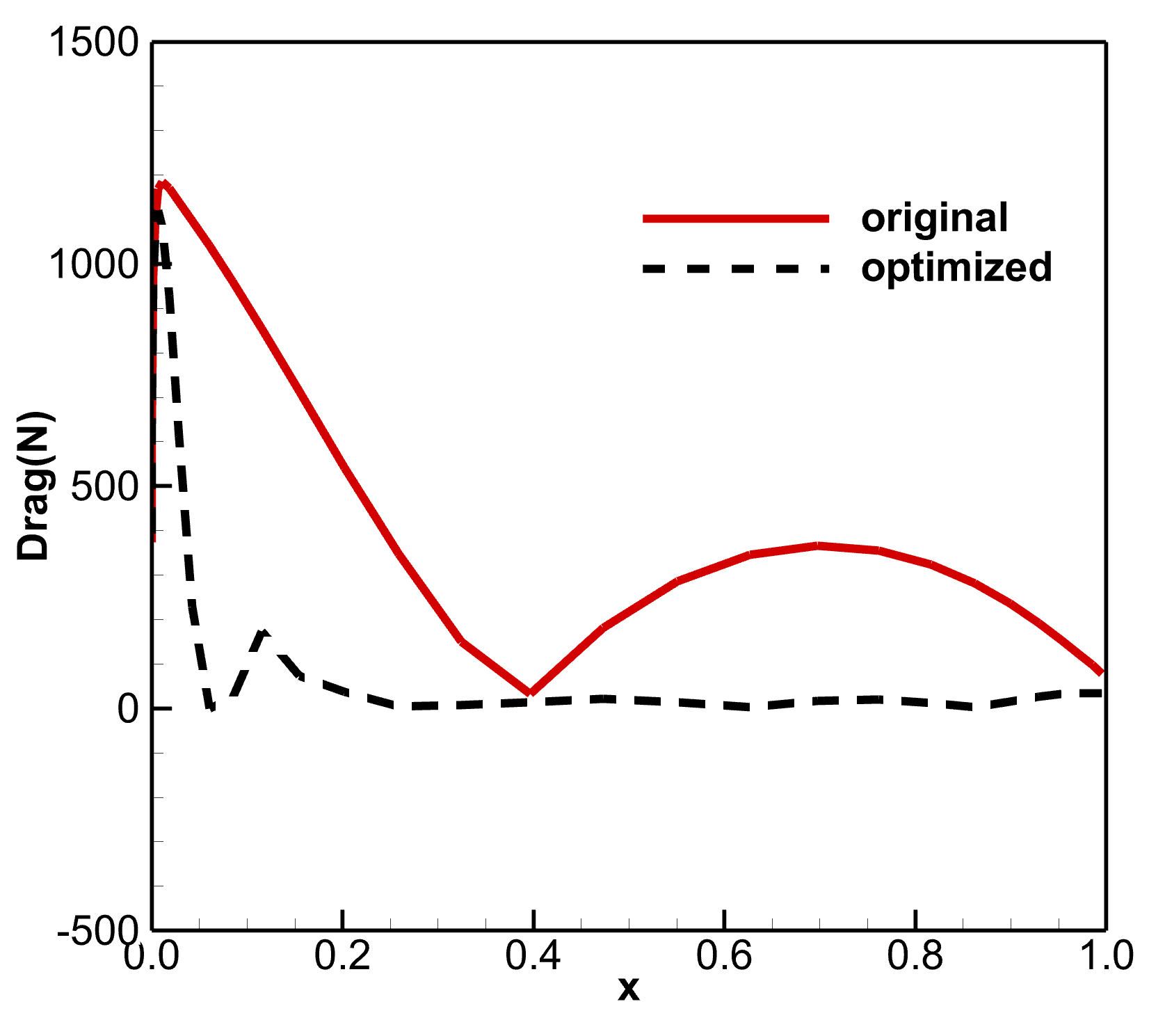}
 \label{naca_Dcomp}
 }
	\caption{Comparison of static pressure and drag on airfoil surfaces for the NACA 0012: a) pressure; b) drag.}
\end{figure}
In this optimization, the objective function is defined as the drag on the airfoil surface. Figures~\ref{naca_FDM} and~\ref{naca_sen} present the verification of the adjoint sensitivities, demonstrating good agreement between the finite difference method and the adjoint approach.
Figure~\ref{naca_obj} shows the evolution of the objective function, with the drag reduced by more than 50\% after 20 design cycles. This reduction can be attributed to the development of a more slender airfoil shape, as illustrated in Fig.~\ref{naca_airfoil}. The slender configuration alleviates the pressure-difference drag, as evidenced by the pressure distributions shown in Figs.~\ref{naca_porg}, ~\ref{naca_popt}, and ~\ref{naca_Pcomp}. Figure~\ref{naca_Dcomp} compares the surface drag distributions. After optimization, the drag is significantly reduced over the entire airfoil surface.

\subsection{Hypersonic Case 4: Flow Past a Semi-Cylinder}

The final test case considers hypersonic flow past a semi-cylinder. The cylinder has a radius of 0.5, while the outer computational boundary is located at three times the cylinder radius. The computational domain is discretized using a structured mesh of 51 $\times$61 cells. The freestream Mach number and Reynolds number are set to 8.0 and 1000, respectively. A kinetic boundary condition is imposed on the solid wall, whereas non-reflecting boundary conditions based on Riemann invariants are applied at the far-field boundaries.

\begin{figure}[h!]
\centering
\subfigure[Flow]{
	\includegraphics[width=2.2in]{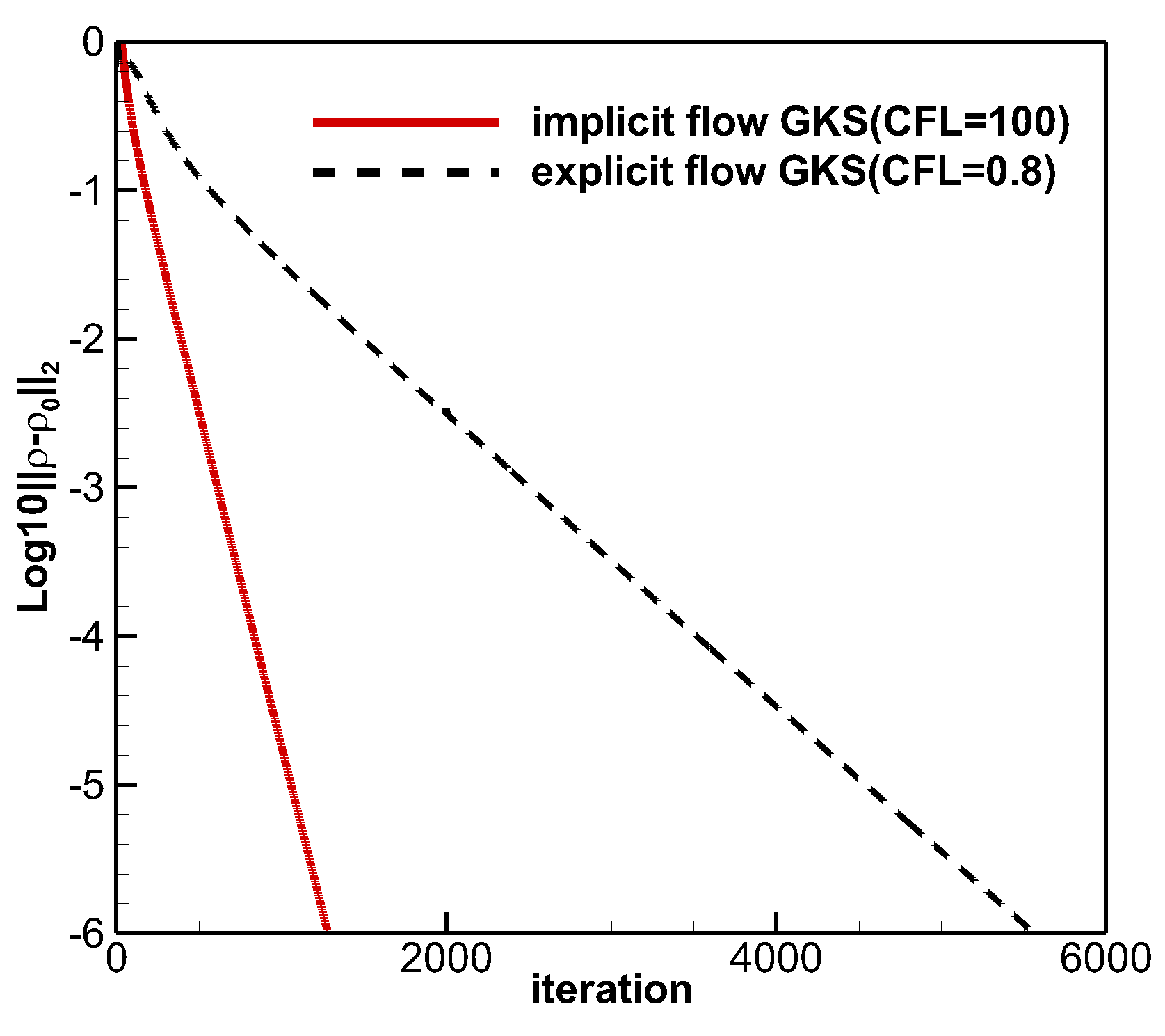}
 \label{semi-cylinder_flowres}
 }
 \subfigure[Adjoint]{
	\includegraphics[width=2.2in]{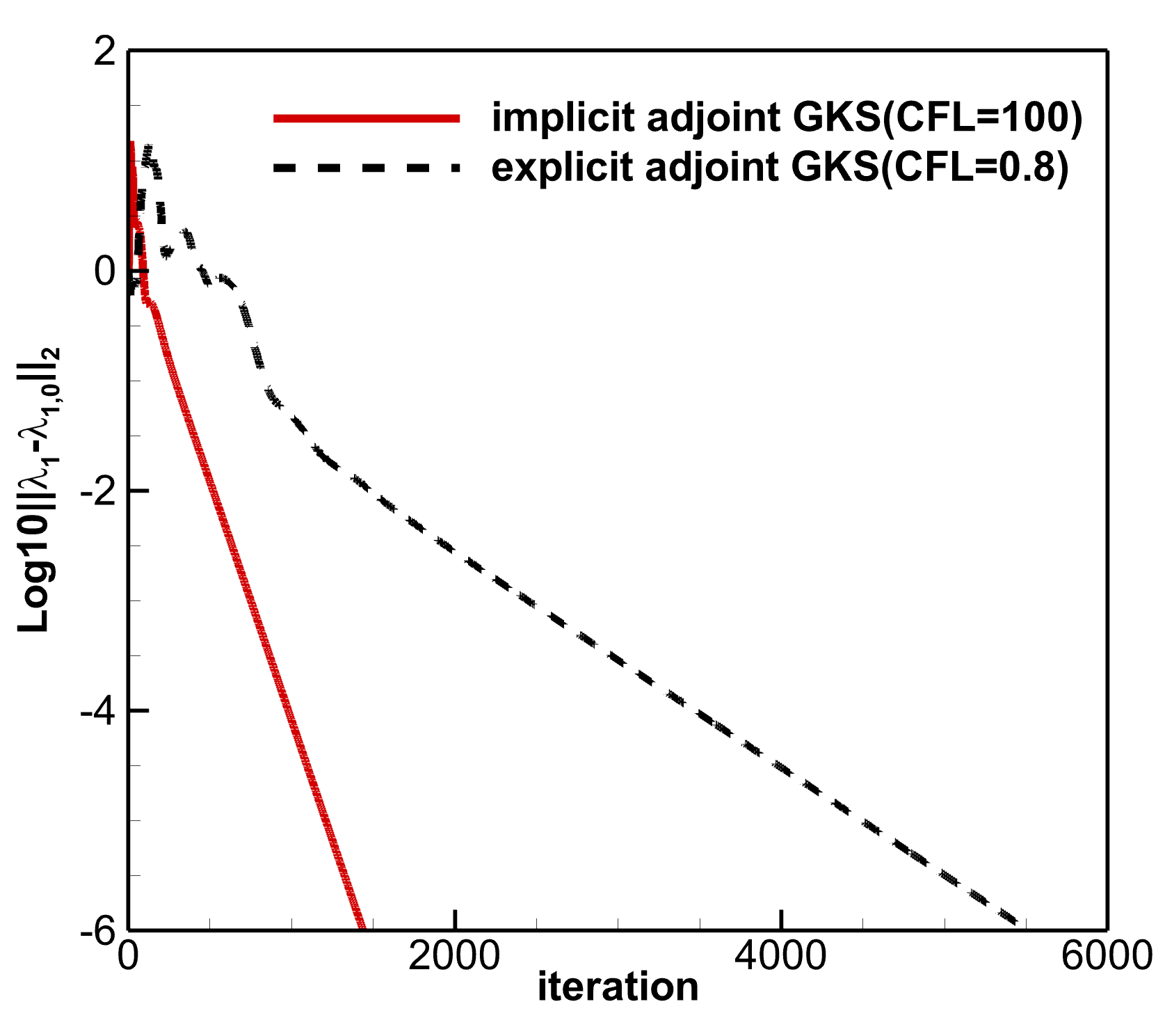}
 \label{semi-cylinder_adjres}
 }
	\caption{Convergence histories of the residuals related to the mass equation: a) flow GKS solver; b) adjoint GKS solver.}
\end{figure} 
First, the effectiveness of the implicit LU-SGS scheme in accelerating convergence is assessed. Figures~\ref{semi-cylinder_flowres} and \ref{semi-cylinder_adjres} present the convergence histories of the flow and adjoint solutions, respectively. The maximum allowable CFL number is 0.8 for the explicit flow and adjoint GKS solvers, whereas it reaches 100 for their implicit counterparts. To achieve a reduction of six orders of magnitude in the flow residual, the explicit flow GKS requires approximately 4800 iterations, while the implicit solver converges within 1200 iterations. This demonstrates that the implicit LU-SGS approach reduces the computational cost by a factor of about four for the flow solver. For the adjoint GKS solver, the implicit formulation yields a computational cost reduction by a factor of approximately 3.5. Based on these advantages, the implicit adjoint GKS solver is subsequently employed for sensitivity analysis and design optimization.

In the design optimization, ten design variables are uniformly distributed along the upper surface of the semi-cylinder, while the deformation of the lower surface follows an opposite trend relative to that of the upper surface. The objective function is defined as the drag coefficient normalized by its initial value, which is given by
\begin{equation}
I = \frac{D}{D_0}.
\end{equation}

\begin{figure}[h!]
\centering
\subfigure[Flow]{
	\includegraphics[width=1.5in]{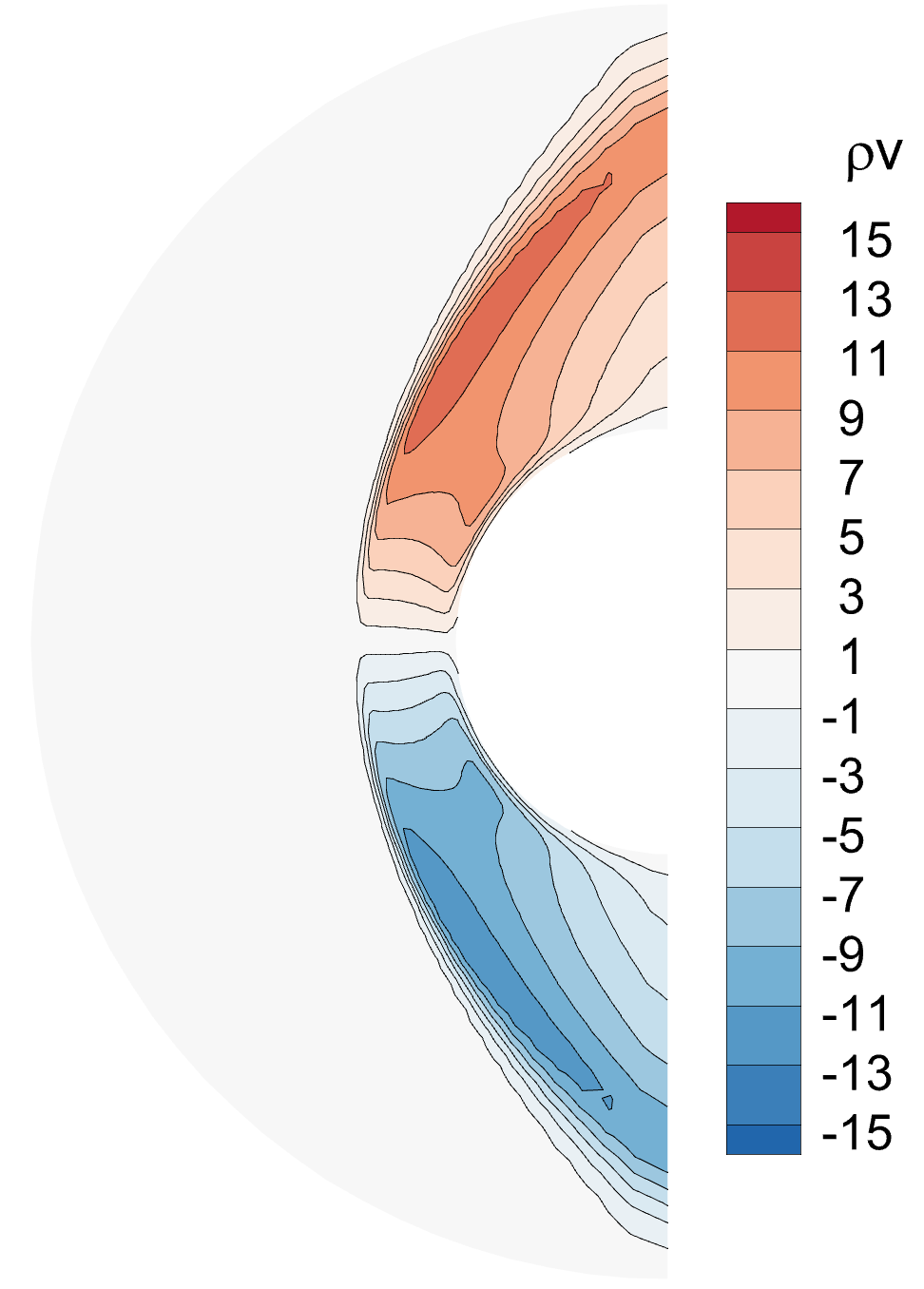}
 \label{semi-cylinder_flow}
 }
 \subfigure[Adjoint]{
	\includegraphics[width=1.7in]{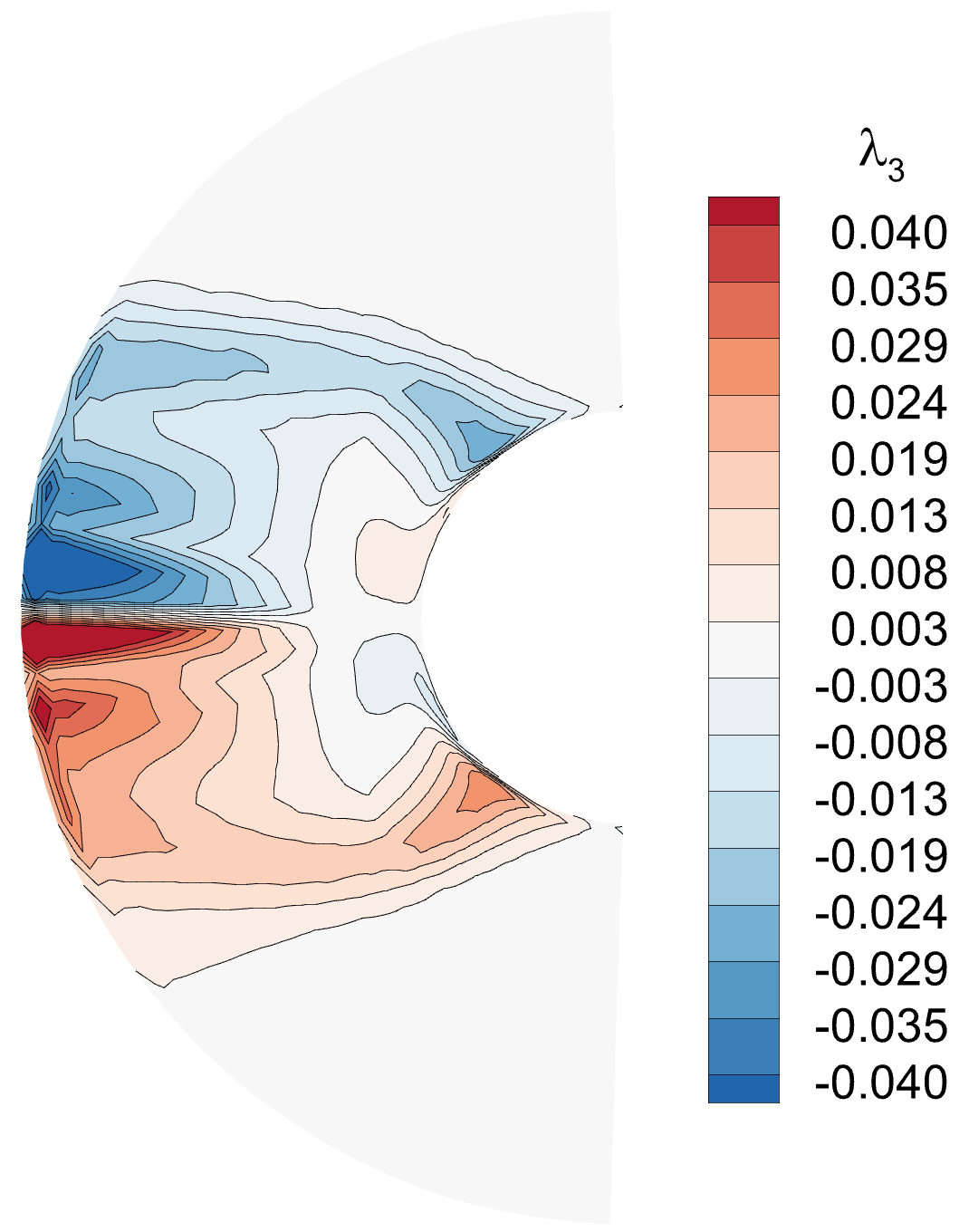}
 \label{semi-cylinder_adj}
 }
	\caption{Flow and adjoint variable contours related to the y-moment equation for the semi-cylinder case.}
\end{figure}

Second, the adjoint variable contours associated with the y-momentum equation are compared with the corresponding flow fields, as shown in Figs.~\ref{semi-cylinder_flow} and \ref{semi-cylinder_adj}. It is observed that the adjoint fields exhibit an opposite distribution to the flow fields; for example, regions of positive flow velocity correspond to negative adjoint variables. This behavior provides a qualitative verification of the correct implementation of the discrete adjoint GKS solver. 

\begin{figure}[h!]
\centering
\subfigure[]{
	\includegraphics[width=2.2in]{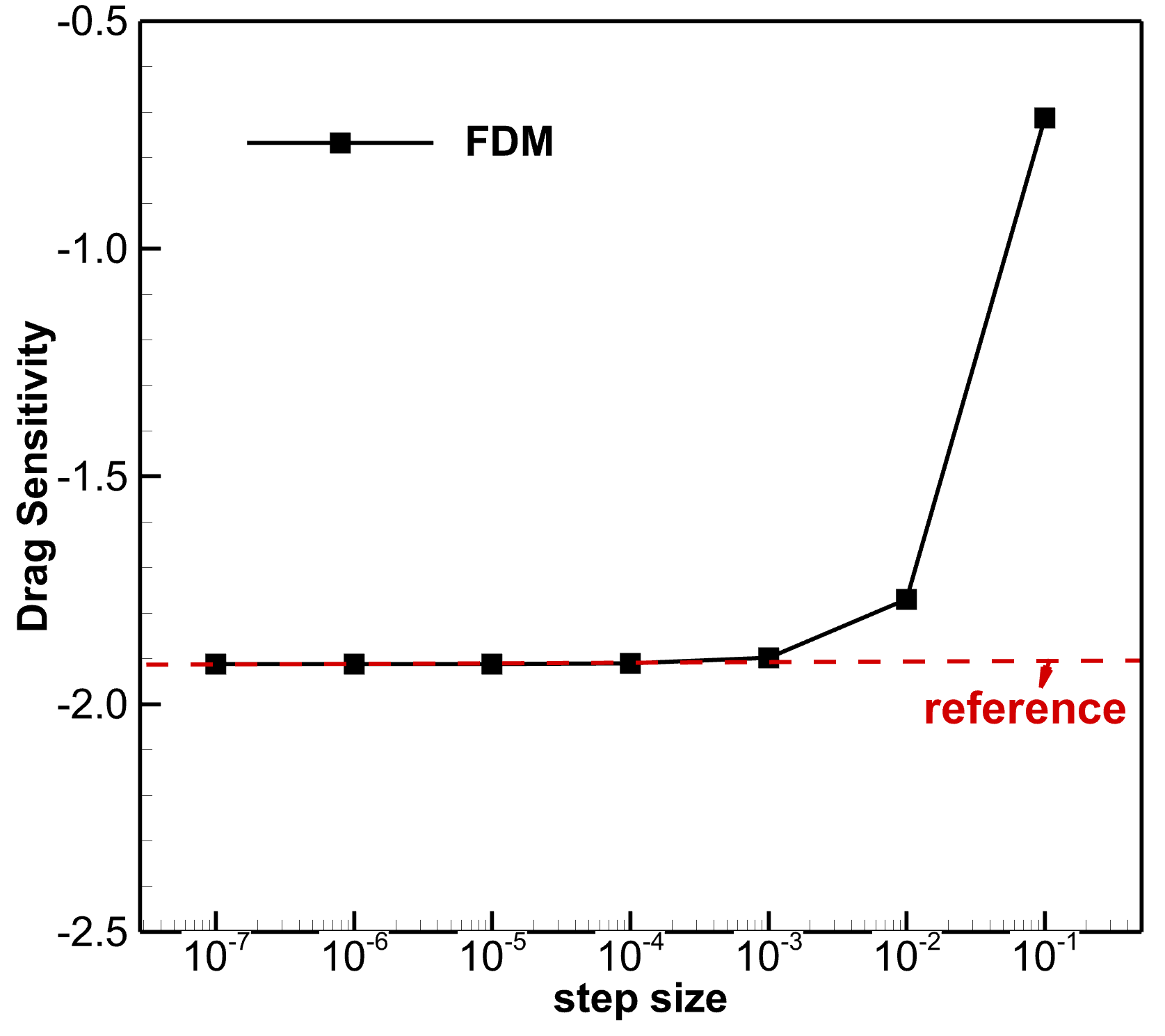}
 \label{semi-cylinder_FDM}
 }
 \subfigure[]{
	\includegraphics[width=2.2in]{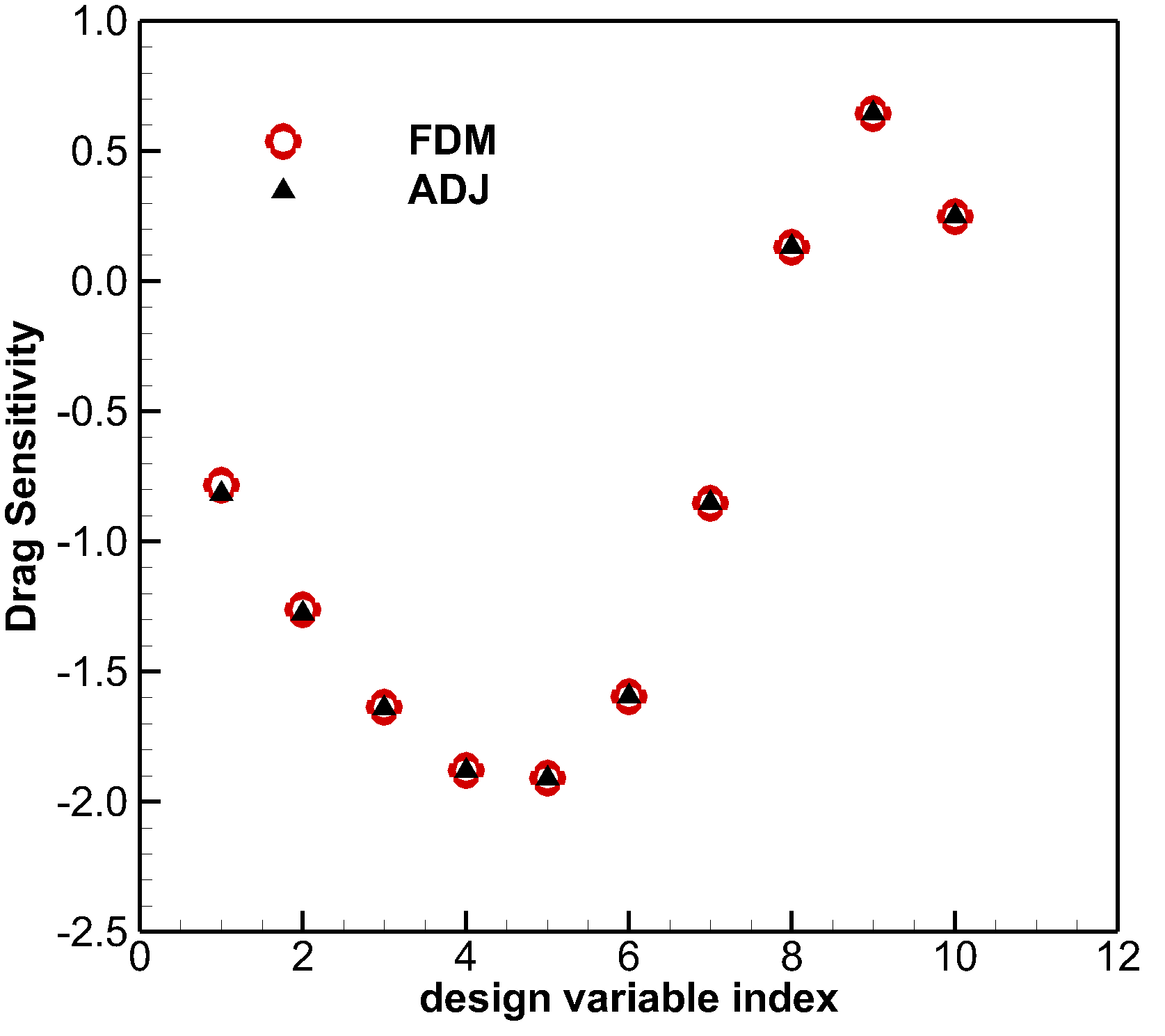}
 \label{semi-cylinder_sen}
 }
	\caption{Verification of the adjoint sensitivities for the semi-cylinder case: a) step size independence study; b) comparison between the adjoint method and FDM.}
\end{figure}
Furthermore, the adjoint sensitivities are quantitatively verified against finite-difference method results. A step-size independence study is first conducted for the fifth design variable, as illustrated in Fig.~\ref{semi-cylinder_FDM}. Within the range of $[10^{-2}, 10^{-7}]$, the FDM sensitivities remain essentially unchanged, whereas larger step sizes introduce significant errors. Based on this observation, a step size of $10^{-4}$ is adopted for all design variables to ensure accuracy. Figure~\ref{semi-cylinder_sen} compares the sensitivity results obtained from the FDM and the adjoint GKS solver, showing excellent agreement between the two approaches.

\begin{figure}[h!]
\centering
\subfigure[Objective Function]{
	\includegraphics[width=2.2in]{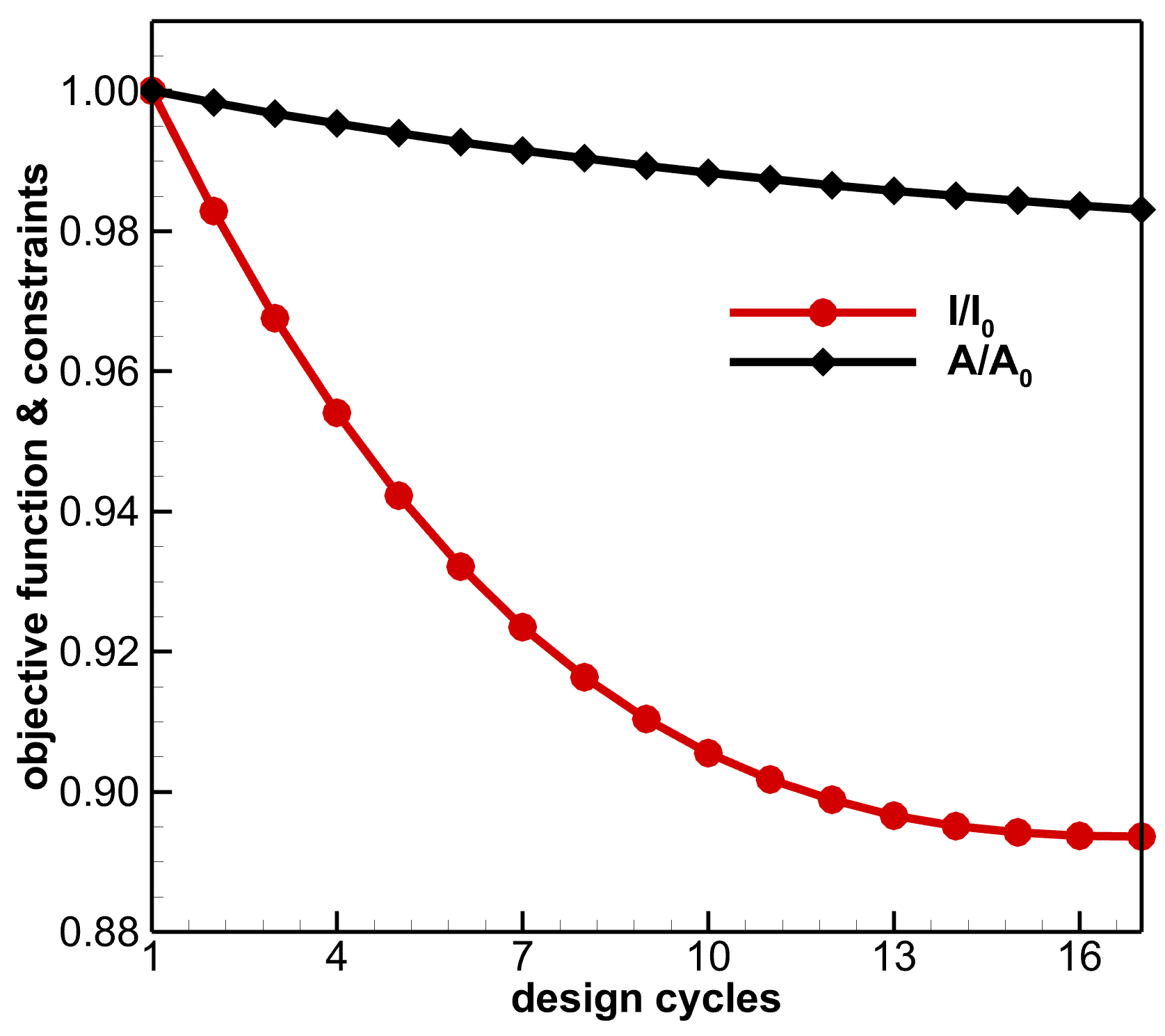}
 \label{semi-cylinder_obj}
 }
 \subfigure[Shape]{
	\includegraphics[width=1.2in]{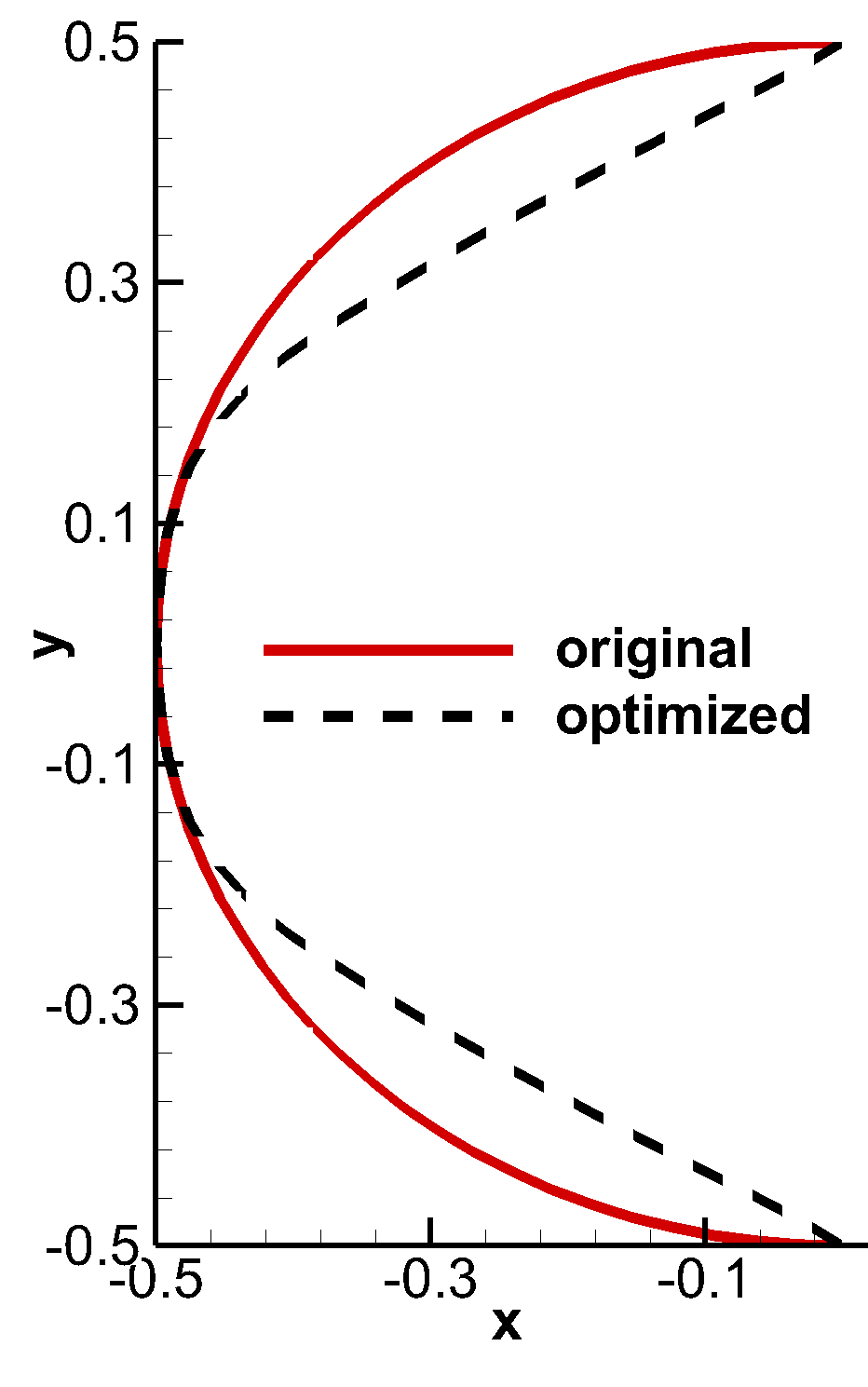}
 \label{semi-cylinder_shape}
 }
	\caption{Aerodynamic shape optimization of the semi-cylinder: a) evolutionary histories; b) original and optimized shapes.}
\end{figure}
\begin{figure}[h!]
\centering
\subfigure[]{
	\includegraphics[width=1.4in]{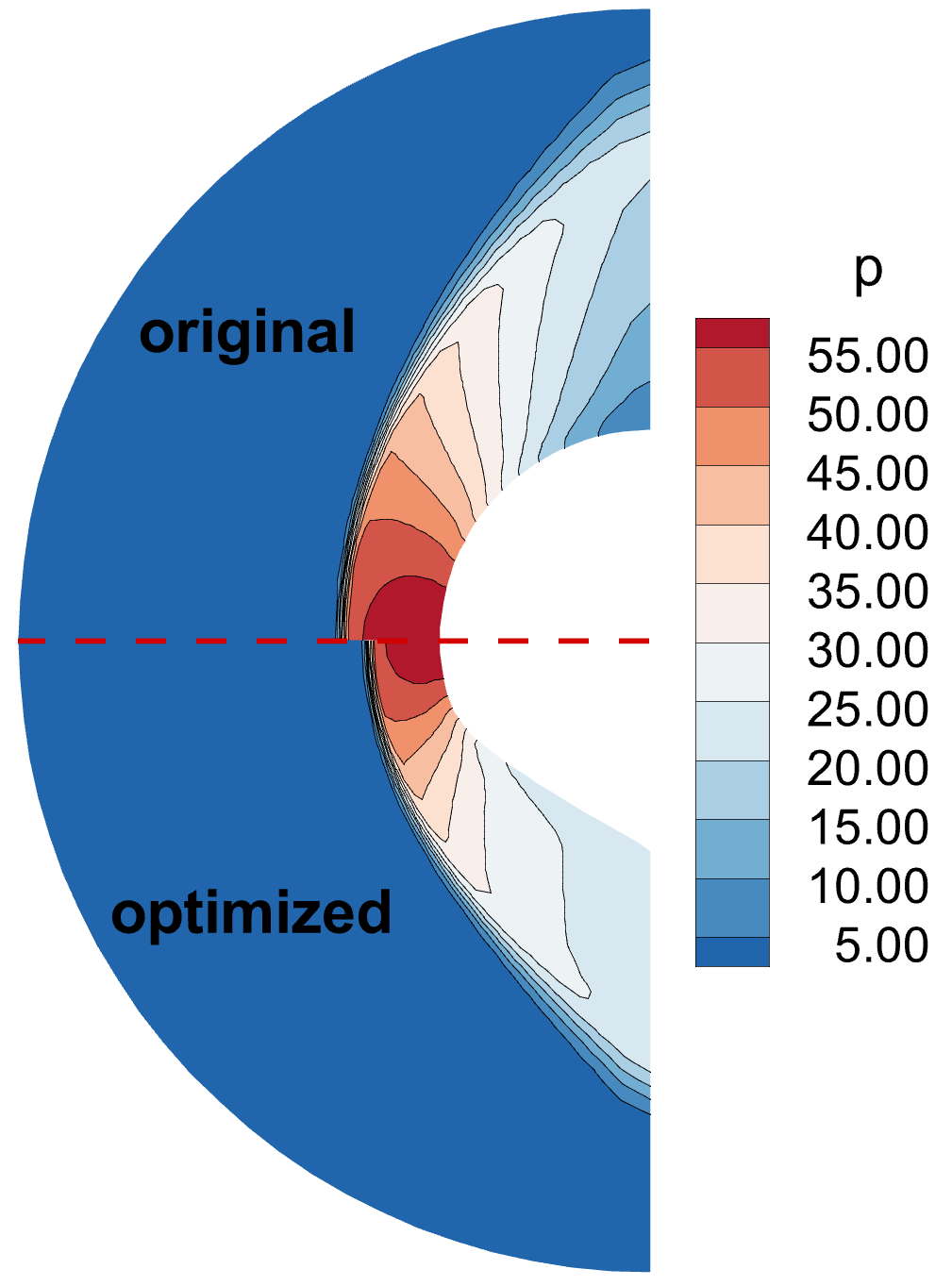}
 \label{semi-cylinder_p}
 }
 \subfigure[]{
	\includegraphics[width=3.0in]{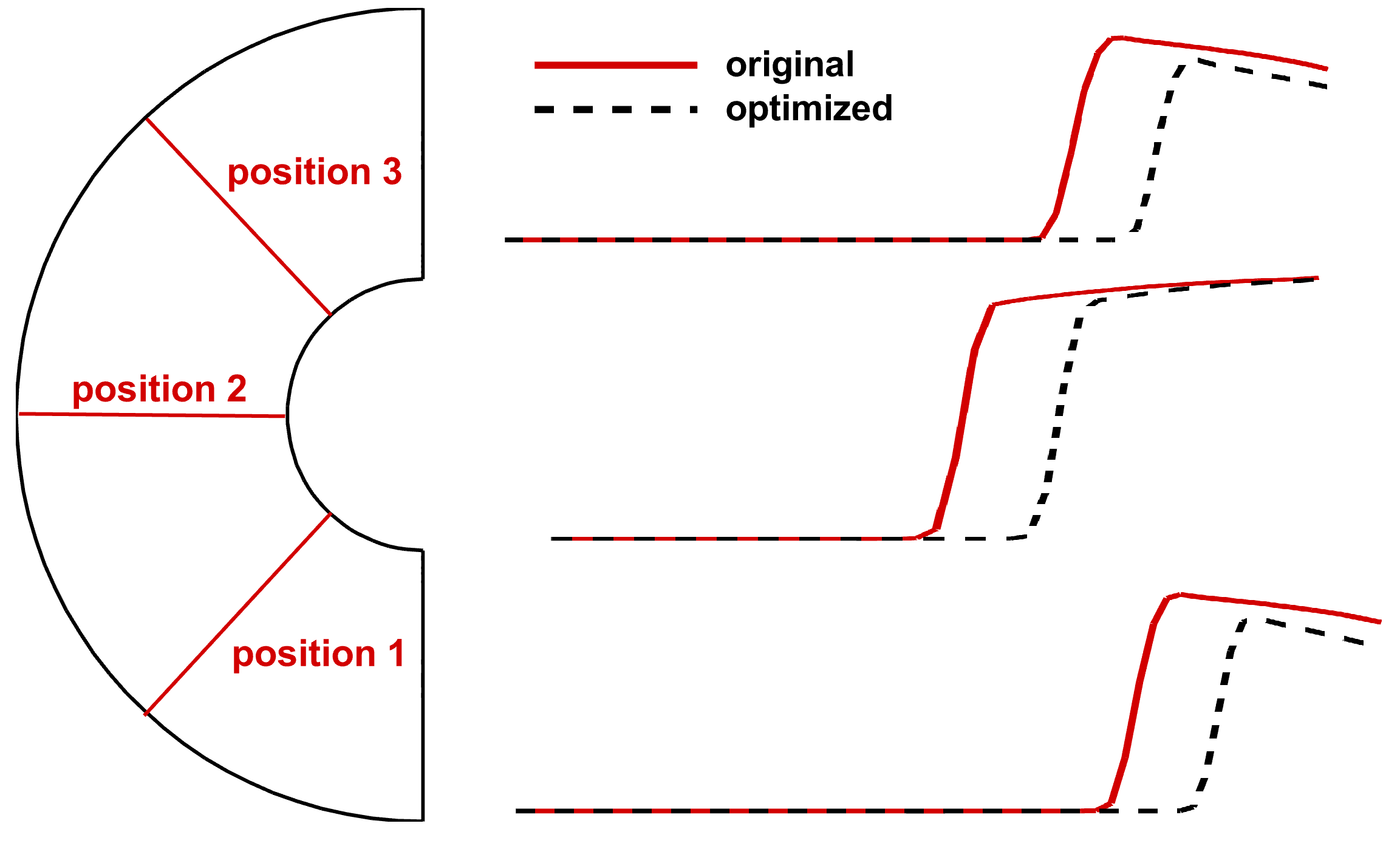}
 \label{semi-cylinder_p-axial}
 }
	\caption{Comparison of the static pressure: a) in the whole computational domain; b) at three different locations.}
\end{figure}

Finally, design optimization is performed. Figure~\ref{semi-cylinder_obj} presents the evolution histories of the objective function and the semi-cylinder area. After seventeen design cycles, the drag is reduced by 11\%, while the reduction in area remains within 2\%.
Figure~\ref{semi-cylinder_shape} compares the original and optimized geometries. After optimization, the blunt leading edge becomes sharper, which is favorable for reducing the normal-shock-induced drag ahead of the semi-cylinder. Figure~\ref{semi-cylinder_p} shows the pressure contours for both configurations, where the upper half corresponds to the original shape and the lower half to the optimized one. It is observed that the shock wave moves downstream relative to the original configuration. This trend is further confirmed in Fig.~\ref{semi-cylinder_p-axial}, which presents the static pressure distributions at three axial locations.

\begin{figure}[h!]
\centering
\subfigure[Cd]{
	\includegraphics[width=2.2in]{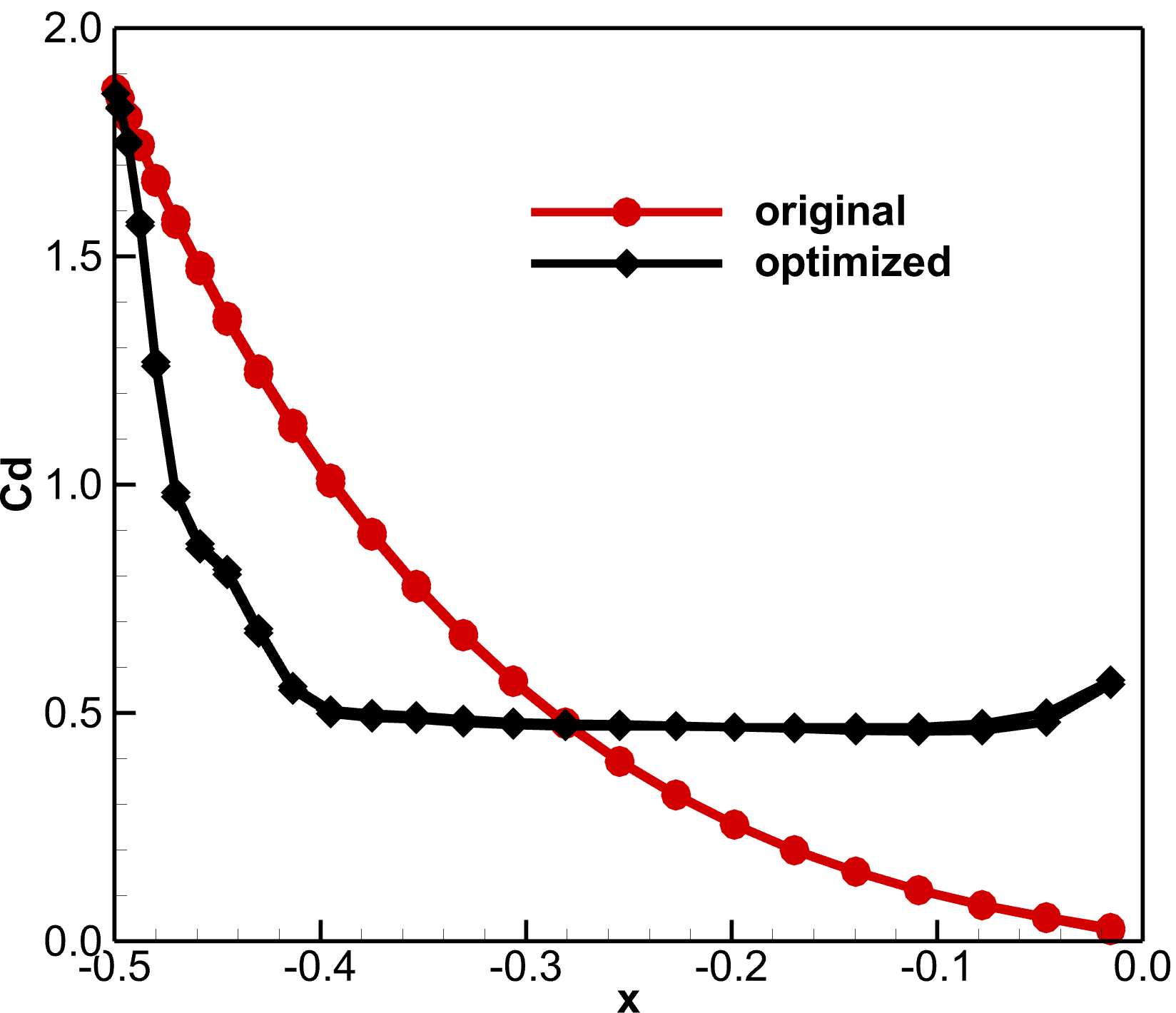}
 \label{semi-cylinder_drag}
 }
 \subfigure[Cp]{
	\includegraphics[width=2.2in]{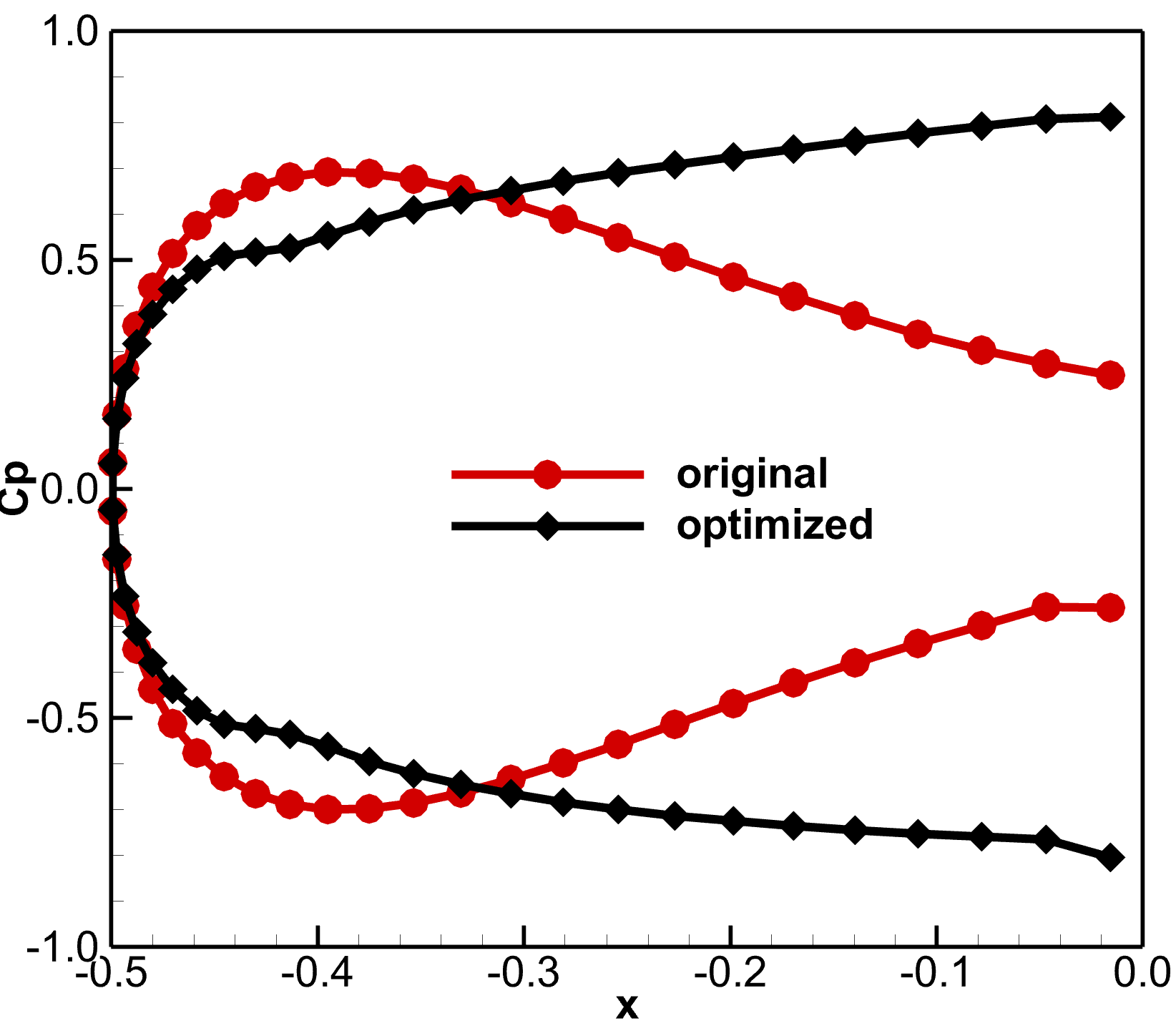}
 \label{semi-cylinder_cp}
 }
	\caption{Comparison of the drag (Cd) and pressure (Cp) coefficients on the semi-cylinder surfaces: a) Cd; b) Cp.}
\end{figure}

Figures~\ref{semi-cylinder_drag} and \ref{semi-cylinder_cp} compare the distributions of drag and pressure coefficients along the semi-cylinder surface for both configurations. For the original geometry, the dominant contribution to drag arises from the leading edge due to the presence of a strong normal shock. After optimization, the drag near the leading edge is significantly reduced. Although the drag exhibits a slight upstream increase from the midsection toward the trailing edge, the overall level is reduced compared to the original case. The pressure coefficient distribution also shows noticeable changes after optimization, with a decrease near the leading edge and an increase over the remaining surface. 

\section{Conclusion}\label{conclusion}

In this work, an implicit discrete adjoint gas-kinetic scheme with kinetic boundary conditions is developed for aerodynamic shape optimization across all Mach-number regimes. The proposed method is validated through four representative test cases, including subsonic flow past a cylinder, transonic flow over the RAE 2822 airfoil, supersonic flow over the NACA 0012 airfoil, and hypersonic flow past a semi-cylinder.
Compared with the explicit approach, which is limited by a maximum Courant number of approximately 0.8, the implicit time-marching method allows for significantly larger Courant numbers (up to 100), leading to a reduction in computational cost by approximately a factor of four for both the flow and adjoint solvers. Moreover, the adjoint fields exhibit structures that are opposite to those of the corresponding flow fields, providing qualitative validation of the correctness of the discrete adjoint formulation.
From a quantitative perspective, the computed adjoint sensitivities show excellent agreement with finite-difference results, with negligible discrepancies, thereby confirming the accuracy of the proposed adjoint GKS solver. In the optimization studies, the adjoint-based framework efficiently yields improved designs within a limited number of optimization iterations, demonstrating high efficiency across all test cases.
Overall, the developed implicit adjoint GKS solver provides an accurate and efficient framework for aerodynamic shape optimization over a wide range of Mach-number regimes in continuum flows.  

 \section*{Acknowledgments} 
This work was supported by the National Key R\&D Program of China (Grant No. 2022YFA1004500), the National Natural Science Foundation of China (Nos. 92371107), and the Hong Kong Research Grant Council (Nos. 16208324). 

\bibliographystyle{elsarticle-num} 
\bibliography{main}

\end{document}